\documentclass[apj,iop,numberedappendix]{emulateapj}
\usepackage{bm}
\usepackage[colorlinks,citecolor=blue,linkcolor=blue,urlcolor=blue]{hyperref}
\usepackage{lineno}
\usepackage{graphicx}
\usepackage[normalem]{ulem}

\newcommand{\be}{\begin{equation}}
\newcommand{\ee}{\end{equation}}
\newcommand{\bea}{\begin{eqnarray}}
\newcommand{\eea}{\end{eqnarray}}

\begin{document}

\title{Kinetic simulations of~instabilities and~particle acceleration in~cylindrical magnetized relativistic jets}
\shorttitle{Kinetic simulations of~instabilities in~magnetized jets}

\author{Jos\'e~Ortu\~no-Mac\'ias$^{\rm 1}$,
Krzysztof~Nalewajko$^{\rm 1}$,
Dmitri~A.~Uzdensky$^{\rm 2}$,
Mitchell~C.~Begelman$^{\rm 3,4}$,
Gregory~R.~Werner$^{\rm 2}$,
Alexander~Y.~Chen$^{\rm 3}$,
Bhupendra~Mishra$^{\rm 5}$}
\shortauthors{Ortu\~no-Mac\'ias et~al.}

\affiliation{$^{\rm 1}$ Nicolaus Copernicus Astronomical Center, Polish Academy of~Sciences, Bartycka 18, 00-716 Warsaw, Poland\\{\tt jortuno@camk.edu.pl, knalew@camk.edu.pl}
\\
$^{\rm 2}$Center for Integrated Plasma Studies, Department of~Physics, University of~Colorado, 390 UCB, Boulder,  CO 80309-0390, USA
\\
$^{\rm 3}$JILA, University of~Colorado and National Institute of~Standards and Technology, 440 UCB, Boulder, CO 80309-0440, USA
\\
$^{\rm 4}$Department of~Astrophysical and Planetary Sciences, University of~Colorado, 391 UCB, Boulder, CO 80309, USA
\\
$^{\rm 5}$Los Alamos National Laboratory, Los Alamos, NM 87545, USA}

\begin{abstract}
Relativistic magnetized jets, such as those from AGN, GRBs and XRBs, are susceptible to current- and pressure-driven MHD instabilities that can lead to particle acceleration and non-thermal radiation.  Here we investigate the development of these instabilities through 3D kinetic simulations of
cylindrically symmetric equilibria involving toroidal magnetic fields with electron-positron pair plasma.
Generalizing recent treatments by~\cite{2018PhRvL.121x5101A} and \cite{2020ApJ...896L..31D}, we~consider a~range of~initial structures in~which the~force due~to~toroidal magnetic field is~balanced by~a~combination of~forces due~to~axial magnetic field and~gas pressure.
We~argue that the~particle energy limit identified by~\cite{2018PhRvL.121x5101A} is~due to~the~finite duration of~the~fast magnetic dissipation phase.
We~find a~rather minor role of~electric fields parallel to~the~local magnetic fields in~particle acceleration.
In~all investigated cases a~kink mode arises in~the~central core region with a~growth timescale consistent with the~predictions of~linearized MHD models.
In the~case of~a~gas-pressure-balanced (Z-pinch) profile, we identify a~weak local pinch mode well outside the~jet core.
 We argue that pressure-driven modes are important for relativistic jets, in regions where sufficient gas pressure is produced by other dissipation mechanisms.
\end{abstract}

\keywords{acceleration of~particles --- instabilities --- magnetic reconnection --- methods: numerical --- plasmas --- relativistic processes}

\section{Introduction}

Magnetic fields are thought to~play a~decisive role in~many of~the~most energetic astrophysical phenomena.
Compact accreting objects tend to~accumulate (or generate locally) magnetic fields and sort them out from the~matter \citep[e.g.,][]{2003PASJ...55L..69N}.
This can produce extreme environments in~which
the magnetic energy density locally dominates the~rest-mass density of~matter.
Such relativistic magnetizations can be converted to high Lorentz factors, driving relativistic outflows in~the~form of~collimated jets \citep[e.g.,][]{1984RvMP...56..255B,1992ApJ...394..459L,1994ApJ...426..269B,2006MNRAS.367..375B,2007MNRAS.380...51K,2009ApJ...699.1789T}.
Luminous non-thermal emission with photon energies extending into the~gamma-ray band is a~key observational signature of~such environments \citep[e.g.,][]{2011Sci...331..736T,2011Sci...331..739A,2016ARA&A..54..725M};
it is evidence of~efficient non-thermal acceleration of~particles to~ultra-relativistic energies.
A likely mechanism for particle acceleration in~such environments is relativistic magnetic reconnection
\citep[e.g.,][]{1971CoASP...3...80M,1992A&A...262...26R,1996A&A...311..172L,2002A&A...391.1141D,2004PhRvL..92r1101K,2011ApJ...737L..40U,2015MNRAS.450..183S,2016ApJ...816L...8W,2018MNRAS.473.4840W,2021JPlPh..87f9013W},
developing from large gradients or~reversals in~the~magnetic field.
Dissipation of~magnetic energy depends crucially on the~magnetic topology and~the~stability of~the~plasma configuration.

Axially symmetric magnetic field configurations
may involve ordered poloidal and toroidal components.
As they expand over several orders of~magnitude in~distance, the~poloidal component decays faster than the~toroidal component.
Even if the configurations are initially dominated by~the~poloidal component, the~toroidal component can in~principle become dominant at some point.
Toroidal magnetic fields are well known to~be unstable to~either current-driven or pressure-driven  pinch (sausage) and kink modes \citep[e.g.,][]{1954RSPSA.223..348K,1966RvPP....2..153K,1982RvMP...54..801F}.
These instabilities have been proposed to~be the~trigger of~magnetic dissipation in relativistic jets of active galactic nuclei (AGNs) and gamma-ray bursts (GRBs), and in~pulsar wind nebulae \citep{1998ApJ...493..291B,2002A&A...391.1141D,2006A&A...450..887G}.

Most analytic studies of~the~stability of~relativistic magnetized jets have been performed in~cylindrical geometry \citep[e.g.,][]{1994MNRAS.267..629I,1998ApJ...493..291B,1999MNRAS.308.1006L,2000A&A...355..818A,2001PhRvD..64l3003T,2012MNRAS.427.2480N,2013MNRAS.434.3030B,2019MNRAS.482.2107D,2019MNRAS.485.2909B}\footnote{However, the~effects of~jet collimation can be important \citep[e.g.,][]{2009ApJ...697.1681N}.}.
This allows one to~introduce a~cylindrical coordinate system $(r,\phi,z)$, in~which all equilibrium parameters depend solely on $r$.
Various assumptions have been adopted on the~radial profile of~the~toroidal magnetic field $B_\phi(r)$, the~presence of~the~axial (poloidal) magnetic field $B_z(r)$,
and crucially on the~radial force balance, which may involve contributions from the magnetic field, gas pressure, centrifugal force, and radial shear of~the~axial velocity.

One line of~research has been to~adopt the~force-free (FF) approximation, in~which contributions from the~gas pressure or inertia are neglected, and 
 the $(\bm{j}\times\bm{B})/c$ force density due to $B_{\phi}(r)$ is balanced by that due to $B_z(r)$
(the \emph{screw-pinch} configuration).
In the~FF limit, it has typically been found that the~pinch mode (with azimuthal wavenumber $m = 0$) is stable, and that the~dominant unstable mode is the~\emph{global} (or \emph{external}, with axial wavelength $\lambda_z$ comparable to~the~jet radius~$R_{\rm j}$) $m = 1$ kink mode \citep[e.g.,][]{1994MNRAS.267..629I,1999MNRAS.308.1006L,2000A&A...355..818A,2013MNRAS.434.3030B}.

Another line of~research has been
 to~balance the force due to $B_\phi(r)$ with
gas pressure gradients, even without any axial magnetic fields ($B_z = 0$).
In this \emph{Z-pinch} configuration, the~most unstable modes are \emph{internal} (with short axial wavelengths $\lambda_z \ll R_{\rm j}$), they can also be \emph{local}
(localized at large radii as compared with the~wavelength $\lambda_z \ll r$),
and can~be either pinch or~kink modes (as~the~growth rate depends weakly on~$m$) \citep{1998ApJ...493..291B,2012MNRAS.427.2480N,2019MNRAS.482.2107D}.

The stability and non-linear evolution of~relativistic jets have~been investigated numerically using 3D relativistic MHD (RMHD) simulations.
The~simplest approach has been to~consider a~static, cylindrically symmetric column representing the~innermost jet region in~its co-moving frame.
Such simulations have been performed in~both 
the~FF limit
\citep{2009ApJ...700..684M,2019ApJ...884...39B,2021MNRAS.505.2267M,2022MNRAS.510.2391B}
and~the~Z-pinch limit \citep{2011ApJ...728...90M}.
These two regimes have been compared 
in~the~work of~\cite{2012MNRAS.422.1436O}, which emphasized a~dramatically more disruptive outcome of~instability in~the~Z-pinch case,
a~result well known in~fusion plasma physics \citep[e.g.,][]{1982RvMP...54..801F}.
Further studies in~the~FF regime considered the~effects of~radial shear of~the~axial velocity \citep{2011ApJ...734...19M,2014ApJ...784..167M} or rotation about the~symmetry axis \citep{2012ApJ...757...16M,2016ApJ...824...48S}.

Global 3D MHD and RMHD simulations resulting in~the~development of~current-driven instabilities have been performed for non-relativistic and relativistic jets
 in~AGNs \citep{2004ApJ...617..123N,2008A&A...492..621M,2009MNRAS.394L.126M,2010MNRAS.402....7M,2016MNRAS.461L..46T},
gamma-ray bursts \citep{2016MNRAS.456.1739B},
and pulsar wind nebulae \citep{2013MNRAS.436.1102M,2014MNRAS.438..278P}.
These simulations demonstrated consistently that the~dominant modes are either pinch or kink.

MHD simulations are able to~provide limited information about non-thermal particle acceleration, e.g., in the test particle approximation \citep[e.g.,][]{2021MNRAS.508.2771P}.
However, in order to fully account for kinetic effects, the particle-in-cell (PIC) algorithm is the method of choice.
Recently, the~results of~the~first 3D kinetic collisionless PIC simulations of~static cylindrical columns with  relativistically strong toroidal magnetic fields in~pair plasmas have been reported.
In the~work of~\cite{2018PhRvL.121x5101A}, a~radial profile of~toroidal magnetic field with an exponentially decaying outer tail was balanced entirely by~gas pressure (the Z-pinch or the~screw-pinch with uniform axial field $B_z$), which resulted in~internal unstable modes.\footnote{A~related work of~\cite{2019PhPl...26g2105A} investigated the~same magnetic configuration in~non-relativistic electron-ion plasmas, in~some cases including the~effect of~Coulomb collisions.}
On the~other hand, in~the~work of~\cite{2020ApJ...896L..31D}, toroidal magnetic field with a~power-law tail (approximately $B_\phi \propto r^{-1}$) was balanced by~a~non-uniform axial field $B_z$ in~the~FF screw-pinch configuration with subrelativistically-warm plasma, which resulted in~external unstable modes.

\cite{2018PhRvL.121x5101A,2019PhPl...26g2105A} demonstrated non-thermal particle
acceleration associated with pressure-driven modes.
 They obtained power-law energy distributions
${\rm d}N/{\rm d}\gamma \propto \gamma^{-p}$
\footnote{Here, $\gamma = (1-\beta^2)^{-1/2} = \mathcal{E}/mc^2$ is the~Lorentz factor of~a~particle with mass $m$, energy $\mathcal{E}$, and dimensionless velocity $\bm\beta = \bm{v}/c$, with $c$ the~speed of~light.}
with the index $p \sim 2\,\text{---}\,3$.
 These distributions extended to~the~maximum energy of~$\gamma_{\rm max} \simeq (1.6\,\text{---}\,1.9)\gamma_{\rm lim}$, where
\be
\gamma_{\rm lim} \equiv \frac{|q|B_0R_0}{mc^2}
\ee
is the~energy limit\footnote{Here, $q$ the~particle electric charge, and $B_0$ roughly the~peak value of~$B_\phi(r)$ attained at the~characteristic radius $R_0$.}
(corresponding to~the~Hillas criterion; \citealt{1984ARA&A..22..425H}),
referred to~as the~\emph{confinement energy} (see Appendix~\ref{app_conf}).
This limit has been tentatively confirmed by~\cite{2020ApJ...896L..31D}, who also found power-law particle energy distributions with $p \sim 3\,\text{---}\,5$ and maximum energy of~$\gamma_{\rm max} \simeq \gamma_{\rm lim}/6$.

Both \cite{2018PhRvL.121x5101A,2019PhPl...26g2105A} and \cite{2020ApJ...896L..31D} investigated the~nature of~electric fields accelerating particles in~unstable cylindrical jets, using the~electric field component $\bm{E}_\parallel = (\bm{E}\cdot\bm{B})\bm{B}/B^2$ parallel to~the~local magnetic field $\bm{B}$ as a~proxy for the~non-ideal
 electric field component $\bm{E}_{\rm nonid} = \bm{E} - \bm{B} \times \bm\beta_{\rm b}$, where $\bm\beta_{\rm b} = \bm{v}_{\rm b}/c$ is the~bulk velocity $\bm{v}_{\rm b}$ in~units of~$c$.
In the
gas-pressure-balanced
configurations investigated by~\cite{2018PhRvL.121x5101A,2019PhPl...26g2105A}, 
perpendicular electric fields $\bm{E}_\perp = \bm{E} - \bm{E}_\parallel$ dominate particle acceleration.
On the~other hand, in~the~$B_z(r)$-supported FF configurations investigated by~\cite{2020ApJ...896L..31D}, it~has~been argued that both perpendicular and~parallel fields contribute
to~particle acceleration,
the~latter due to~finite-guide-field reconnection.

In this work, we~introduce a~new radial profile of~toroidal magnetic field that approximates a power-law,
$B_\phi(r) \propto r^{\alpha_{\rm B\phi}}$,
with the~toroidal field index $\alpha_{\rm B\phi} \le 0$.
 The force due to $B_\phi(r)$ is~balanced initially by~combinations of~forces due to~axial magnetic field and~gas pressure, using a~single pressure mixing parameter $f_{\rm mix}$ to~transition between the~FF screw-pinch configuration with no~gas pressure gradients ($f_{\rm mix} = 0$) and~the~Z-pinch configuration with $B_z = 0$ ($f_{\rm mix} = 1$).
We~perform a~series of~3D kinetic PIC simulations
of~relativistic collisionless pair plasmas
using this setup on a~regular Cartesian grid with~periodic boundaries.
We investigate the~effects of~the~key parameters $f_{\rm mix}$ and $\alpha_{\rm B\phi}$ on the~development of~instabilities and the~resulting particle acceleration.

Section~\ref{sec_config} describes the~initial configuration used in~our simulations.
 The~presentation of our results begins from introducing basic details on~our reference simulation in~Section~\ref{sec_res_ref}.
Section~\ref{sec_res_modes} presents 
the~results at~the~fluid level focused on~the~instability modes.
In~Section~\ref{sec_res_modes_m} we~compare the~strengths and~linear growth time scales of~the~fundamental azimuthal
modes in~the~distributions of~the~axial electric field component~$E_z$.
In~Section~\ref{sec_res_modes_fftz} we~compare the~effective axial wavelengths
of~the~$E_z$ distributions.
Section~\ref{sec_res_accel} presents the~results at~the~kinetic level focused on~the~particle acceleration.
In~Section~\ref{sec_res_accel_glim} we~investigate the~maximum energies achieved by~particles in~our simulations.
In~Section~\ref{sec_res_accel_Epara} we~investigate the~role of~parallel electric fields in~particle acceleration.
Section~\ref{sec_disc} contains a~discussion of~our results,
and~Section~\ref{sec_conc} provides the~conclusions.

\section{Initial configuration}
\label{sec_config}

We performed a~set of~3D kinetic PIC simulations in~electron-positron pair plasma, using a~modified version of~the~public numerical code {\tt Zeltron} \citep{Cer13}.

In our standard collisionless implementation of the PIC method, the electric and magnetic fields $\bm{E},\bm{B}$ are represented on a staggered Cartesian Yee grid, and the gas is represented by individual macroparticles, from which a current density field $\bm{j}$ is calculated and deposited on the Yee grid using a charge conserving scheme of \cite{Esi01}.
The $\bm{E},\bm{B}$ fields are advanced in time by solving directly the Amp\`ere-Maxwell and Maxwell-Faraday equations:
\bea
\frac{\partial\bm{E}}{\partial t} &=& c\bm\nabla\times \bm{B} - 4\pi\bm{j}\,,
\\
\frac{\partial\bm{B}}{\partial t} &=& -c\bm\nabla \times\bm{E}\,,
\eea
using a simple leapfrog algorithm (assuring the satisfaction of $\bm\nabla\cdot\bm{B} = 0$ with numerical accuracy).
The particle positions and momenta are advanced in time in a leapfrog scheme, using the Vay pusher algorithm for advancing the momenta under the local Lorentz force \citep{Vay08}.
Our implementation is nominally of second-order accuracy, although the main source of inaccuracy is the limited number of macroparticles per grid cell.

\begin{figure*}
\includegraphics[width=\textwidth]{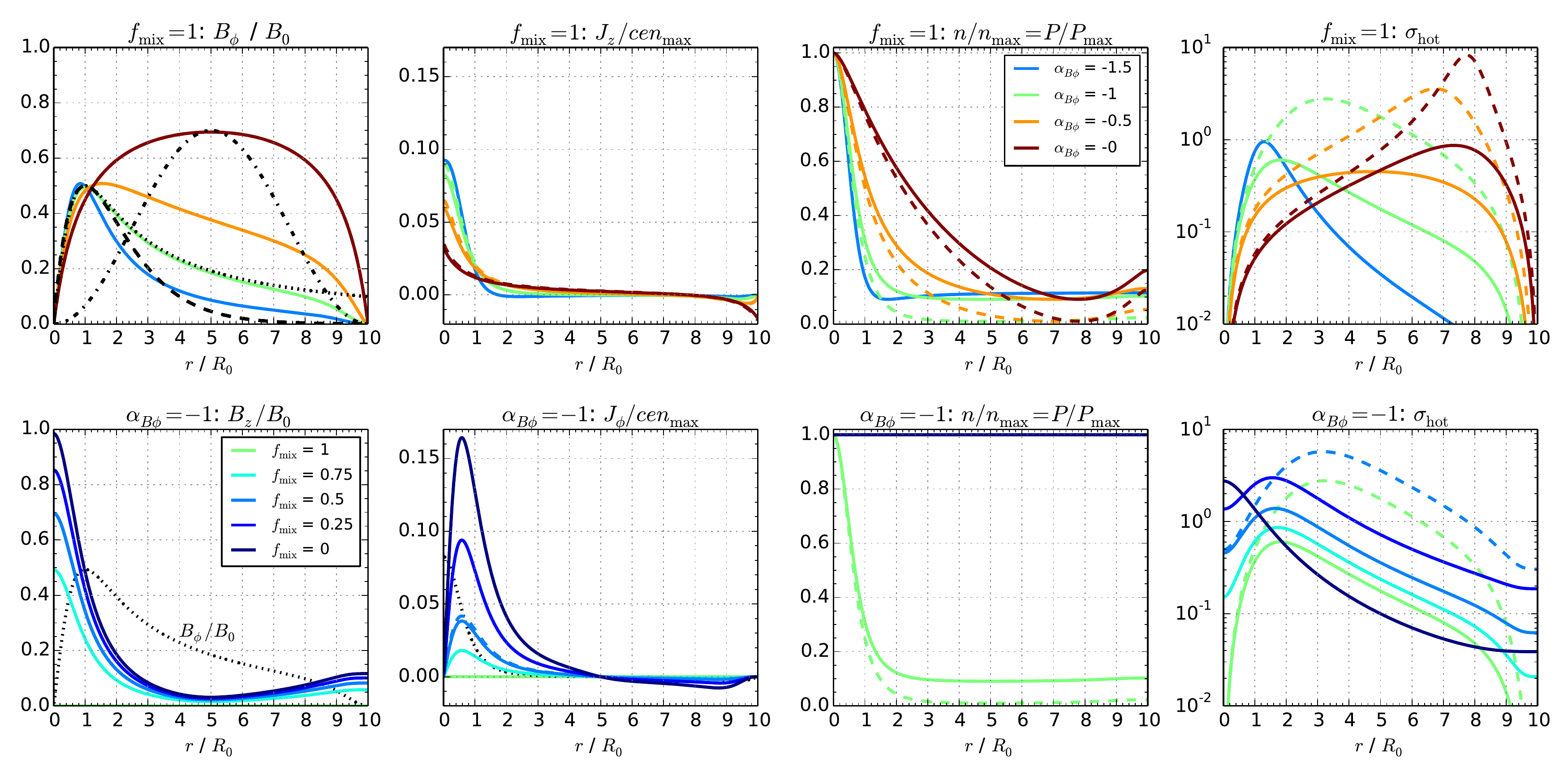}
\caption{\emph{Upper row of~panels}: initial configurations compared for the~series of~simulations with gas pressure  balance ($f_{\rm mix} = 1$) and different toroidal field indices $\alpha_{\rm B\phi}$ (indicated by~line colors defined in~the~legend).
\emph{From left to~right:} radial profiles of~the~initial toroidal magnetic field $B_\phi(r)$ in~units of~$B_0$; radial profiles of~the~axial current density $j_z(r)$ normalized to~$cen_{\rm max}$; radial profiles of~the~gas density $n(r)$ normalized to~$n_{\rm max}$ (proportional to~the~gas pressure $P(r)/P_{\rm max}$); and radial profiles of~the~initial hot magnetization $\sigma_{\rm hot}(r)$.
 For comparison with the configuration of our simulations, in the~upper left panel we also show the~`constant pitch' profile simulated by~\cite{2009ApJ...700..684M}, \cite{2019ApJ...884...39B} and \cite{2020ApJ...896L..31D} (dotted black line); the~exponentially decaying profile of~\cite{2018PhRvL.121x5101A,2019PhPl...26g2105A} (dashed black line); and the~`sinusoidal' profile of~\cite{2012MNRAS.422.1436O} (dash-dotted black line).
\emph{Lower row of~panels:} initial configurations compared for the~series of~simulations with the~same toroidal field index $\alpha_{\rm B\phi} = -1$ and different pressure mixing parameters $f_{\rm mix}$ (indicated by~line colors defined in~the~legend).
\emph{From left to~right:} radial profiles of~the~initial axial magnetic field $B_z(r)/B_0$ (and the~common profile of~$B_\phi(r)/B_0$ with the~black dotted line); radial profiles of~the~toroidal current density $j_\phi(r)/cen_{\rm max}$ (and the~common profile of~$j_z(r)/cen_{\rm max}$ with the~black dotted line); radial profiles of~the~gas density $n(r)/n_{\rm max} = P(r)/P_{\rm max}$; and radial profiles of~the~initial hot magnetization $\sigma_{\rm hot}(r)$ based on the~total magnetic field.
In both series, the~solid color lines correspond to~the~moderate density ratio of~$\xi_n = 10$, and the~dashed color lines correspond to~the~high density ratio of~$\xi_n = 100$.}
\label{fig_config_fmix1}
\end{figure*}

Our simulations were performed in~a~cubic domain of~physical size $L$ ($-L/2 \le x,y \le L/2$; $0 \le z \le L$).
Introducing a~cylindrical coordinate system ($0 \le r \le R_{\rm out}$; $0 \le \phi < 2\pi$) centered along  the~$x = y = 0$ axis,
one can fit in~this domain an
 axially and translationally symmetric equilibrium with the~outer radius $R_{\rm out} = L/2$.
The equilibrium is based on the~radial profile of~toroidal magnetic field $B_\phi(r)$ in~the~form of~a~power-law with inner and outer cutoffs (see the~upper left panel of~Figure~\ref{fig_config_fmix1}):
\be
B_\phi(r) = B_0 \left(\frac{r}{R_0}\right)^{\alpha_{\rm B\phi}}\,C(\alpha_{\rm B\phi},r)\,C(\alpha_{\rm B\phi},R_{\rm out}-r)\,,
\label{eq_Bphi}
\ee
\bea
C(\alpha_{\rm B\phi},r) = \frac{(r/R_0)^{1-\alpha_{\rm B\phi}}}{1 + (r/R_0)^{1-\alpha_{\rm B\phi}}}\,,
\nonumber
\eea
with the~\emph{toroidal field index} $\alpha_{\rm B\phi} \le 0$ and the~\emph{core radius} $R_0 = R_{\rm out}/10 = L/20$.
These profiles peak at radii $R_{\rm peak}/R_0 \simeq 0.84,1,1.55,5$ for $\alpha_{\rm B\phi} = -1.5,-1,-0.5,0$, respectively.
In order to~achieve the~most consistent scaling of our results with $\alpha_{\rm B\phi}$,
we introduce a~characteristic radius $R_{\rm B\phi}$ that is equal to~$R_{\rm peak}$ for $\alpha_{\rm B\phi} = -1.5,-1,-0.5$ and equal to~$R_{\rm peak}/2$ for $\alpha_{\rm B\phi} = 0$.

The initial equilibrium is provided by~the~axial electric current $j_z(r) = (c/4\pi r)\,{\rm d}(rB_\phi)/{\rm d}r$,
as well as by~a combination of~the~axial magnetic field $B_z(r)$ and the~radial gas pressure $P_{rr}(r)$:
\be
\frac{B_\phi}{4\pi r}\frac{{\rm d}(rB_\phi)}{{\rm d}r}
+\frac{1}{8\pi}\frac{{\rm d}B_z^2}{{\rm d}r}
+\frac{{\rm d}P_{rr}}{{\rm d}r}
= 0\,.
\ee
We introduce a~constant \emph{pressure mixing parameter} $f_{\rm mix} \in [0:1]$,
which is the~fraction of~the~toroidal magnetic stress
that is balanced by~the~gas pressure
(with the~remainder balanced by~the~axial magnetic pressure):
\bea
\frac{{\rm d}P_{rr}}{{\rm d}r} &=& -f_{\rm mix}\frac{B_\phi}{4\pi r}\frac{{\rm d}(rB_\phi)}{{\rm d}r}\,,
\\
\frac{{\rm d}B_z^2}{{\rm d}r} &=& -(1-f_{\rm mix})\frac{2B_\phi}{r}\frac{{\rm d}(rB_\phi)}{{\rm d}r}\,.
\eea
The case 
$f_{\rm mix} = 0$ means that $B_\phi$ is balanced entirely by~$B_z$ (with uniform gas pressure; the~FF screw-pinch),
and the~case 
$f_{\rm mix} = 1$ means that $B_\phi$ is balanced entirely by~the~gas pressure
($B_z = 0$; the~Z-pinch).
For~$\alpha_{B\phi} = -1$ and~$f_{\rm mix} < 1$ ($B_z \ne 0$),
 the~magnetic pitch profile $\mathcal{P}(r) = rB_z(r)/B_\phi(r)$ is almost constant at~$\mathcal{P}_0 = R_0\sqrt{1-f_{\rm mix}}$ for~$r < R_0$, then slightly decreasing with $r$, and then strongly increasing with $r$ for $r > 4.5R_0$.
 In the FF limit $f_{\rm mix} = 0$, we have $\mathcal{P}_0 = R_0$, and since $L/\mathcal{P}_0 = 20 > 2\pi$, the Kruskal-Shafranov stability criterion is not satisfied, hence this configuration is predicted to be unstable \citep[e.g.,][]{2019ApJ...884...39B}.

For~$f_{\rm mix} < 1$, we also introduce a~toroidal component of~electric current density:
\be
j_\phi(r) = (1-f_{\rm mix})\frac{c}{4\pi r}\frac{B_\phi}{B_z}\frac{{\rm d}(rB_\phi)}{{\rm d}r}\,.
\ee

The initial profile of~gas pressure $P_{rr}(r)$
 determines
the~initial profile of~gas density $n(r) = P_{rr}(r) / \Theta_0m_{\rm e}c^2$
 for~a~uniform initial relativistic temperature $\Theta_0 = k_{\rm B}T_0/m_{\rm e}c^2$,
where $m_{\rm e}$ is the~electron mass,
$k_{\rm B}$ is the~Boltzmann constant,
and $T_0$ is the~initial temperature in~kelvins.
In our simulations we have adopted  an ultra-relativistic temperature with $\Theta_0 = 10^4$.
In~this~work we~neglect the~effects of~radiative cooling, leaving this important topic for~a~future study.

In~the~case of~non-uniform gas pressure ($f_{\rm mix} > 0$), the~gas density profile is normalized  by~adding a~constant value to~obtain the~desired \emph{density contrast} $\xi_n \equiv n_{\rm max} / n_{\rm min}$ --- the~ratio of~maximum (central) $n_{\rm max}$ to~minimum $n_{\rm min}$  density values.
Two values have been adopted for the~density contrast: a~moderate value of~$\xi_n = 10$ and a~high value of~$\xi_n = 100$.
The density contrast is particularly important for determining the~radial profile of~the~\emph{hot magnetization} based on the~total magnetic field $\sigma_{\rm hot}(r) = B^2(r) / 4\pi w(r)$, where $w(r) \simeq 4\Theta_0n(r)m_{\rm e}c^2$ is the~relativistic enthalpy density.
 A higher density contrast results in~a~lower gas density and higher magnetization outside the~central core region (see Figure~\ref{fig_config_fmix1}).

The~case $f_{\rm mix} = 0$ requires a~uniform gas pressure, and hence a~uniform gas density, which corresponds to~a~density contrast of~$\xi_n = 1$.
In this case, the~gas density is normalized to~a~value such that the~drift velocity profile $\bm{\beta}_{\rm d}(r) = \bm{j}(r) / cen(r)$ satisfies the~condition that $\max(\beta_{\rm d}) = 0.5$\footnote{ Note that the~axial current density peaks at~$r=0$ at~the~level of~$j_{\rm max} \simeq cB_0/2\pi R_0$, which implies a~lower limit on~the~particle density $n_{\rm max} \simeq j_{\rm max}/ce\beta_{\rm d} \gtrsim B_0/\pi eR_0$, which in~turn implies an~upper limit on~the~magnetization of~$\sigma(r) \lesssim (B/B_0)^2(n/n_{\rm max})^{-1}(R_0/16\rho_0)$, which depends primarily on~the~scale separation between $\rho_0$ and~$R_0$.}.
For all other cases, we make sure that $\beta_{\rm d}(r) \le 0.5$.

The minimum value of~gas pressure is given by~$P_{\rm min} = \Theta_0 n_{\rm min} m_{\rm e}c^2$.
In the~cases that involve axial magnetic field ($f_{\rm mix} < 1$), we also add a~small constant to~$B_z(r)$, such that it 
 satisfies the~relation $\min(B_z^2) = 10^{-3} \max(B_z^2)$.

The key parameters of~the~initial configuration are thus $f_{\rm mix}$, $\alpha_{\rm B\phi}$ and $\xi_n$.
Their values for our main simulations are listed in~Table \ref{table}.
 The~initial radial profiles of~$B_\phi(r), B_z(r), j_\phi(r), j_z(r), n(r) \propto P(r)$ and $\sigma_{\rm hot}(r)$ are compared in~Figure~\ref{fig_config_fmix1}.

Our simulations were performed on a~numerical grid size of~$N = 1152$, with the~numerical resolution 
${\rm d}x = {\rm d}y = {\rm d}z = L/N = \rho_0/1.28$, where $\rho_0 = \Theta_0m_{\rm e}c^2/eB_0$ is the~nominal particle gyroradius with $e$ the~positron charge  and $R_0/\rho_0 = 45$. 
The number of~particles of~both species per grid cell is $16$.
 The~actual set of~macroparticles in~each cell is drawn from 
Lorentz-boosted Maxwell-J\"{u}ttner distributions, the~appropriate moments of~which are~consistent with~$n(r)$, $\bm{\beta}_{\rm d}(r)$ and~$\Theta_0$.
The simulation time~step was~set as~${\rm d}t = 0.99{\rm d}x/\sqrt{3}c$, according to~the~Courant-Friedrichs-Lewy (CFL) condition.
Digital filtering was~applied at~every time~step to~the~current and~charge densities deposited on~the~Yee grid.
Periodic boundary conditions were adopted at~all faces of~the~Cartesian domain.
The~edge regions that would correspond to~$r > R_{\rm out}$ were~initialized with~$B_\phi = 0$, $\bm{j} = 0$ and~uniform $B_z,n,P$.

Our simulations have a~typical duration of~$\sim 5 L/c$. In some cases we interrupt them at~an~earlier time, once the~perturbations produced by~the~instability reach the~$x,y$ domain boundaries.
This is formally defined by~considering whether the~root-mean-square of~the~axial electric field component ${\rm rms}(E_z)$, calculated within an~outer cylindrical shell $9 < r/R_0 < 10$ (cf.~Section~\ref{sec_res_modes_m}), exceeds the~level of~$2\times 10^{-3}B_0$.
Once that happens during an~episode of~rapid growth, the~simulation is~interrupted at~the~end of~that episode.

\begin{table}
\caption{List of~performed simulations with the~key parameters of~the~initial configurations.}
\label{table}
\centering
\vskip 1ex
\begin{tabular}{lcccc}
\hline\hline
label & $f_{\rm mix}$ & $\alpha_{\rm B\phi}$ & $\xi_n$ \\
\hline
  f0\_$\alpha$-1\_$\xi$1    & 0    & -1   & 1 \\
f025\_$\alpha$-1\_$\xi$10   & 0.25 & -1   & 10 \\
 f05\_$\alpha$-1\_$\xi$10   & 0.5  & -1   & 10 \\
 f05\_$\alpha$-1\_$\xi$100  & 0.5  & -1   & 100 \\
f075\_$\alpha$-1\_$\xi$10   & 0.75 & -1   & 10 \\
  f1\_$\alpha$-15\_$\xi$10  & 1    & -1.5 & 10 \\
  f1\_$\alpha$-1\_$\xi$10 (ref) & 1    & -1   & 10 \\
  f1\_$\alpha$-1\_$\xi$100  & 1    & -1   & 100 \\
  f1\_$\alpha$-05\_$\xi$10  & 1    & -0.5 & 10 \\
  f1\_$\alpha$-05\_$\xi$100 & 1    & -0.5 & 100 \\
  f1\_$\alpha$0\_$\xi$10    & 1    &  0   & 10 \\
  f1\_$\alpha$0\_$\xi$100   & 1    &  0   & 100 \\
\hline\hline
\end{tabular}
\end{table}

\section{Results: introducing the~reference case}
\label{sec_res_ref}

\begin{figure*}
\includegraphics[width=0.197\textwidth]{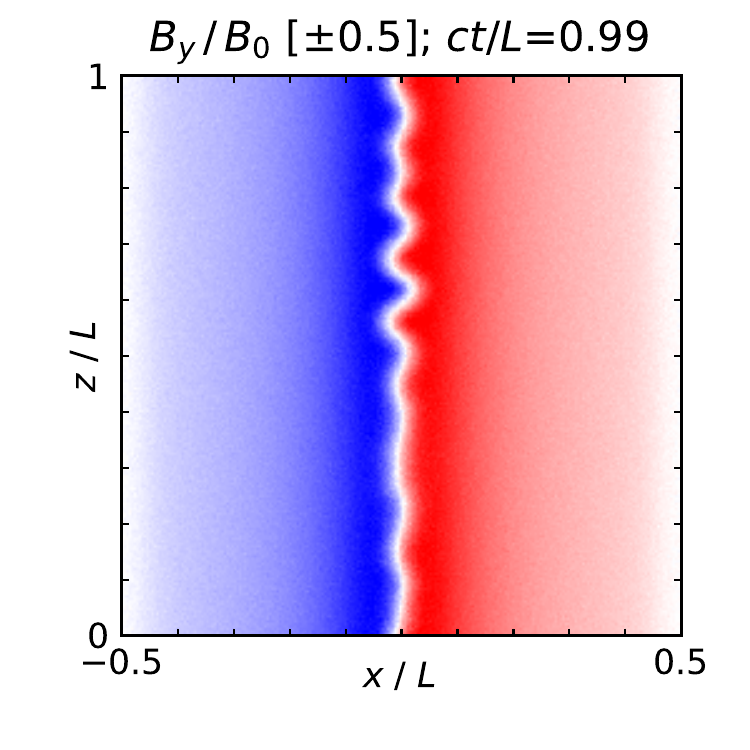}
\includegraphics[width=0.197\textwidth]{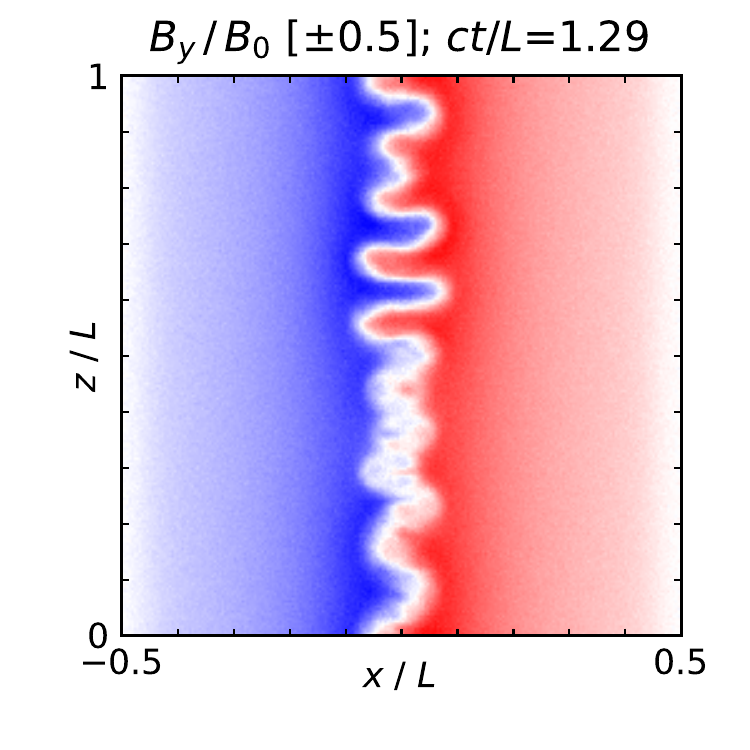}
\includegraphics[width=0.197\textwidth]{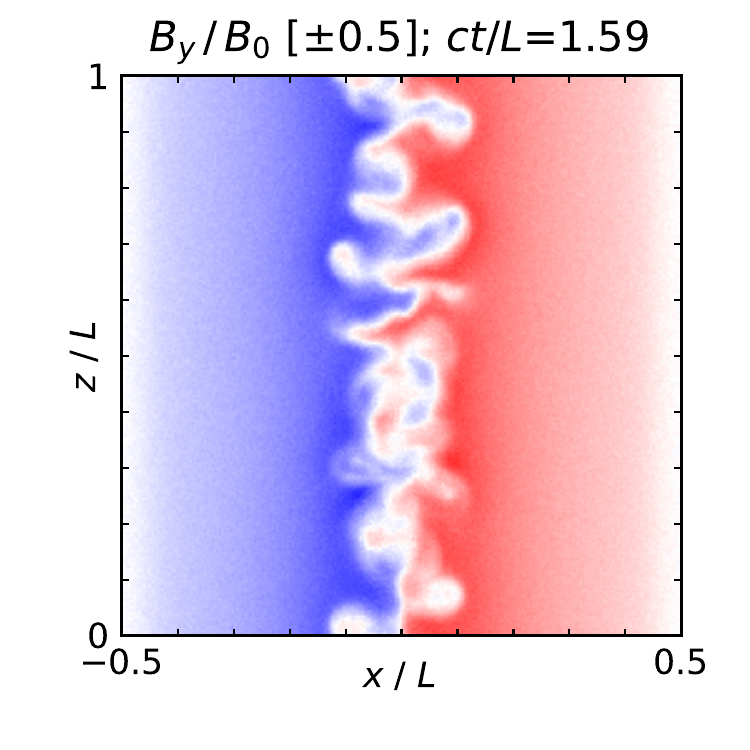}
\includegraphics[width=0.197\textwidth]{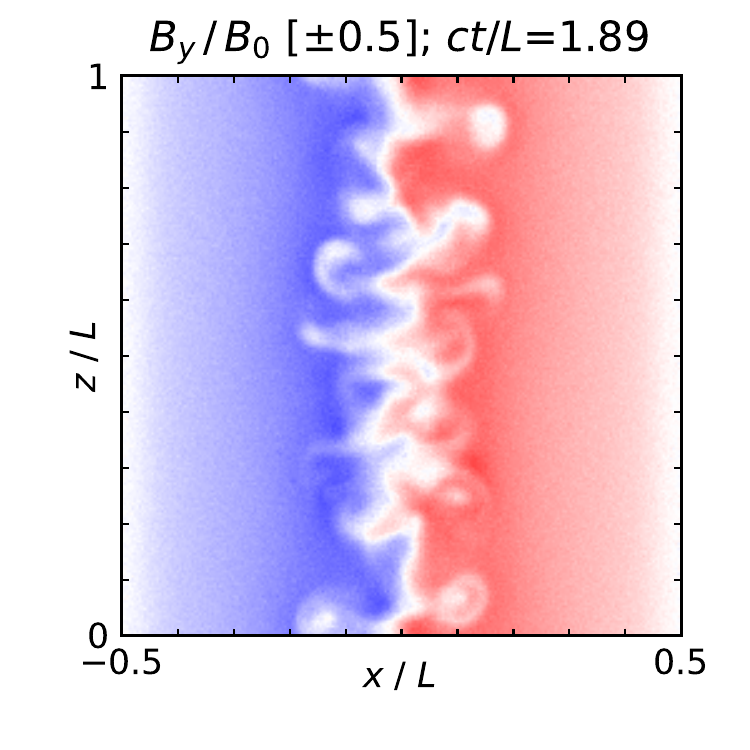}
\includegraphics[width=0.197\textwidth]{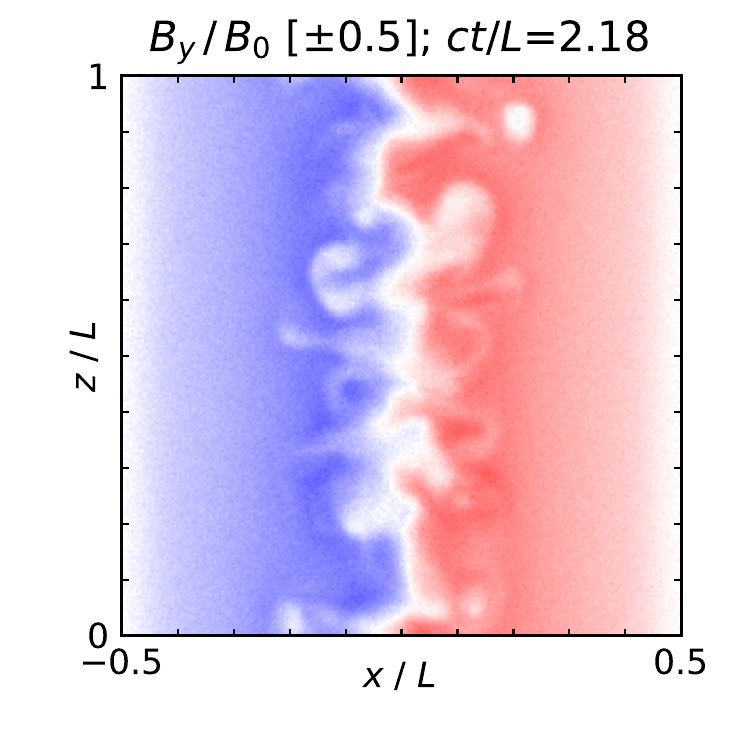}
\\
\includegraphics[width=0.197\textwidth]{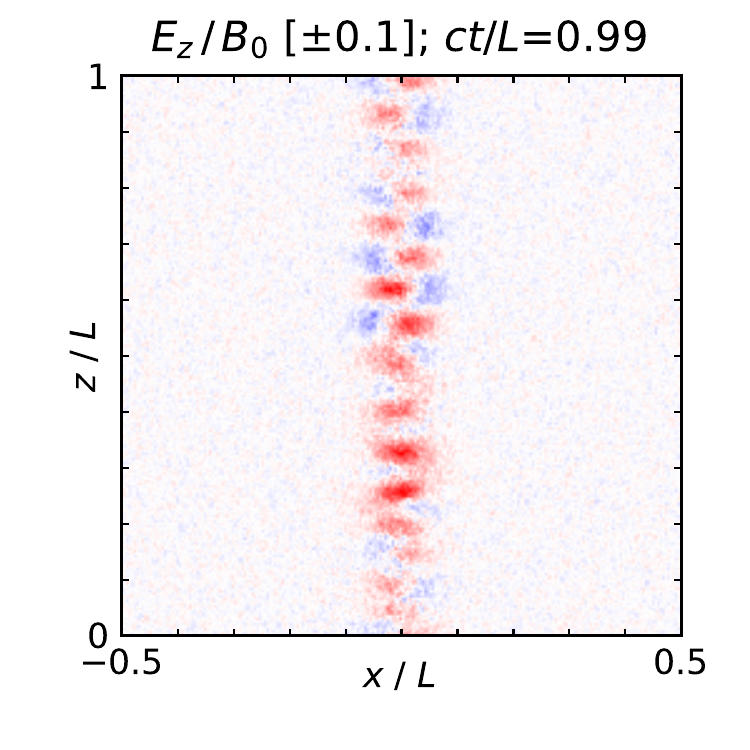}
\includegraphics[width=0.197\textwidth]{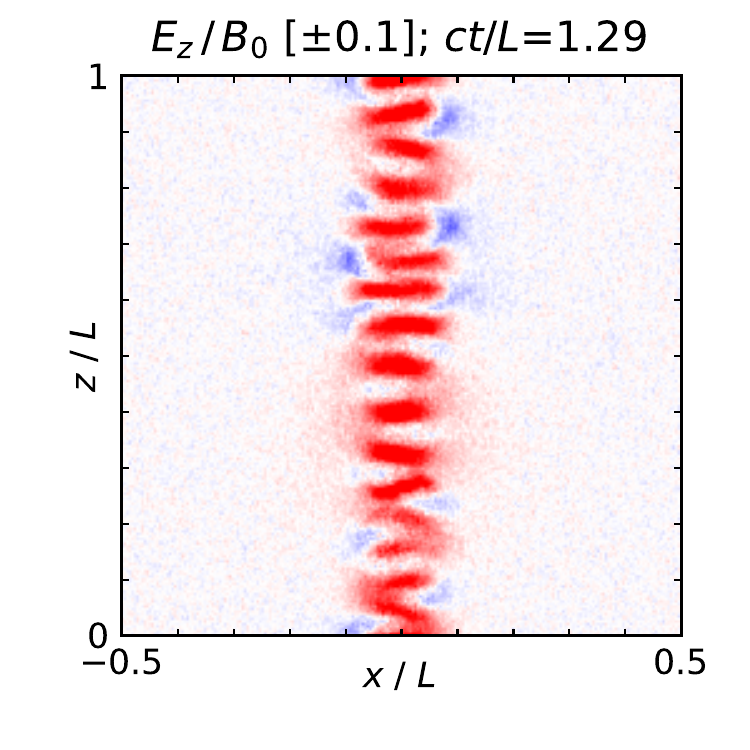}
\includegraphics[width=0.197\textwidth]{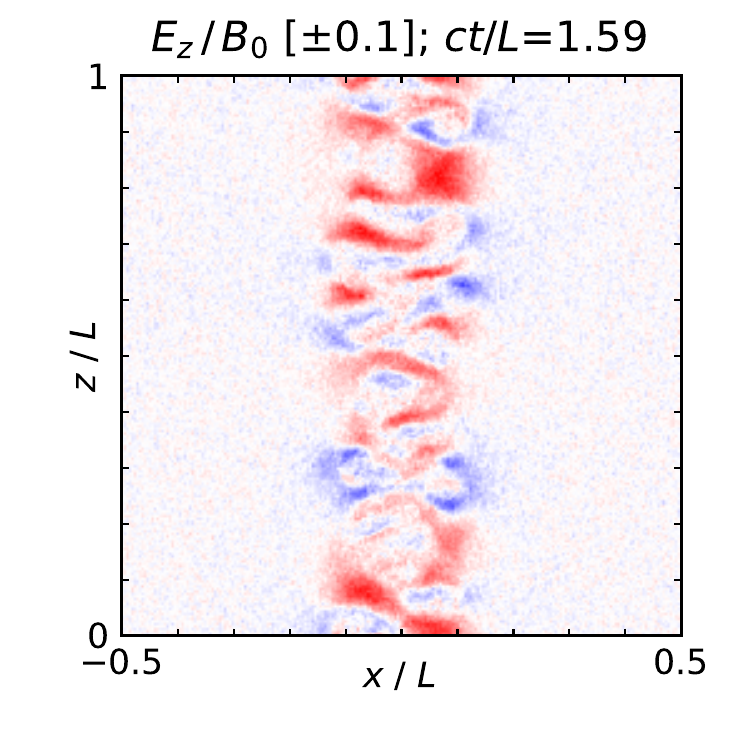}
\includegraphics[width=0.197\textwidth]{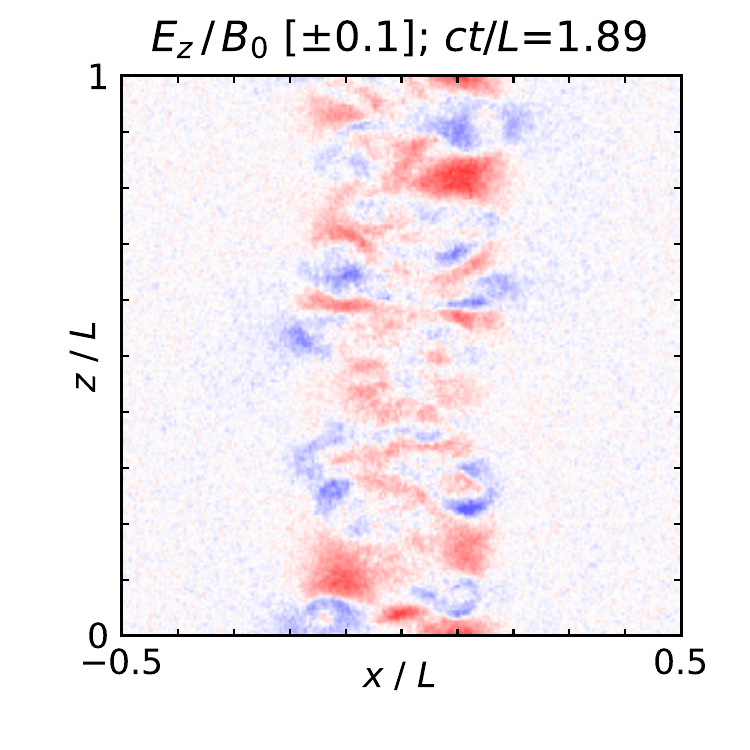}
\includegraphics[width=0.197\textwidth]{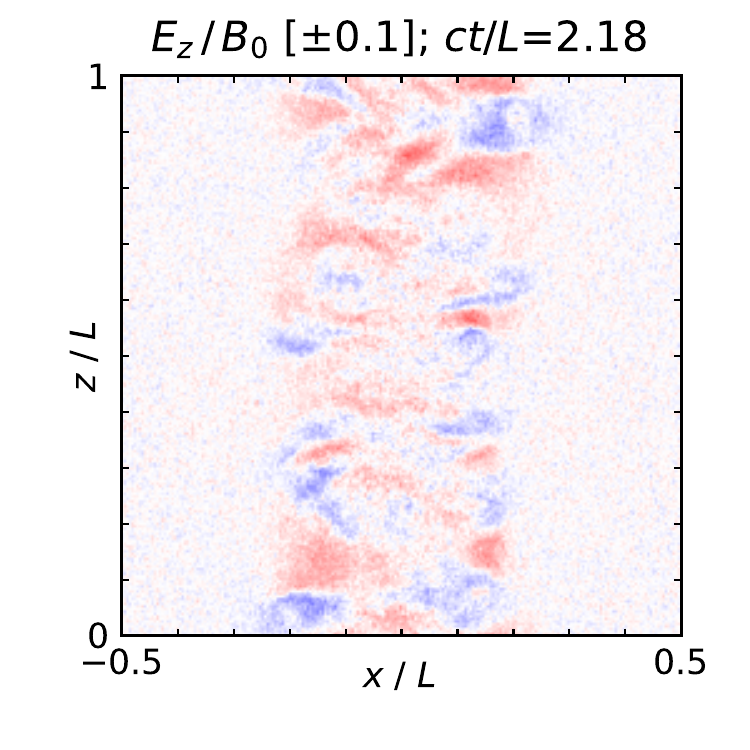}
\\
\includegraphics[width=0.197\textwidth]{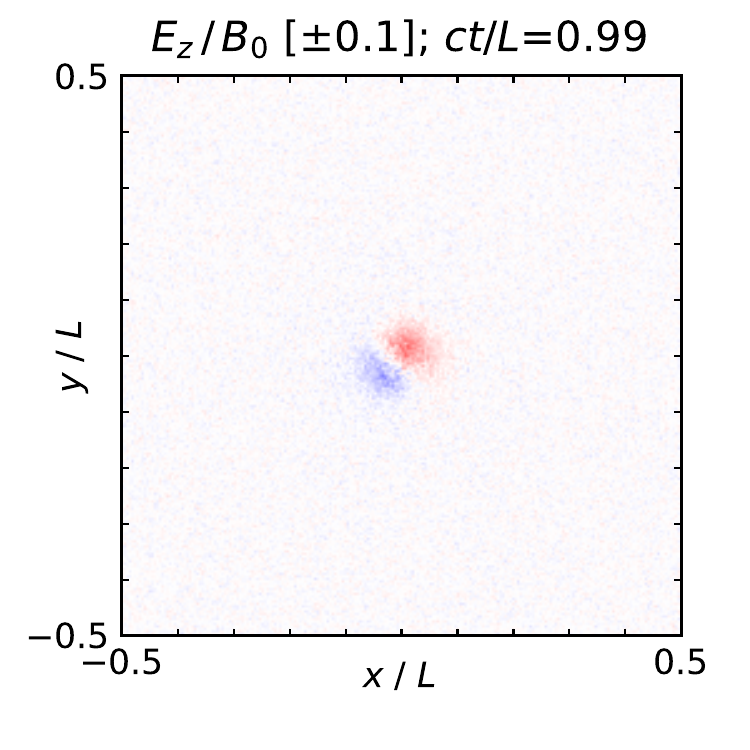}
\includegraphics[width=0.197\textwidth]{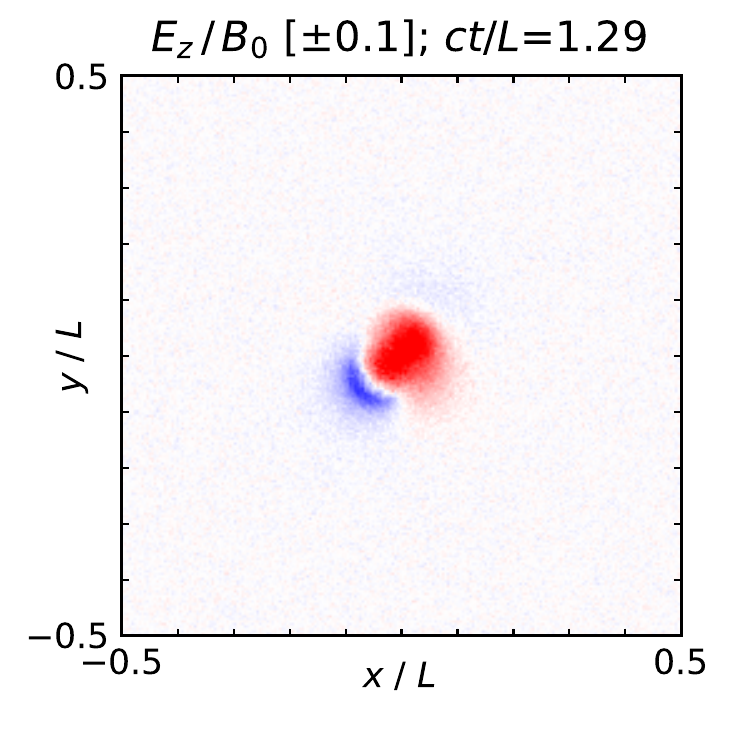}
\includegraphics[width=0.197\textwidth]{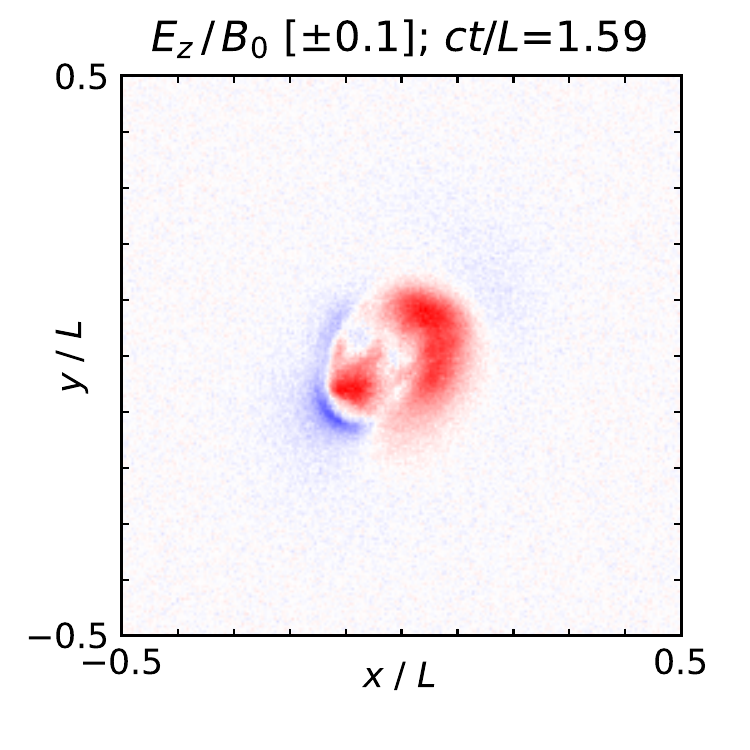}
\includegraphics[width=0.197\textwidth]{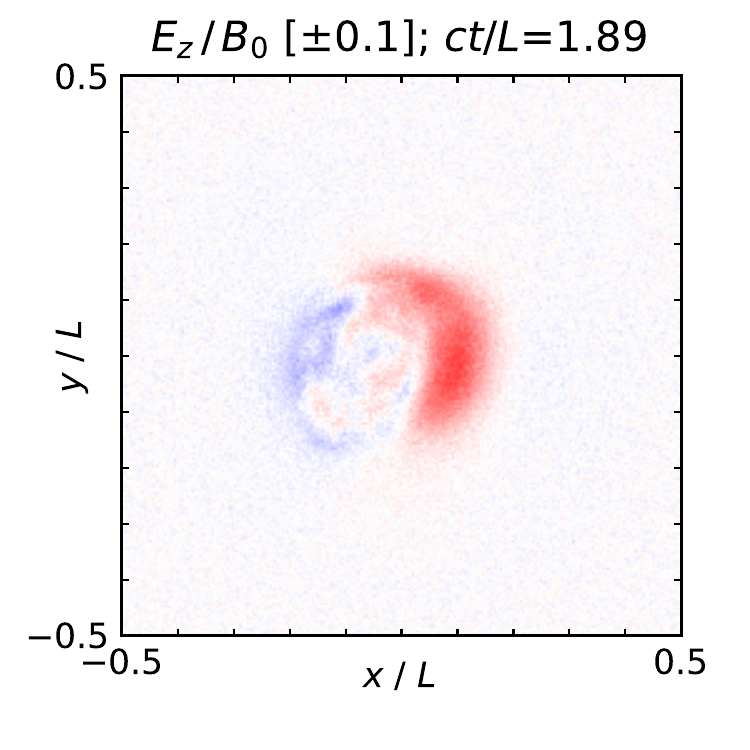}
\includegraphics[width=0.197\textwidth]{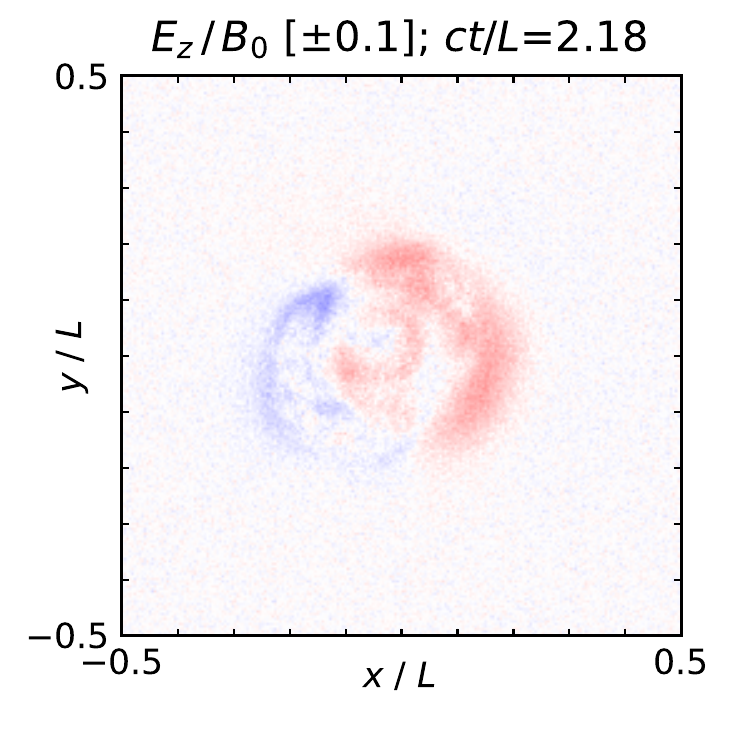}
\caption{Maps of~the~magnetic field component $B_y$ (top row of~panels) and the~electric field component $E_z$ (middle row of~panels) in~the~$y = 0$ plane, as well as the~$E_z$ component in~the~$z = 0$ plane (bottom row of~panels), all in~units of~$B_0$ (positive values in~red, negative in~blue), at regular time intervals (from left to~right) for the~reference simulation f1\_$\alpha$-1\_$\xi$10.}
\label{fig_xzmaps_f1_nr10_aBp1}
\end{figure*}

We begin the~presentation of our results by~introducing some basic details on~the~reference case, which we choose to~be the~simulation f1\_$\alpha$-1\_$\xi$10 with toroidal field index $\alpha_{\rm B\phi} = -1$ balanced entirely by~gas pressure ($f_{\rm mix} = 1$),
which is the~closest case to~the~configurations investigated by~\cite{2018PhRvL.121x5101A}.

\begin{figure*}
\includegraphics[width=0.499\textwidth]{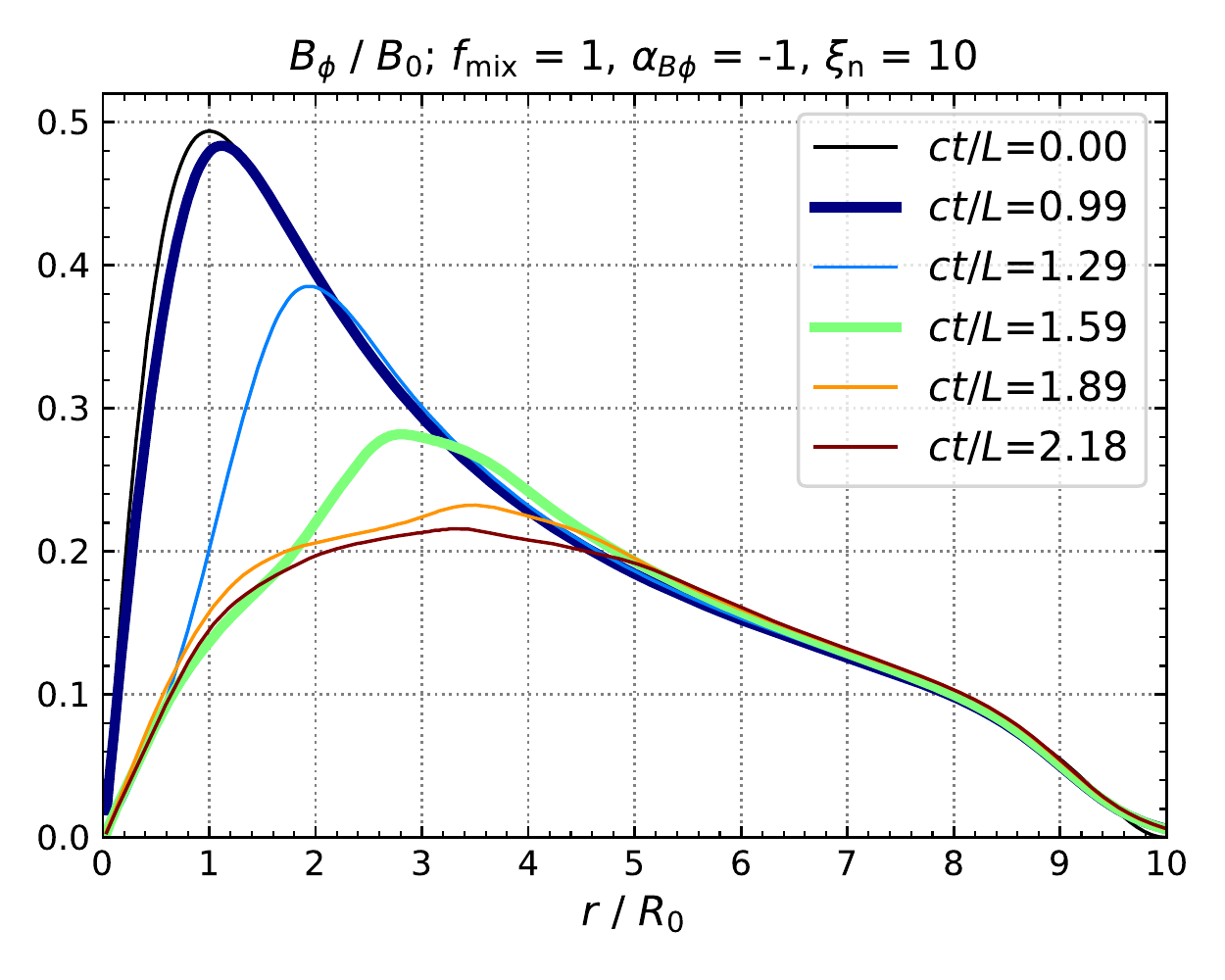}
\includegraphics[width=0.499\textwidth]{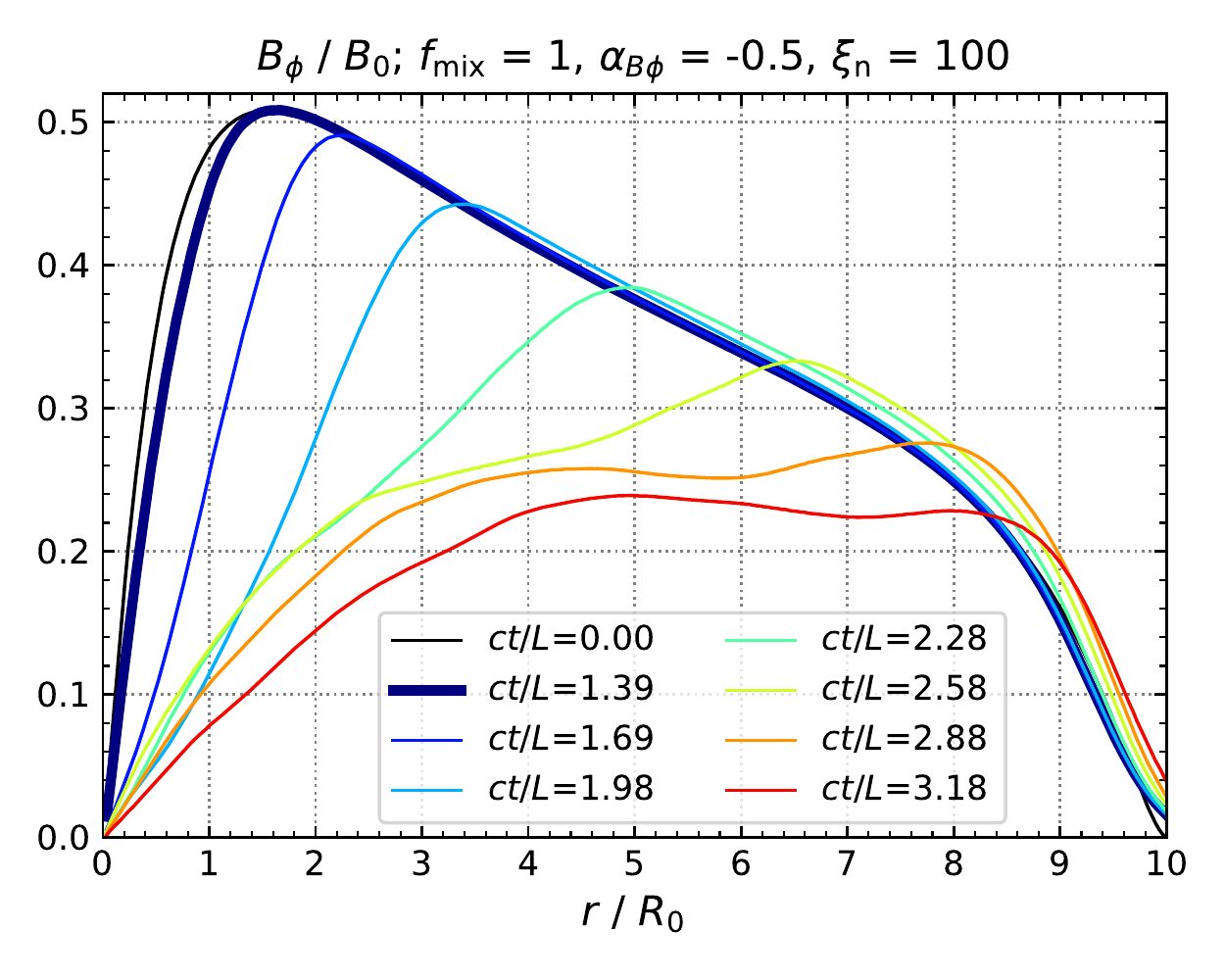}
\\
\includegraphics[width=0.499\textwidth]{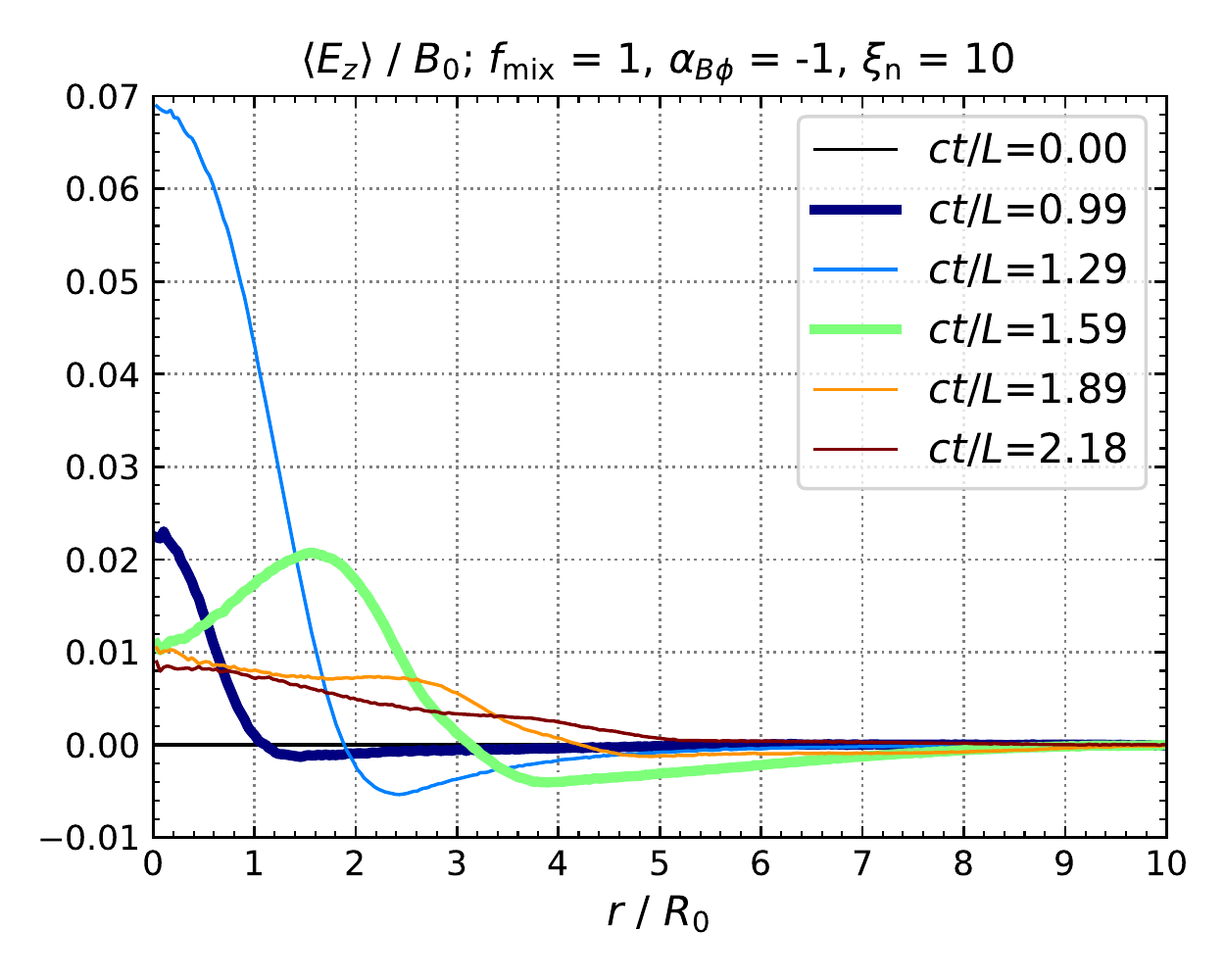}
\includegraphics[width=0.499\textwidth]{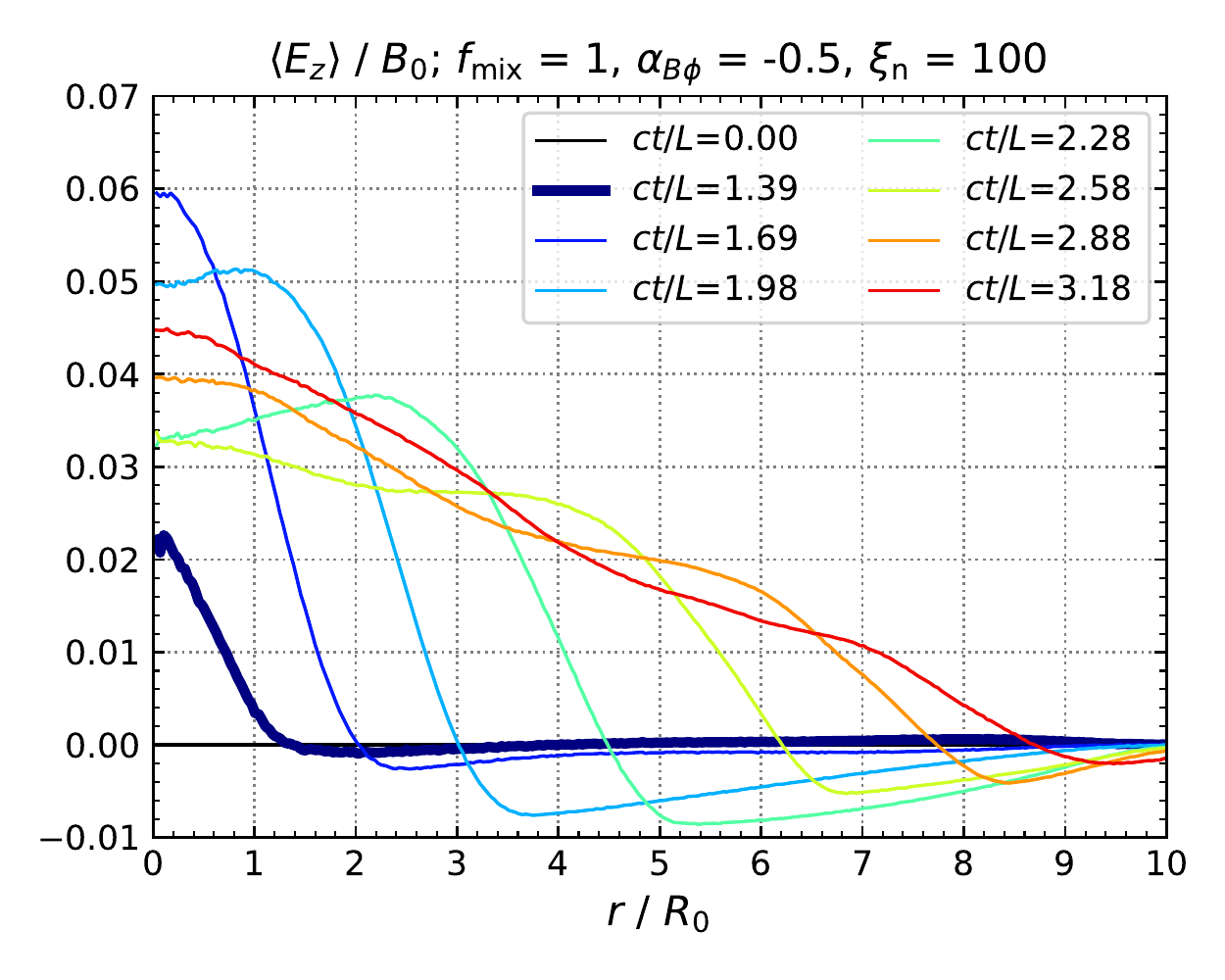}
\caption{\emph{Top left panel}: radial profiles of~the~mean toroidal magnetic field $\left<B_\phi\right>(r)$ (averaged over $z$ and $\phi$) for the~5 moments in~time of~the~reference simulation f1\_$\alpha$-1\_$\xi$10 presented in~Figure~\ref{fig_xzmaps_f1_nr10_aBp1} (and for $t=0$).
\emph{Top right panel}: radial profiles of~$\left<B_\phi\right>(r)$ for the~simulation f1\_$\alpha$-05\_$\xi$100.
\emph{Bottom panels}: radial profiles of~the~mean axial electric field $\left<E_z\right>(r)$ (averaged over $z$ and $\phi$) normalized to~$B_0$ for the~same  moments in~time of~the~same simulations as in~the~respective top panels.}
\label{fig_rprof_Bphi_meanEz}
\end{figure*}

Figure~\ref{fig_xzmaps_f1_nr10_aBp1} presents selected snapshots from the~reference simulation.
We show 2D $(x,z)$ slices along the~$y = 0$ plane (containing the~symmetry axis $r = 0$ of~the~initial configuration) for the~out-of-plane magnetic field component $B_y$ (dominated by~the~toroidal component $B_\phi$) and for the~axial electric field component $E_z$,
 as well as $(x,y)$ slices along the~$z = 0$ plane for $E_z$, at 5 uniformly spaced moments in~time.
The sequence illustrates the~development of~unstable modes beginning in~the~central core region and growing outwards.
These are characterized by~short wavelength along $z$
($\lambda_z \simeq 2.7R_0$, which means $\simeq 7.5$ full wavelengths per $L$; see Section~\ref{sec_res_modes_fftz})
and varying levels of asymmetry in the $(x,y)$ plane, suggesting the dominance of the $m=1$ kink mode\footnote{A further analysis reveals that this mode is circularly polarized, i.e., the phase $\phi_1$ is a strong quasi-periodic function of $z$ and a weak function of simulation time.}.
They break into~a~non-linear phase by~$t \simeq 1.6L/c$.
The~axial electric field appears to~be the~strongest and~largely positive around $t \simeq 1.3L/c$.

The~left panels of~Figure~\ref{fig_rprof_Bphi_meanEz} present the~radial profiles of~the~mean toroidal magnetic field $\left<B_\phi\right>(r)$ (averaged over $z$ and $\phi$) and 
the~mean axial electric field $\left<E_z\right>(r)$ for the~same 5 moments in~time.
At $t \simeq L/c$, the~toroidal magnetic field profile is still very similar to~the~initial one, while a~net positive axial electric field builds up along the~axis.
For $1.0 < ct/L < 1.6$, we observe rapid decay of~the~toroidal magnetic field within $r < 3R_0$.
 Note from Figure~\ref{fig_xzmaps_f1_nr10_aBp1} that by $ct/L \simeq 1.6$ a~turbulent structure of both $B_y$ and $E_z$ develops exactly within these radii ($|x| < 0.15L$).
Therefore, we~consider this decay of~$B_\phi$ as an~irreversible magnetic dissipation.
We will refer to~this period of~time as the~\emph{fast magnetic dissipation phase};
its beginning and ending are~indicated with thick lines.
At the~end of~the~fast magnetic dissipation phase ($t \simeq 1.6L/c$), $\left<B_\phi\right>(r)$ peaks at the~level of~$\simeq 0.28B_0$ at a~radius of~$r \simeq 2.7R_0$.
Using these numbers, we can estimate the~post-dissipation energy limit as $e(0.28B_0)(2.7R_0) \simeq 0.76\gamma_{\rm lim}$, a~rather minor decrease from the~initial value.
The net axial electric field shoots up to~$\simeq 0.07B_0$ in~the~middle of~the~fast magnetic dissipation phase and decays to~$\simeq 0.02B_0$ by~$t \simeq 1.6L/c$.
For $t > 1.6L/c$, dissipation of~toroidal magnetic field is significantly slower
 and $\left<E_z\right>(r)$
does not exceed $\simeq 0.01B_0$.

\begin{figure}
\includegraphics[width=\columnwidth]{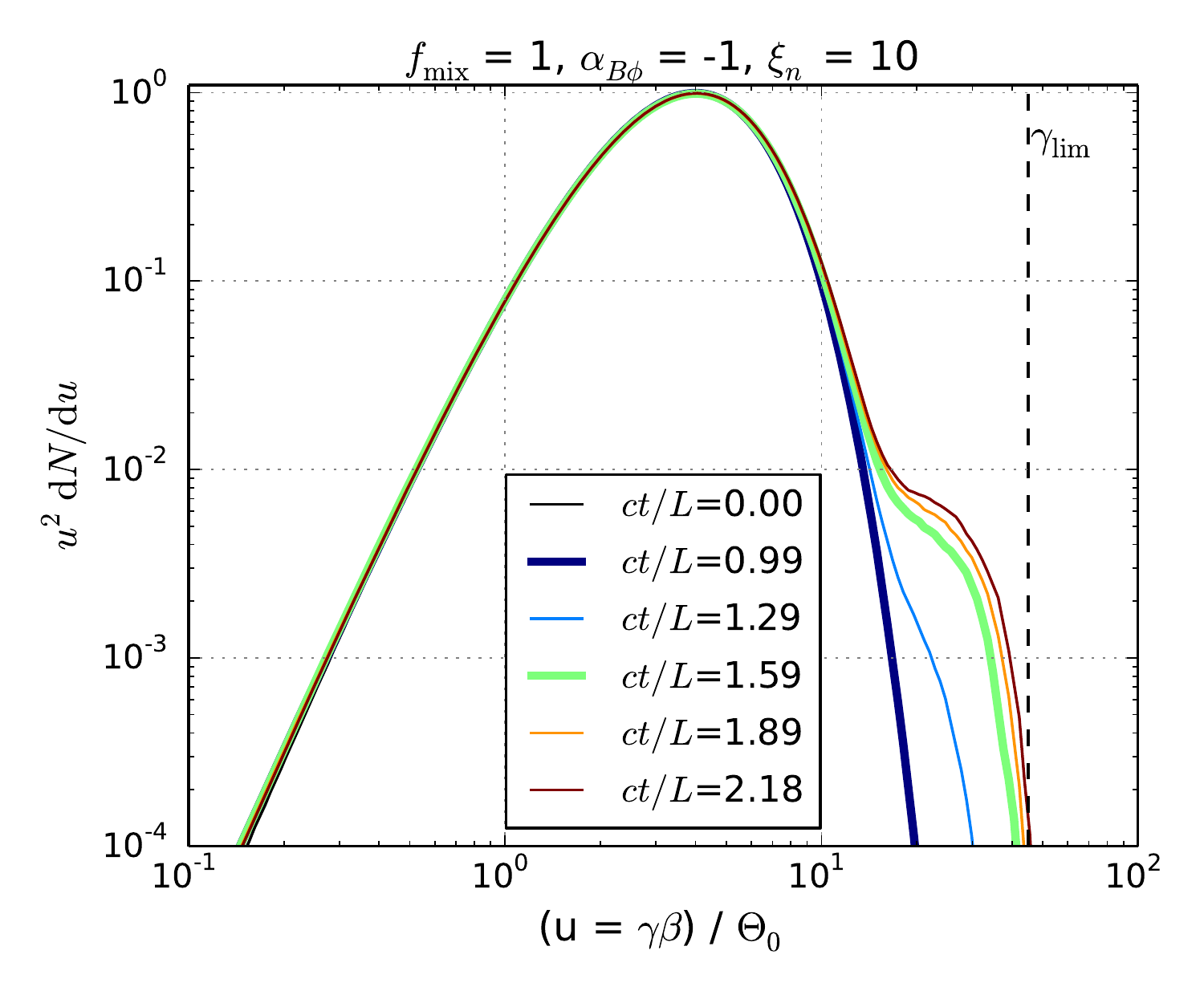}
\caption{Particle momentum distributions $u^2\,{\rm d}N/{\rm d}u$ (equivalent to~energy distributions since $\Theta_0 \gg 1$), combined for both electrons and positrons  and normalized to peak at unity, for the~5 moments in~time of~the~reference simulation f1\_$\alpha$-1\_$\xi$10 presented in~Figure~\ref{fig_xzmaps_f1_nr10_aBp1}.
The vertical dashed line indicates the~\emph{confinement} energy limit $\gamma_{\rm lim} = 45\Theta_0$.}
\label{fig_spec_f1_nr10_aBp1}
\end{figure}

Figure~\ref{fig_spec_f1_nr10_aBp1} compares the~particle momentum $u = \gamma\beta$ ($u \simeq \gamma$ for $\gamma \gg 1$)
 distributions ${\rm d}N/{\rm d}u$ compensated
by~$u^2$ for the~same 5 moments in~time.
The distribution at $t \simeq L/c$ is indistinguishable from the~initial Maxwell-J\"{u}ttner distribution.
The fast magnetic dissipation phase corresponds to~a~rapid build-up of~a~high-energy component of~the~distribution.
The high-energy component evolves much more slowly for $t > 1.6L/c$.
The most energetic particles approach the~energy limit, which in~our simulations amounts to~$\gamma_{\rm lim} = 45\Theta_0$.
Based on such momentum distributions, we define the~\emph{maximum particle energy} $\gamma_{\rm max}$  corresponding to~the~$u_{\rm max}$ value at which $u^2\,{\rm d}N/{\rm d}u$ equals $10^{-4}$ of~the~peak level determined at $t=0$.

\section{Results: instability modes}
\label{sec_res_modes}

\subsection{Pinch vs. kink modes}
\label{sec_res_modes_m}

Here we compare the~relative strengths of~the~two fundamental azimuthal instability modes -- the~$m = 0$ pinch vs. the~$m = 1$ kink --  in both the linear and non-linear stages.
We~have analyzed the~3D spatial distributions of~the~axial electric field component~$E_z$.
Using the~cylindrical coordinates $(r,\phi,z)$ centered at~the~initial symmetry axis, we define a~series of~cylindrical shells $\mathcal{S}_n$ for $n \in \{1,...,9\}$  delimited by~$R_{n-1} < r < R_n$, where $R_n = (n+1)R_0$, so that $R_9 = R_{\rm out}$.
For the~pinch mode, we have also analyzed the~central core region $r < R_0$ (the $\mathcal{S}_0$ shell).
Within each shell, the~values of~$E_z$ have been averaged over $r$.
Then, for every value of~$z$, the~function $\left<E_z\right>(\phi,z)$ has been decomposed into a~Fourier series $\sum_m E_m(z) \cos[m\phi+\phi_m(z)]$ with real amplitudes $E_m(z)$ and phases $\phi_m(z)$.
These amplitudes have been averaged over $z$ and will be presented as functions of~simulation time.

\begin{figure}
\includegraphics[width=\columnwidth]{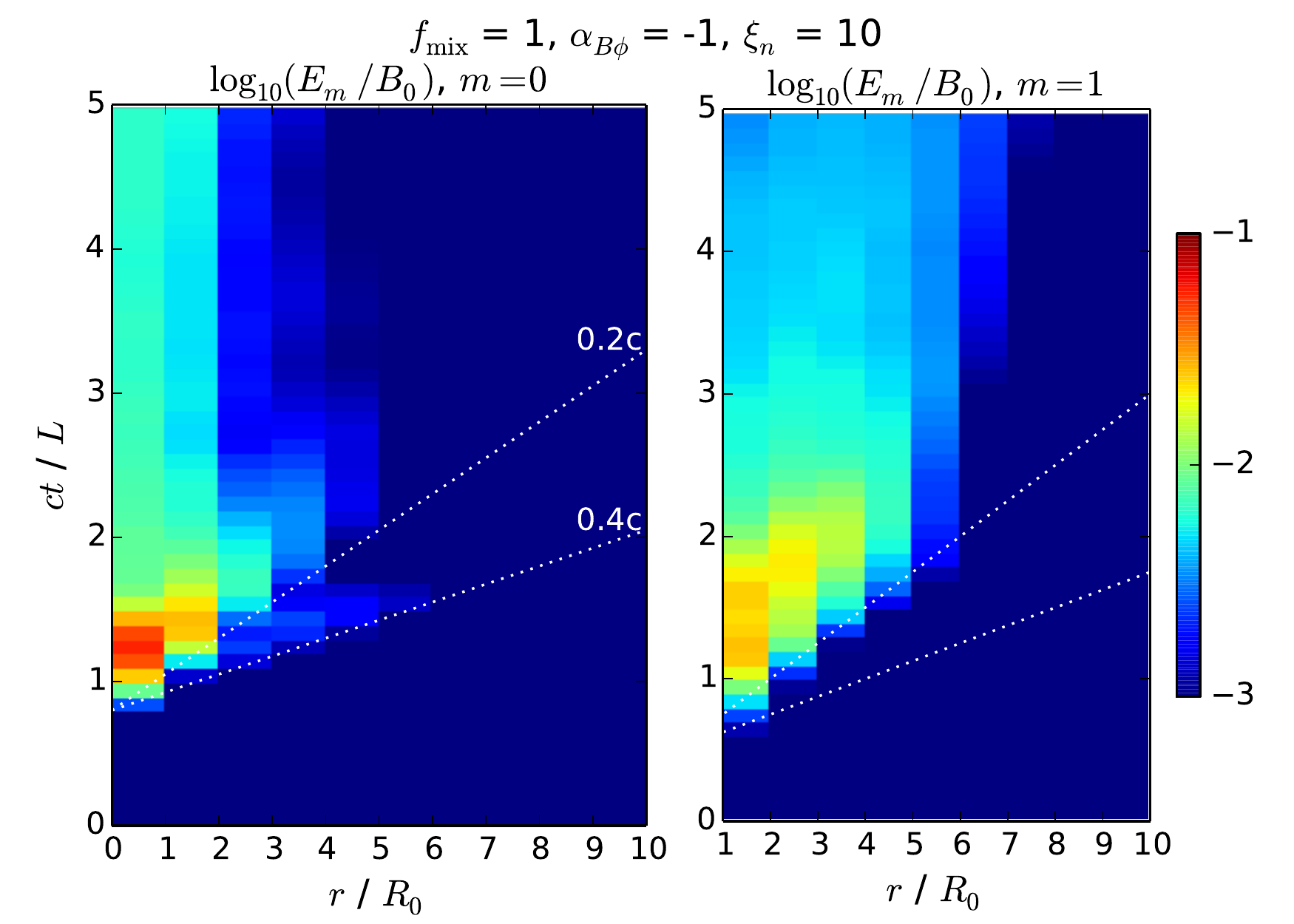}
\\
\includegraphics[width=\columnwidth]{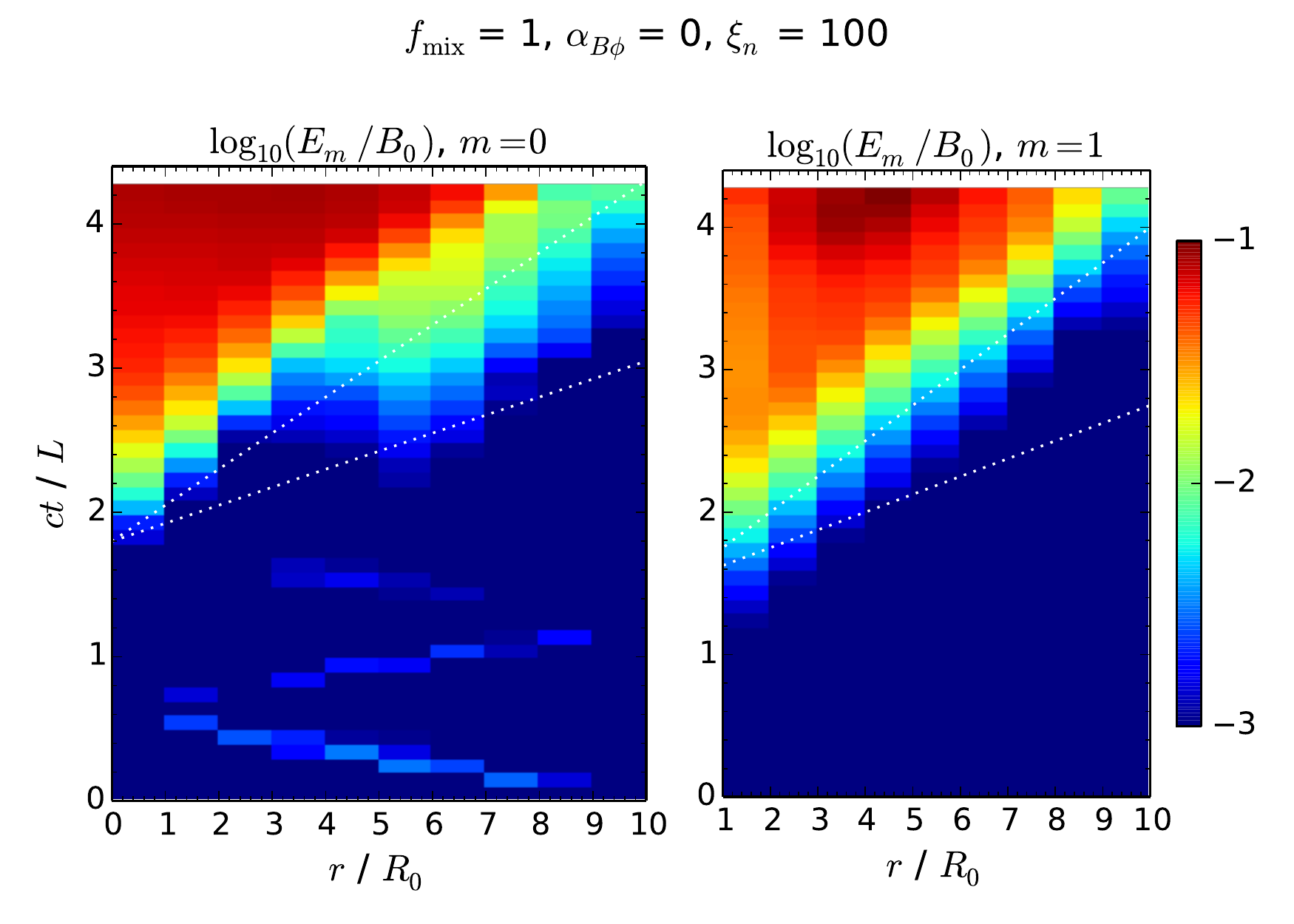}
\caption{Spacetime diagrams of~the~amplitudes $E_m$ (averaged over $z$) of~the~azimuthal modes $m$ of~the~$E_z(\phi,z)$ distributions compared for two simulations.
 The~white dotted lines indicate the speed levels of~$0.2c$ and~$0.4c$.
In the~case f1\_$\alpha$0\_$\xi$100,
 the simulation was interrupted at $ct/L \simeq 4.27$ when the~perturbations reached
the~boundary according to~the~definition described in~Section~\ref{sec_config}.
}
\label{fig_phimodes_r0-10}
\end{figure}

Figure~\ref{fig_phimodes_r0-10} presents the~spacetime diagrams $(r,t)$ of~the~amplitudes of~the~pinch and kink modes for two simulations, including all 9 shells $\{\mathcal{S}_1,...,\mathcal{S}_9\}$, and for the~pinch mode also the~core region $\mathcal{S}_0$.
In~the~reference case f1\_$\alpha$-1\_$\xi$10 (two left panels in~Figure~\ref{fig_phimodes_r0-10}), 
we~find a~strong pinch mode (with~$E_0 \sim 0.1B_0$) highly localized in~both radius and~time.
In~the~core region it~is found for $1.0 < ct/L < 1.5$, which is during the~fast magnetic dissipation phase.
It~is also found in~the~first shell $\mathcal{S}_1$ during a~slightly later period of~$1.3 < ct/L < 1.6$.
A~slightly weaker kink mode is~more extended in~both time and~radius.
In~the~first shell, it~can be~seen for~$1.1 < ct/L < 1.8$; in~the~next two shells ($2 < r/R_0 < 4$) it~is successively delayed at~the~rate similar to~that for~the~pinch mode.
These modes appear to~originate in~the~central core region and~propagate radially outwards.
At~lower amplitude levels of~$\sim 10^{-3}B_0$, the~pinch mode appears to~propagate at~the~velocity of~$\sim 0.4c$, faster than the~kink mode ($\sim 0.2c$).

In~the~case f1\_$\alpha$0\_$\xi$100 (two right panels in~Figure~\ref{fig_phimodes_r0-10}), both the~pinch and~the~kink modes achieve higher amplitudes of~$\sim 0.1B_0$, starting around $t \simeq 2L/c$ and~lasting at~least until $t \simeq 4.3L/c$, at~which point the~simulation has~been interrupted because the~modes reached the~outer boundary according to~the~criterion described in~Section~\ref{sec_config}.
Both modes appear to~originate in~the~central core and~propagate outwards with~similar velocities $\sim 0.2c$.
Looking at the~amplitude levels of~$\sim (10^{-2.5}\,\text{---}\,10^{-2.0})B_0$ of~the~pinch mode, we see an excess signal in~the~$\mathcal{S}_5,\mathcal{S}_6$ shells ($5 < r/R_0 < 7$) at $ct/L \simeq 2.5\,\text{---}\,3$.
Such a~signal is not seen in~the~kink mode, and also it is not found (at least that clearly) in~our other simulations.
We~discuss the~origin of~this radially localized pinch mode in~Section~\ref{sec_disc}.

\begin{figure*}
\includegraphics[width=\textwidth]{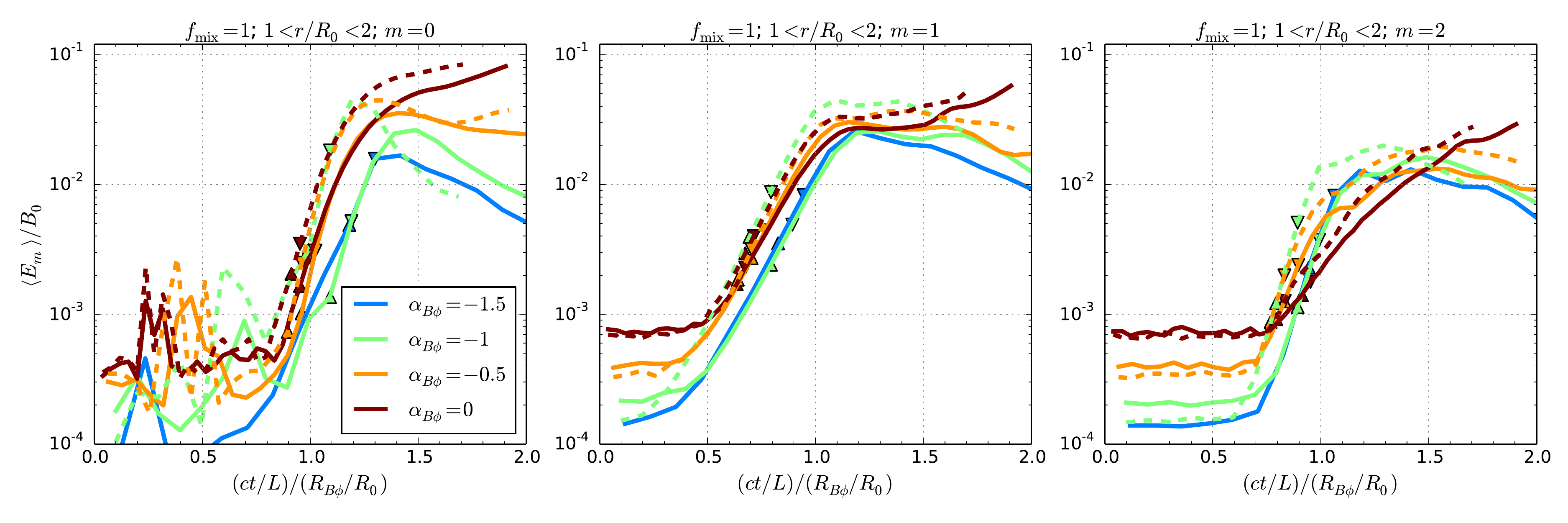}
\includegraphics[width=\textwidth]{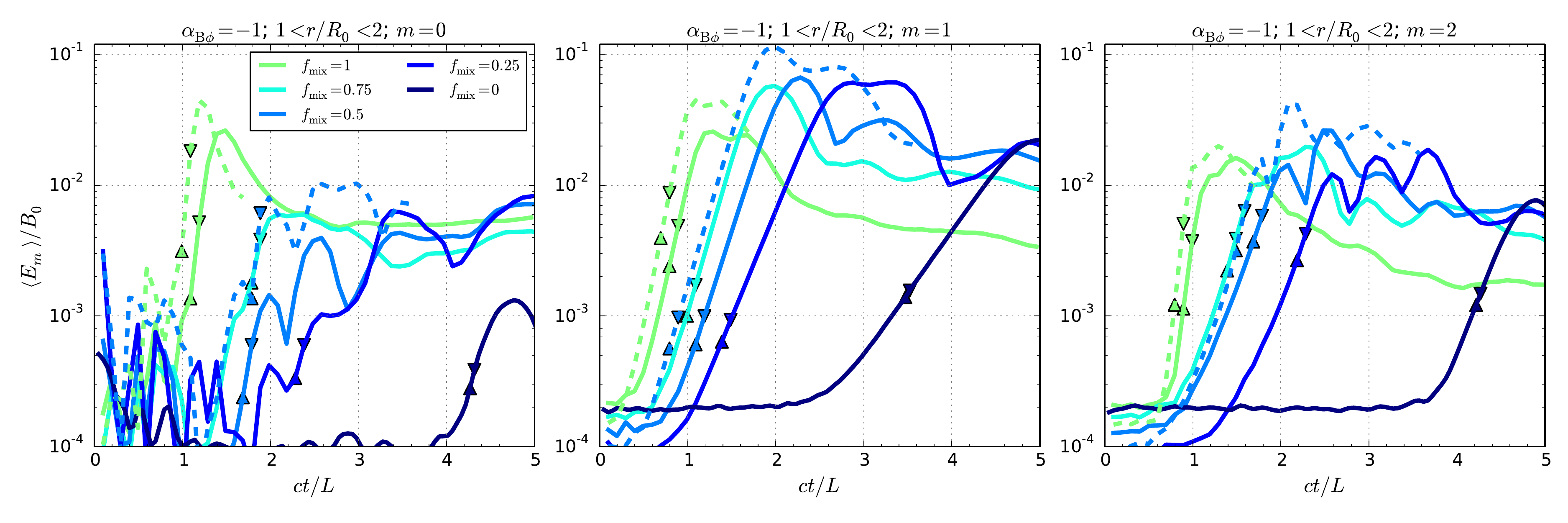}
\caption{Amplitudes $E_m$ of~the~azimuthal modes $m$ of~the~axial electric field $E_z$ extracted from the~$1 < r/R_0 < 2$ cylindrical shell, averaged over $z$ and presented as functions of~the~simulation time $t$  scaled by~the~characteristic radius $R_{\rm B\phi}$.
The upper row of~panels compares the~simulations with $f_{\rm mix} = 1$, different values of~$\alpha_{\rm B\phi}$ (indicated by~the~line color), and different values of~$\xi_n$ (10: solid lines, 100: dashed lines).
The lower row of~panels compares the~simulations with $\alpha_{\rm B\phi} = -1$, different values of~$f_{\rm mix}$ (indicated by~the~line color), and different values of~$\xi_n$ (as before).
The left panels show the~$m = 0$ pinch mode, the~middle panels show the~$m = 1$ kink mode, and the~right panels show the~$m = 2$ mode.
The triangles indicate the~line segments for which the~growth timescale $\tau$ of~the~mode amplitude is minimized.}
\label{fig_phimodes_r1-2}
\end{figure*}

We~now focus on~the~first shell $\mathcal{S}_1$ ($1 < r/R_0 < 2$).
Figure~\ref{fig_phimodes_r1-2} compares the~time evolutions of~the~$m = 0,1,2$ modes for~two~series of~simulations:
$f_{\rm mix} = 1$ (varying $\alpha_{B\phi}$ and~$\xi_n$)
and~$\alpha_{\rm B\phi} = -1$ (varying $f_{\rm mix}$ and~$\xi_n$).
In~the~$f_{\rm mix} = 1$ series (upper panels in~Figure~\ref{fig_phimodes_r1-2}), the~amplitudes of~the~$m = 0$ pinch modes are~comparable  to~the~amplitudes of~the~$m = 1$ kink modes, and~the~$m = 2$ modes are~only slightly weaker.
With the~simulation time $t$ scaled by~the~characteristic radius $R_{\rm B\phi}$, the~evolutions of~the~modes are~similar for~different values of~$\alpha_{\rm B\phi}$.
 At~the same time,
the~mode evolution is~not~sensitive to~the~density contrast $\xi_n$, especially in~the~linear stage.
In~all
simulations, we~find that~the~kink mode is~the~first to~emerge (roughly when $\left<E_1\right>$ exceeds the~$10^{-3}B_0$ level) followed by~the~pinch mode and all the~higher modes ($m \ge 2$).

In~the~$\alpha_{\rm B\phi} = -1$ series (lower panels in~Figure~\ref{fig_phimodes_r1-2}), we~observe a~systematic weakening of~the~pinch mode together with a~strengthening of~the~kink mode with decreasing value of~$f_{\rm mix}$ \citep[in agreement with the linear analysis of][]{2013MNRAS.434.3030B}.
The~reason for~this is~that the~axial magnetic flux resists radial compression, and~hence stabilizes the~$m=0$ pinch mode for~$f_{\rm mix} \to 0$.
For~$f_{\rm mix}<1$, the~evolution of~the~pinch mode does~not even show regular linear stages of~exponential growth, while the~kink mode shows linear stages extending over 2~orders of~magnitude in~$\left<E_1\right>$, reaching values up~to~$0.1B_0$ in~the~case $f_{\rm mix} = 0.5$ and~$\xi_n = 100$.
 Higher modes ($m \ge 2$) are~once again systematically weaker and~delayed.

\begin{figure}
\includegraphics[width=\columnwidth]{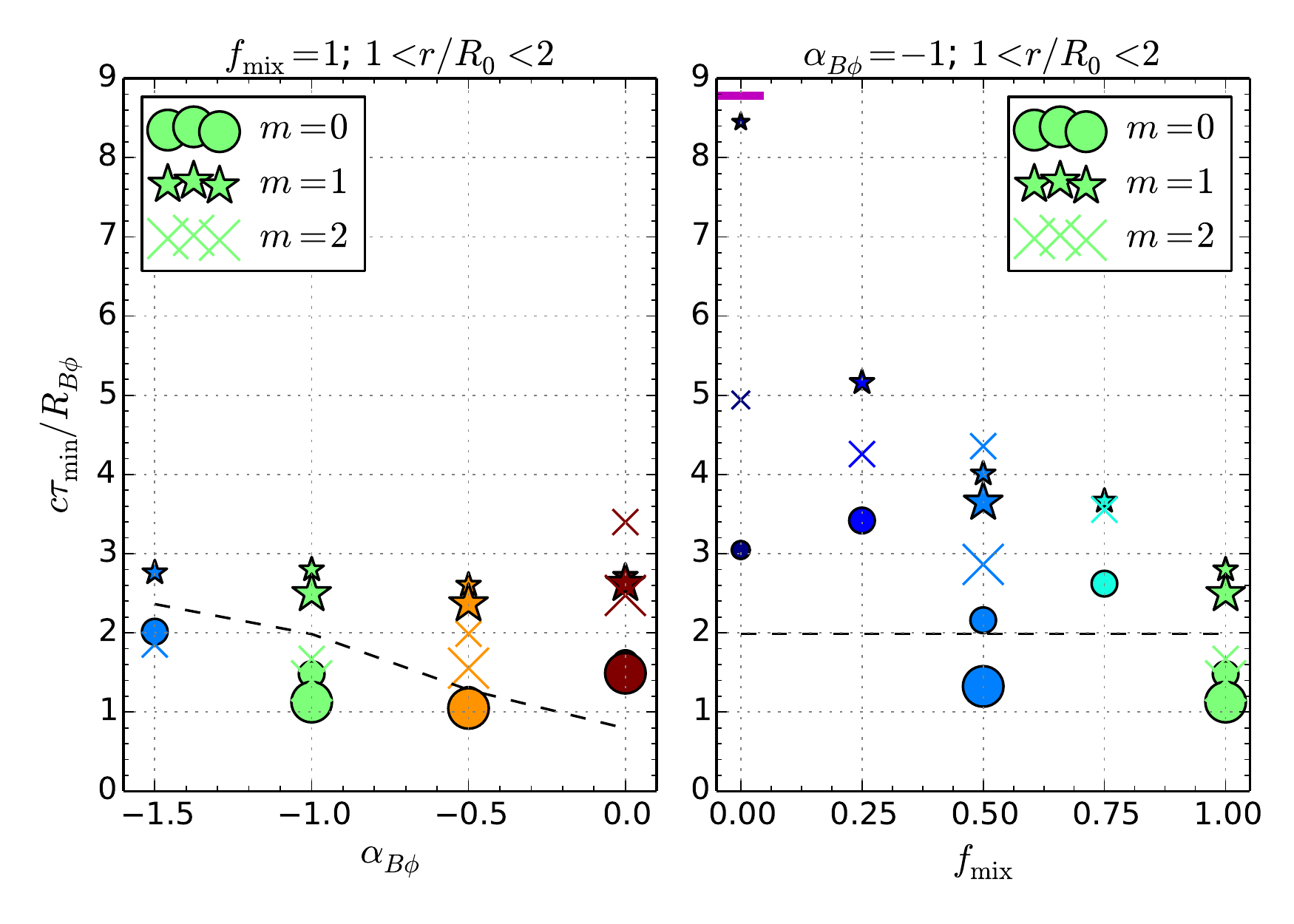}
\caption{Minimum growth timescales $\tau_{\rm min}$ normalized to~$R_{\rm B\phi}/c$ evaluated for the~amplitudes $E_m$ of~the~azimuthal modes of~$E_z$ presented in~Figure~\ref{fig_phimodes_r1-2}.
The~symbol colors correspond to~the~line colors used in~Figure~\ref{fig_phimodes_r1-2}.
Different symbol types indicate the~mode number $m$.
For each symbol type, the~smaller symbols correspond to~$\xi_n = 10$, and the~larger symbols correspond to~$\xi_n = 100$ (except for the~case $f_{\rm mix} = 0$, in~which $\xi_n = 1$).
The short magenta line in~the~right panel indicates the~analytical prediction for the~$|m| = 1$ kink mode in~the~constant-pitch FF configuration \citep{2000A&A...355..818A}.
The black dashed lines mark the~time resolution limit for this analysis.}
\label{fig_tau_min}
\end{figure}

We~have measured the~minimum growth time scales $\tau_{\rm min} = (c\Delta t/R_{\rm B\phi})\min(1/\Delta\ln\left<E_m\right>)$ for~each mode $m$ in~every studied case; their values are~compared in~Figure~\ref{fig_tau_min} (the corresponding segments of~the~$\left<E_m\right>(t)$ functions are indicated in~Figure~\ref{fig_phimodes_r1-2}).
One~should note that the~time resolution for~this analysis is~limited to~$\Delta t \simeq 2R_0/c$, hence any shorter time scales should be considered as~upper limits.
Nevertheless, our results indicate that the~pinch modes are typically the~fastest ones ($\tau_{\rm min} < 2$ for $f_{\rm mix} = 1$), and the~kink modes are typically the~slowest ones ($\tau_{\rm min} > 2$).
In general, the~value of~$\tau_{\rm min}$ decreases
with increasing $f_{\rm mix}$,
and its dependence on $\alpha_{\rm B\phi}$ is captured by~its explicit scaling with $R_{\rm B\phi}$.

\subsection{Effective axial wavelength}
\label{sec_res_modes_fftz}

\begin{figure}
\includegraphics[width=0.495\columnwidth]{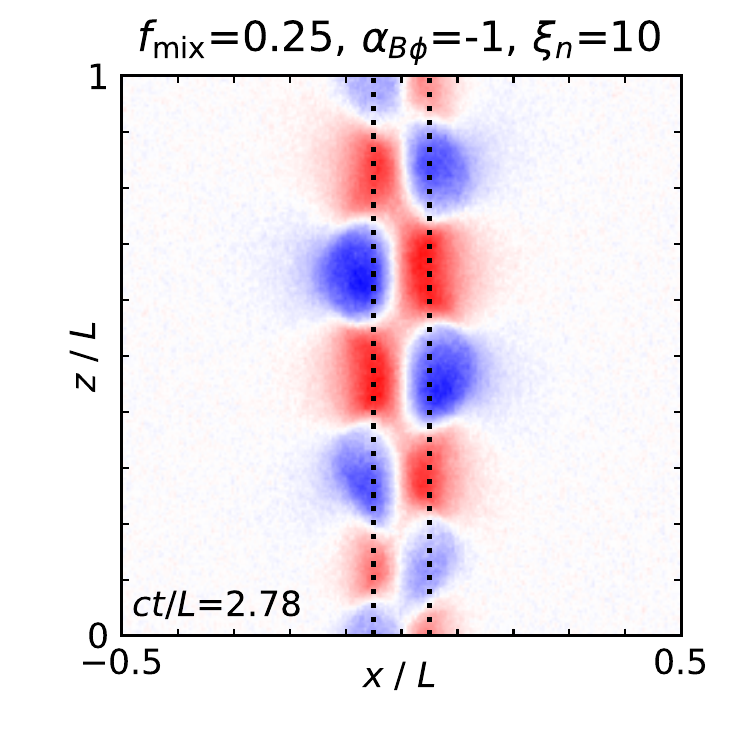}
\includegraphics[width=0.495\columnwidth]{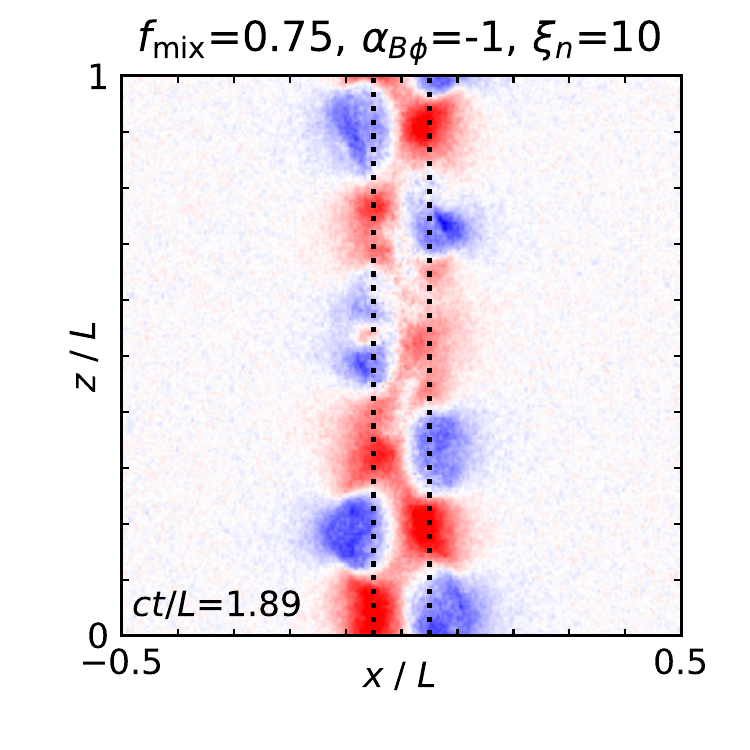}
\includegraphics[width=0.495\columnwidth]{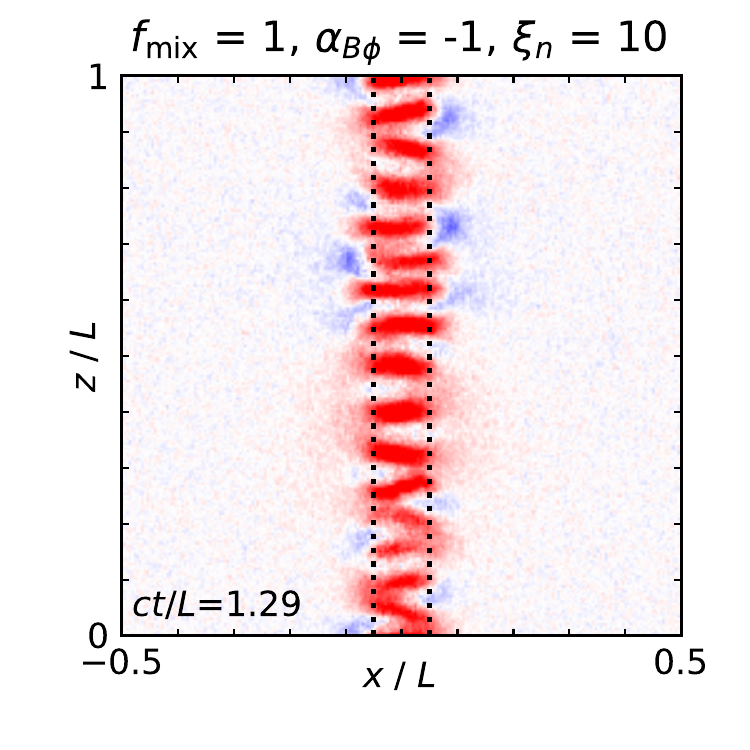}
\includegraphics[width=0.495\columnwidth]{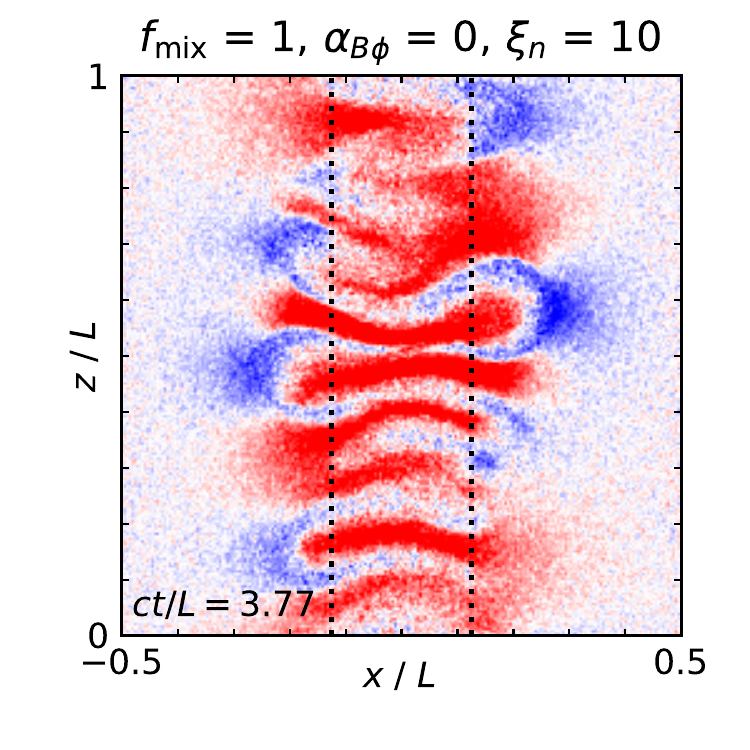}
\caption{Maps of~electric field component $E_z$ in~units of~$B_0$ (positive values in~red, negative in~blue) in~the~$y = 0$ plane compared for several simulations at the~moments (indicated in~the~bottom left corners) of~peak root-mean-square value of~$E_z$ evaluated at $r=R_0$.
The~vertical dotted lines indicate $r = R_{B\phi}$, which equals $R_0$ for~$\alpha_{B\phi} = -1$, and~$2.5R_0$ for~$\alpha_{B\phi} = 0$.}
\label{fig_xzmaps_rmsEz_peak}
\end{figure}

Figure~\ref{fig_xzmaps_rmsEz_peak} compares the~$(x,z)$ maps in~the~$y=0$ plane of~the~axial electric field component $E_z$ for four selected simulations at the~moments when the~root-mean-square values of~$E_z$ evaluated at~the~central core radius $r = R_0$ achieve their peaks.
In~the~case f025\_$\alpha$-1\_$\xi$10 we~observe a~regular structure with the~regions of~$E_z > 0$ (red) slightly more extended than the~regions of~$E_z < 0$ (blue), aligned asymmetrically about the~central axis (indicating the~dominance of~the~kink mode), with 3 full wavelengths over $\Delta z = L$.
In~the~case f075\_$\alpha$-1\_$\xi$10, the~structure of~$E_z(x,z)$ appears to~be very similar to~the~previous case, however, some short-wavelength  kink-like fluctuations appear superposed.
In~the~reference case f1\_$\alpha$-1\_$\xi$10 (cf.~Figure~\ref{fig_xzmaps_f1_nr10_aBp1}), we~find a~very different structure of $E_z$, in~the~form of~a~somewhat irregular stack of~$\simeq 16$ short-wavelength patches of~positive values, separated by~narrow gaps of~weak (but still positive) values.
This can be compared with the~last presented case f1\_$\alpha$0\_$\xi$10, in~which the~$E_z > 0$ patches are~much more extended radially, because in~this case the~evolution of~the~instability is~longer by~a~factor $\simeq R_{B\phi}/R_0 = 2.5$.

\begin{figure}
\includegraphics[width=\columnwidth]{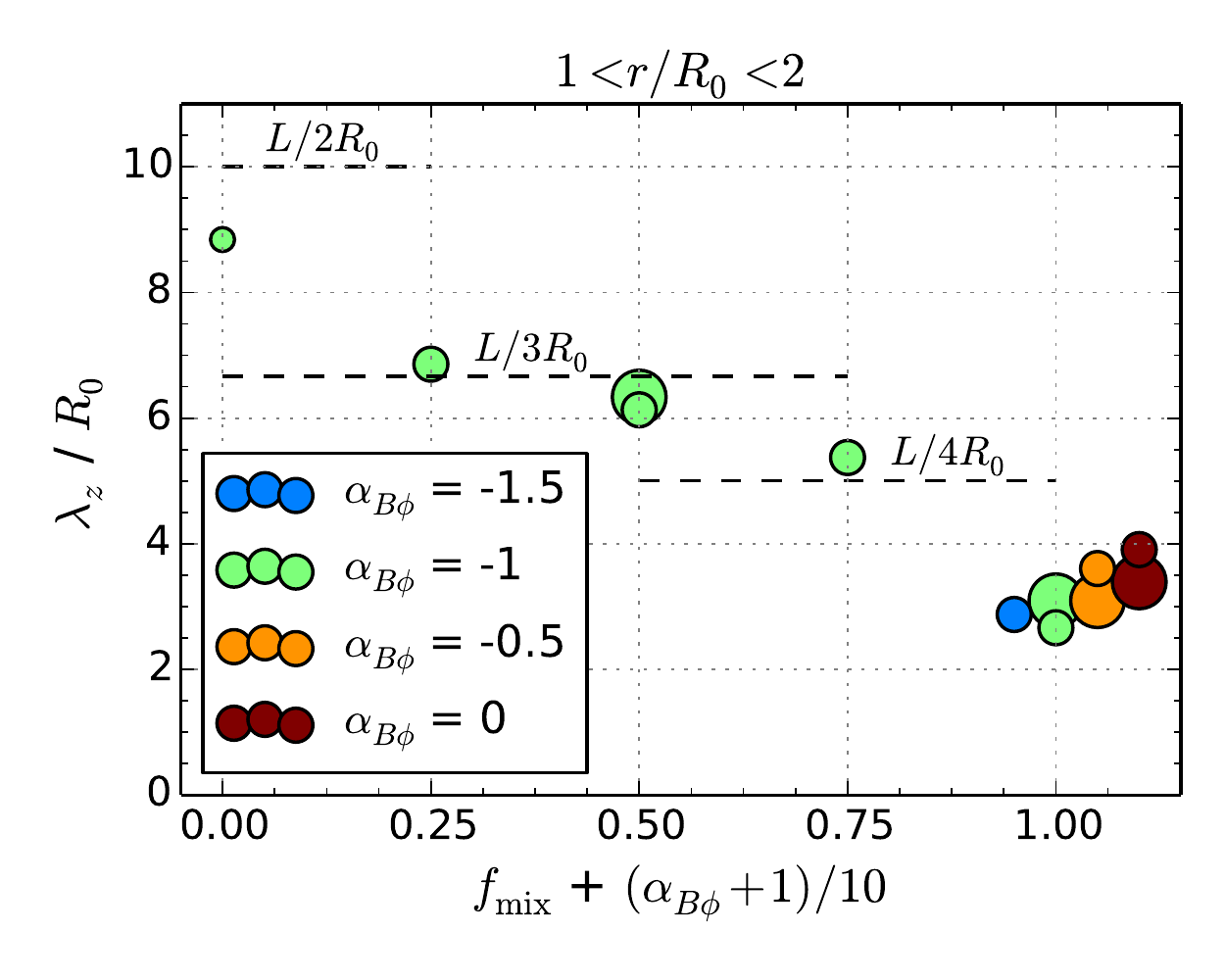}
\caption{Effective axial wavelength $\lambda_z$ of~the~$\left<E_z\right>(z)$ fluctuations measured within the~cylindrical shell region $\mathcal{S}_1$ ($1 < r/R_0 < 2$) at the~moments of~peak ${\rm rms}(E_z)$.
The smaller symbols correspond to~$\xi_n = 10$, and the~larger symbols correspond to~$\xi_n = 100$ (except for the~case $f_{\rm mix} = 0$, in~which $\xi_n = 1$).}
\label{fig_fftz}
\end{figure}

Using the~$(r)$-averaged profiles of~$\left<E_z\right>(\phi,z)$ extracted from the~first cylindrical shell region $\mathcal{S}_1$,
we calculated the~discrete Fourier transform $E_k(\phi) = \sum_j\left<E_z\right>(\phi,z_j)\,\exp(-2\pi ik z_j/L)$ over a~regular grid $0 \le z_j < L$,
averaged the~$E_k(\phi)$ amplitudes over $\phi$,
then calculated the~effective axial wavenumber $\left<k_z\right> = (\sum_k kE_k^2) / (\sum_kE_k^2)$,
and finally the~corresponding effective wavelength $\lambda_z = 2\pi/\left<k_z\right>$.
Figure~\ref{fig_fftz} compares the~values of~$\lambda_z$  for all our simulations.
We find that $\lambda_z$ is decreasing systematically with increasing $f_{\rm mix}$.
 For~$f_{\rm mix} = 0$, we~find that $\lambda_z \simeq 8.8R_0 \simeq L/2.25$, which reflects the~dominance of~2 full wavelengths in~the~$E_z$ structure.
For~$f_{\rm mix} \sim 0.25\,\text{---}\,0.5$,
$\lambda_z$ is~consistent with~$L/3$.
For~$f_{\rm mix} = 0.75$, 
the~effective wavelength is~$\lambda_z \simeq 5.4R_0 \simeq L/3.7$, in~agreement with
a~combination of~long ($L/3$) current-driven modes characteristic of~the~FF screw-pinches and~short pressure-driven modes characteristic of~the~Z-pinches
(cf.~the~upper right panel of~Figure~\ref{fig_xzmaps_rmsEz_peak}).
For~$f_{\rm mix} = 1$, all of~our simulations produce consistent values of~$\lambda_z \simeq (2.7\,\text{---}\,3.9)R_0 \simeq (L/7.5 \, \text{---}\, L/5)$, increasing somewhat with increasing $\alpha_{\rm B\phi}$, but not~scaling clearly with~$R_{B\phi}$.
There~is little dependence of~$\lambda_z$ on~the~density ratio parameter $\xi_n$.

\section{Results: particle acceleration}
\label{sec_res_accel}

\subsection{Fast magnetic dissipation phase and particle energy limit}
\label{sec_res_accel_glim}

In this subsection we show that the Z-pinch configurations with steep toroidal field indices $(\alpha_{\rm B\phi} \le -1)$ satisfy the~$\gamma_{\rm lim}$ energy limit, while those with shallow indices $(\alpha_{\rm B\phi} > -1)$ exceed that limit.
Energetic particles are~well confined within the~jet core in~either case. 
For~$\alpha_{\rm B\phi} \le -1$, efficient magnetic dissipation proceeds over a~limited period of~time, transitioning from a~\emph{fast magnetic dissipation phase} to~a~slow magnetic dissipation phase before the~perturbations reach the~domain boundaries.
We~will argue that particle energies are~limited by~the~finite time duration of~the~fast magnetic dissipation phase.

\subsubsection{The reference case $f_{\rm mix} = 1$ and~$\alpha_{\rm B\phi} = -1$}
\label{sec_res_accel_ref}

\begin{figure}
\includegraphics[width=\columnwidth]{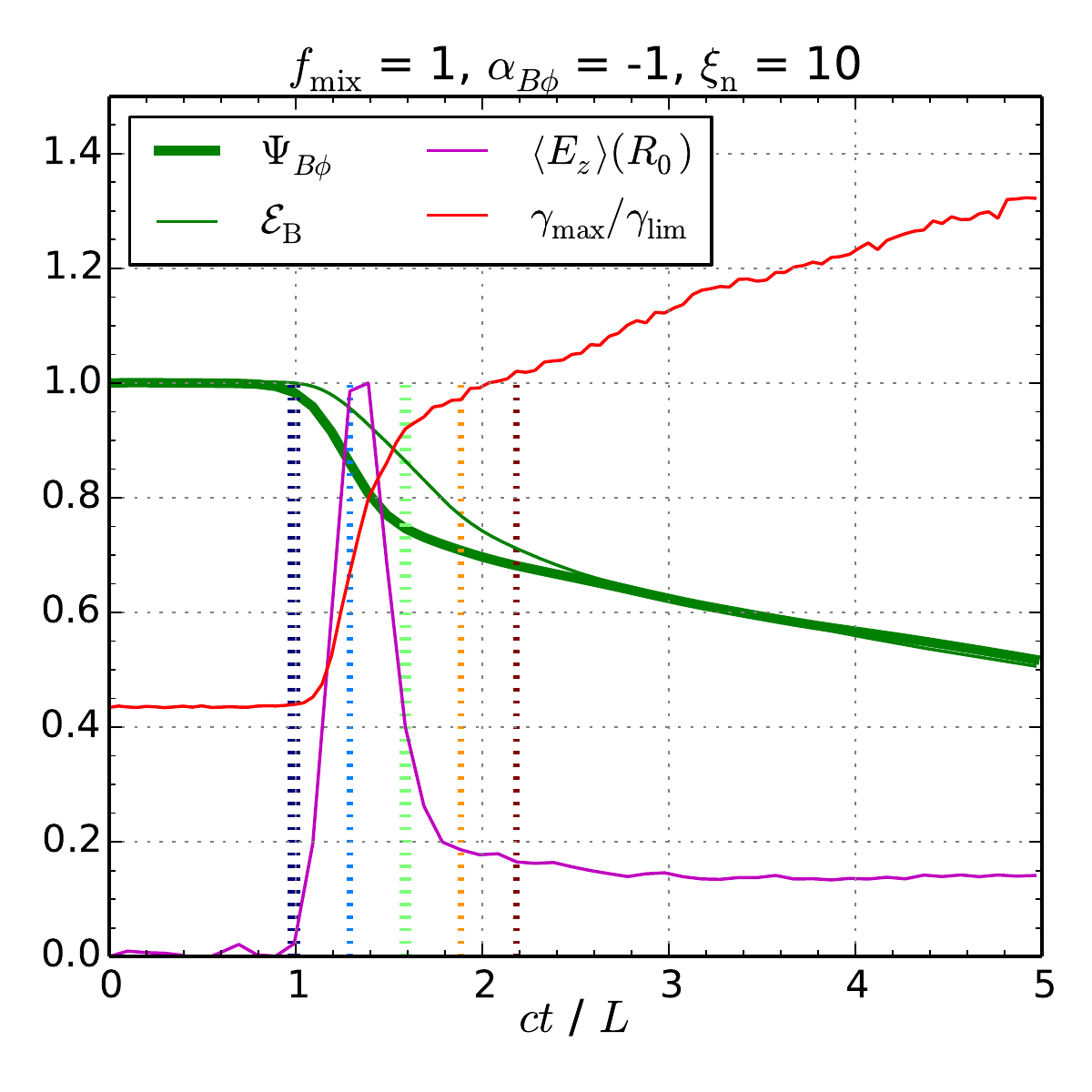}
\caption{Time evolutions of~the~toroidal magnetic field $B_\phi$ and axial electric field $E_z$ for the~reference simulation f1\_$\alpha$-1\_$\xi$10.
The thick green line shows the~toroidal magnetic flux $\Psi_{\rm B\phi}$ normalized to~its initial value;
the thin green line shows the~total magnetic energy $\mathcal{E}_{\rm B}$ normalized to~its initial value;
the red line shows the~maximum particle energy $\gamma_{\rm max}$ normalized to~the~Lorentz factor limit $\gamma_{\rm lim}$;
the magenta line shows the~mean axial electric field $\left<E_z\right>$ (averaged over $z$ and $\phi$) evaluated at $r = R_0$ and normalized to~its peak value of~$E_{\rm z,peak} = 0.043B_0$;
the 5 vertical thick dotted lines indicate the~5  moments in~time presented in~Figures~\ref{fig_xzmaps_f1_nr10_aBp1} -- \ref{fig_spec_f1_nr10_aBp1}.}
\label{fig_flux_Bphi_f1_nr10_aBp1}
\end{figure}

Figure~\ref{fig_flux_Bphi_f1_nr10_aBp1} compares in~detail the~time evolutions of~the~toroidal magnetic flux
(calculated as $\Psi_{\rm B\phi} \propto \left<|B_y|\right>$ with averaging over $x$ and $z$ along the~$y=0$ plane), the~total magnetic energy $\mathcal{E}_B$, the~mean axial electric field at~$r = R_0$, and~the~maximum particle energy as~a~fraction of~the~energy limit,
$\gamma_{\rm max}/\gamma_{\rm lim}$.
The~first thing to~notice is~that the~relative decrease of~toroidal magnetic flux (the thick green line) is~very similar to~the~relative decrease of~total magnetic energy (the thin green line), the~main difference being an~earlier onset of~the~toroidal flux dissipation.
The~toroidal magnetic flux clearly shows two phases of~magnetic dissipation -- the~fast magnetic dissipation phase for~$1.0 < ct/L < 1.6$, followed by~a~slow magnetic dissipation phase
\citep[this is~consistent with the~results of][]{2022MNRAS.510.2391B}.

Let us now discuss the~evolution of~the~mean axial electric field $\left<E_z\right>(R_0)$.
The fast magnetic dissipation phase involves a~temporary  spike of~$\left<E_z\right>(R_0)$
peaking at the~level of~$\left<E_z\right>_{\rm peak} \simeq 0.043B_0$ for $t \sim (1.3\,\text{---}\,1.4)L/c$, simultaneous with the~most rapid increase of~$\gamma_{\rm max}$.\footnote{The fastest e-folding growth time scale of~$\left<E_z\right>(R_0)$ has been estimated as $\tau_{Ez} \simeq 0.046L/c \simeq 0.9R_0/c$ over the~period of~$1.0 < ct/L < 1.1$. Hence, the~duration of~the~fast magnetic dissipation phase corresponds to~$\simeq 13\tau_{Ez}$.}
To illustrate the~connection between the~electric field strength at $r = R_0$ and evolution of~$\gamma_{\rm max}$, consider a~slightly different time period of~$1.2 < ct/L < 1.7$, during which $\left<E_z\right>(R_0)$ exceeds the~level of~$0.2\left<E_z\right>_{\rm peak}$.
During that time period of~$\Delta t \simeq 0.5L/c$, $\gamma_{\rm max}$ increases by~$\Delta\gamma_{\rm max} \simeq 18.8\Theta_0 \simeq 0.42\gamma_{\rm lim}$.
This energy gain corresponds to~linear acceleration 
by~the~average electric field of~$\left<E_{\rm acc}\right>/B_0 = (L/\rho_0)^{-1}\Delta(\gamma_{\rm max}/\Theta_0)/\Delta(ct/L) \simeq 0.042 \simeq \left<E_z\right>_{\rm peak}/B_0$.
Hence, the electric field strength $\left<E_{\rm acc}\right>$ required to~explain the~acceleration of~the~most energetic particles during that~period is~consistent with~$\left<E_z\right>_{\rm peak}$\footnote{One can note that, on~one hand the~average axial electric field at~$R_0$ during that~time period should~be roughly $\simeq 0.6\left<E_z\right>_{\rm peak}$, on~the~other hand the~axial field is~somewhat stronger for~$r < R_0$.}.
Our~analysis of~individually tracked particles confirms that the~most energetic ones are indeed accelerated within the~core region ($r < R_0$) predominantly by~the~positive axial electric field during the~fast magnetic dissipation phase, as~stated by~\cite{2018PhRvL.121x5101A}.

Typical energy gain of~an~energetic~particle
can~be derived directly from the~duration $\Delta t$ of~the~fast magnetic dissipation phase:
\bea
\frac{\Delta\gamma}{\gamma_{\rm lim}} &=& \frac{1}{2}\left(\frac{L}{20R_0}\right)\left(\frac{c\Delta t}{L/2}\right)\frac{\left<E_{\rm acc}\right>}{B_0/20}\,,
\eea
where $\left<E_{\rm acc}\right>$ is the~average electric field component along the~particle velocity vector.
This means that a~relatively short duration $\Delta t \sim L/2c$, in~combination with realistic electric field strengths,
is sufficient to~explain one~half of~the~`confinement' limit (also known as the~Hillas limit) on~the~particle energy gain.

\begin{figure*}
\includegraphics[width=0.497\textwidth]{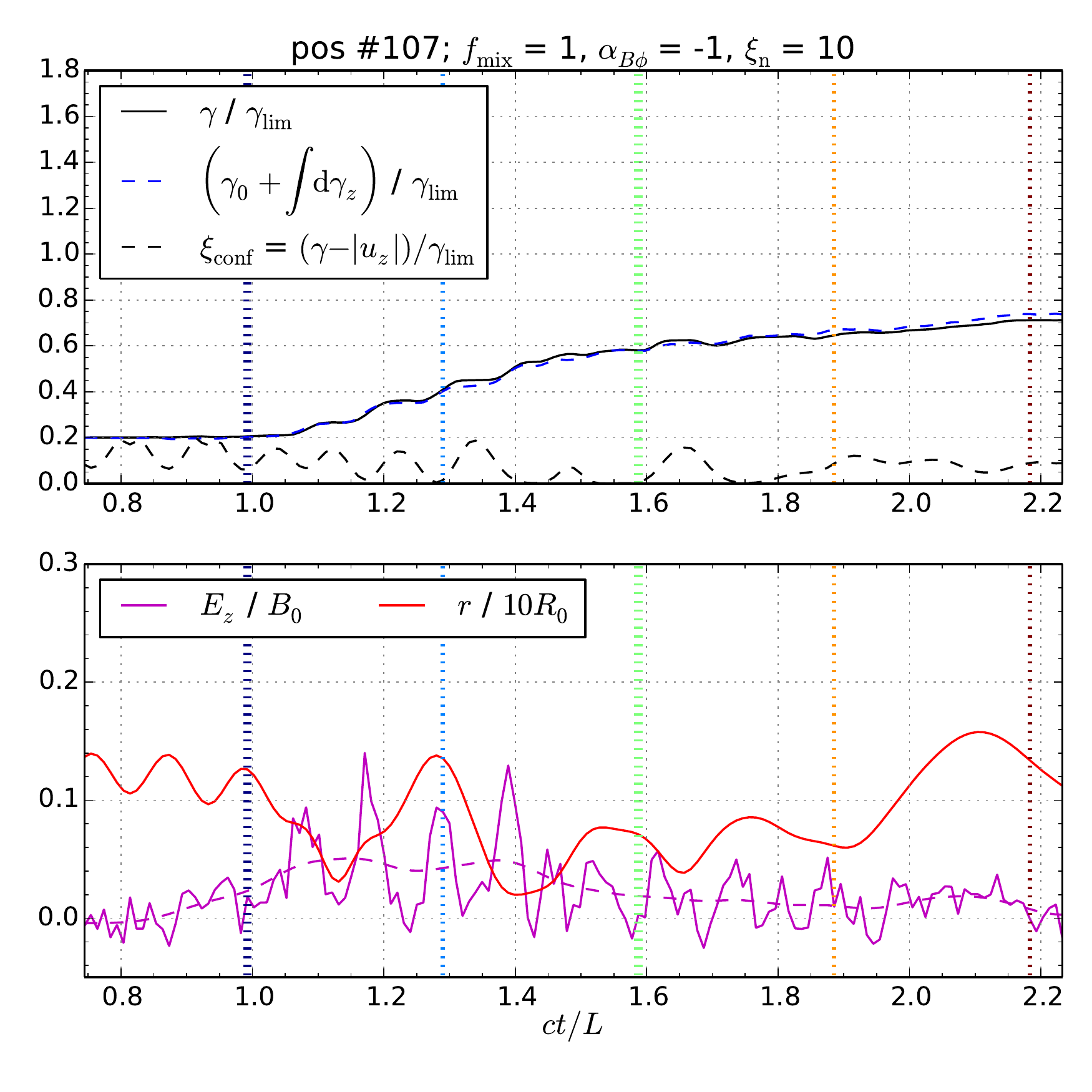}
\includegraphics[width=0.497\textwidth]{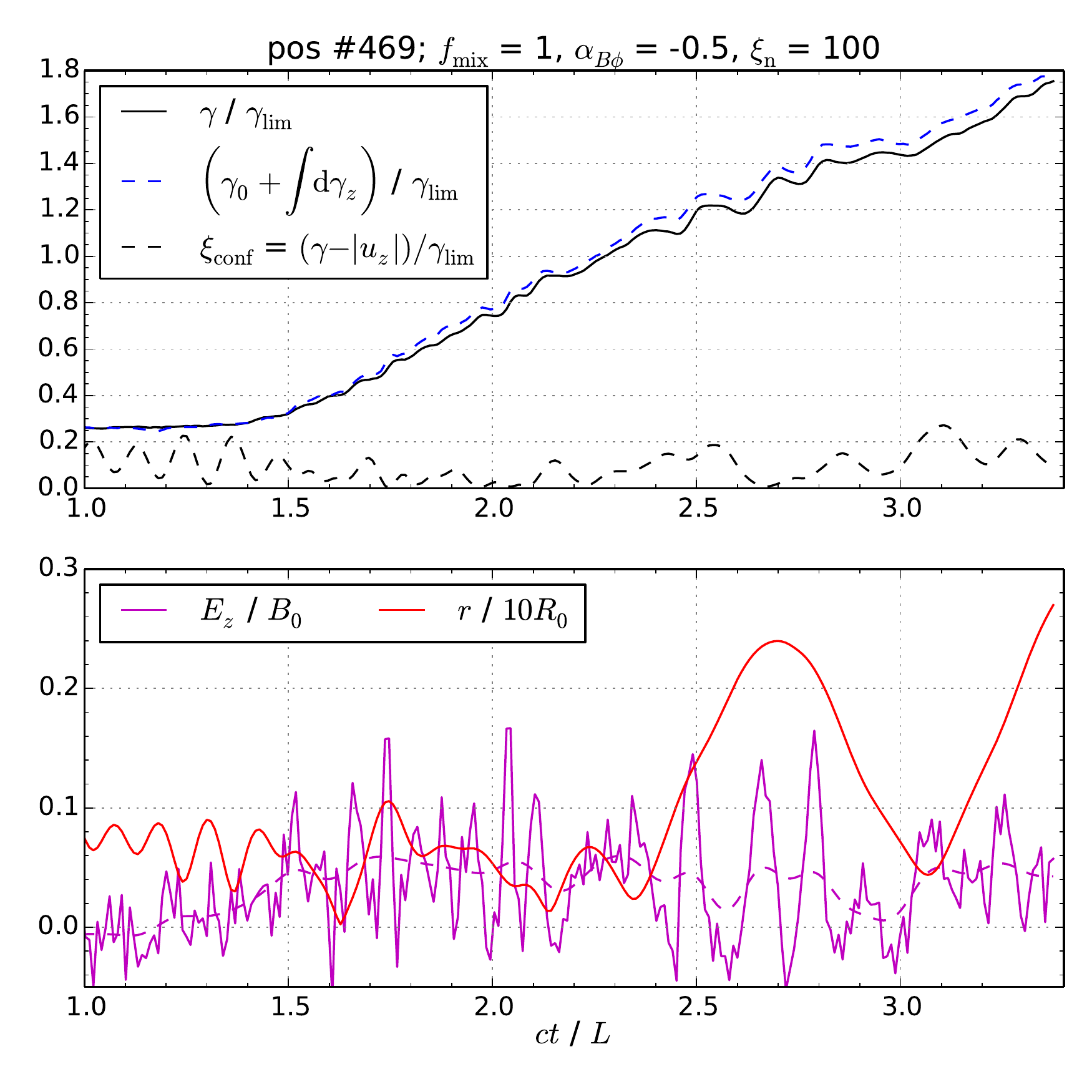}
\caption{Acceleration histories of~single energetic tracked positrons in~the~reference simulation f1\_$\alpha$-1\_$\xi$10 (left panels) and in~the~simulation f1\_$\alpha$-05\_$\xi$100 (right panels).
In the~top panels,
the solid black lines show the~particle Lorentz factor $\gamma$ normalized to~the~energy limit $\gamma_{\rm lim} = eB_0R_0/(mc^2)$;
the dashed black lines show the~particle confinement indicator $\xi_{\rm conf} = (\gamma-|u_z|)/\gamma_{\rm lim}$ (Appendix~\ref{app_conf});
and the~dashed blue lines show what would be the~particle Lorentz factor due to~the~work done by~the~axial electric field component $E_z$.
In the~bottom panels,
the solid magenta lines show the~local axial electric field component $E_z$ (the dashed magenta lines show its moving average);
and the~solid red lines show the~particle radial coordinate $r$ in~units of~$10R_0$.
The 5 vertical thick dotted lines in~the~left panels indicate the~5 moments in~time presented in~Figures \ref{fig_xzmaps_f1_nr10_aBp1} --~\ref{fig_spec_f1_nr10_aBp1}.}
\label{fig_ion_histories}
\end{figure*}

As a~representative example, let us consider the~acceleration history of~an individual energetic particle.
The~left panels of~Figure~\ref{fig_ion_histories} show the~history of~an energetic positron (denoted as `pos \#107') from the~reference simulation.
This  positron is energized from $\simeq 0.2\gamma_{\rm lim}$ to~$\simeq 0.6\gamma_{\rm lim}$ during the~fast magnetic dissipation phase.
The energy gain of~$\Delta\gamma \simeq 0.4\gamma_{\rm lim}$ over the~time scale of~$\Delta t \simeq 0.6L/c$ requires an average electric field of~$\left<E_{\rm acc}\right> \simeq B_0/30$, which is consistent with the axial electric field $E_z$ experienced by the~positron.
In~Figure~\ref{fig_ion_histories} top panels, the~dashed blue line shows that the~acceleration of~this particle can be attributed almost exclusively to~the~action of~the~axial electric field component $E_z$.
Likewise, it can be demonstrated that the~action of~the~electric field component parallel to~the~local magnetic field is negligible in~this case.
During most of~the~fast magnetic dissipation phase and until $t \simeq 2L/c$, this positron is located within the~central core at $r < R_0$.
In order to~determine whether this  positron is confined to~the~central core by~toroidal magnetic field, we calculate the~\emph{particle confinement indicator} defined as $\xi_{\rm conf} = (\gamma - |u_z|) / \gamma_{\rm lim}$ (see Appendix~\ref{app_conf}).
In the~present case we find that $\xi_{\rm conf} < 0.2$ (other energetic particles in~this simulation reached $\xi_{\rm conf} \simeq 0.3$), which means that this particle is indeed confined (as are the~others).
The acceleration rate slows down significantly around the~end of~the~fast magnetic dissipation phase at $t \simeq 1.6L/c$.
The sole reason for slower acceleration is that the~electric fields within the~central core become weaker.

We have thus demonstrated in~our reference case that  the~evolution of~the~toroidal magnetic field includes the~fast magnetic dissipation phase, the~ short duration of~which can explain a~significant part of~the~particle energy limit $\gamma_{\rm lim}$.

\begin{figure*}
\includegraphics[width=\textwidth]{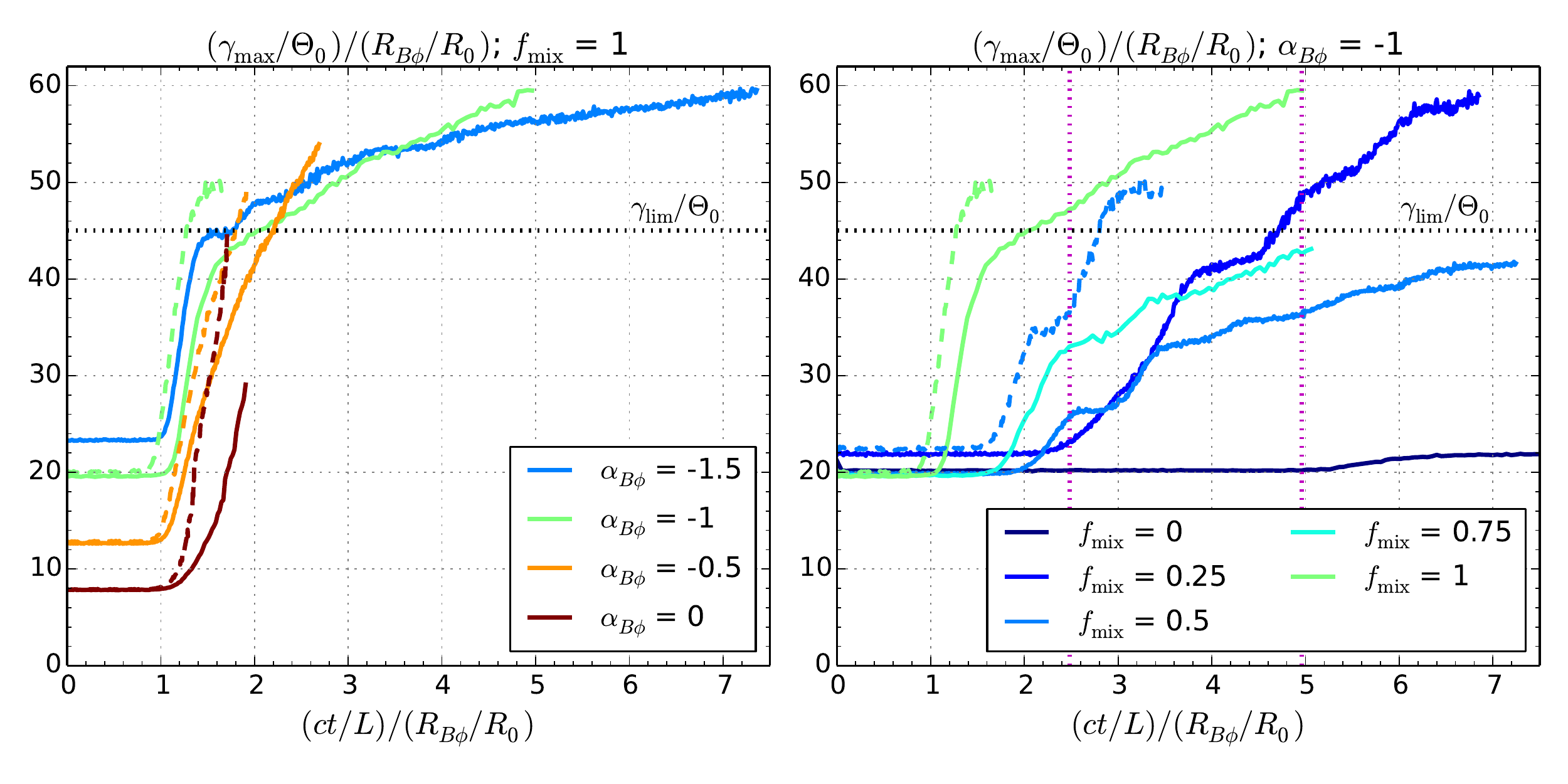}
\caption{
Time evolutions of~the~maximum particle energy $\gamma_{\rm max}$ evaluated at the~level of~$10^{-4}$ of~the~$u^2\,{\rm d}N/{\rm d}u$ particle distributions normalized to~peak at unity at $t=0$.
The results are compared for two series of~simulations:
the left panel shows simulations for $f_{\rm mix} = 1$ and different values of~$\alpha_{\rm B\phi}$,
and the~right panel shows simulations for $\alpha_{\rm B\phi} = -1$ and different values of~$f_{\rm mix}$.
 Both $\gamma_{\rm max}$ and the~simulation time $t$ are scaled by~the~characteristic radius $R_{\rm B\phi}$, which depends on $\alpha_{\rm B\phi}$.
The solid lines indicate the~cases of~$\xi_n = 10$, and the~dashed lines the~cases of~$\xi_n = 100$.
The horizontal dotted black lines indicate the~\emph{confinement} energy limit $\gamma_{\rm lim}/\Theta_0 \equiv R_0/\rho_0 = 45$.
The vertical dotted magenta lines in the right panel indicate the times at which particle distributions are compared in~Figure~\ref{fig_spec_fmix}.
The simulations are interrupted at different times, before the~perturbations reach the~$x,y$ boundaries.}
\label{fig_gamma_max}
\end{figure*}

\subsubsection{The effects of~$\alpha_{\rm B\phi} > -1$ and $f_{\rm mix} < 1$}
\label{sec_res_accel_alpha}

In our other simulations we have explored the~effects of~three parameters: the~pressure mixing ratio $f_{\rm mix}$, the~toroidal field index $\alpha_{\rm B\phi}$, and the~density contrast~$\xi_n$.
Figure~\ref{fig_gamma_max}  presents the~time evolutions of~the~maximum particle energy $\gamma_{\rm max}$
 --- the left panel compares all $f_{\rm mix} = 1$ cases (for different $\alpha_{B\phi}$ and $\xi_n$),
and the right panel compares all $\alpha_{B\phi} = -1$ cases (for different $f_{\rm mix}$ and $\xi_n$).
The~reference simulation f1\_$\alpha$-1\_$\xi$10 is~shown with the~solid green line in~each panel.
The~values of~$\gamma_{\rm max}$, as well as the~simulation time $t$, have been scaled by~the~characteristic radius $R_{\rm B\phi}$, which depends on~$\alpha_{\rm B\phi}$ (see Section~\ref{sec_config}).
One can see that
 $\gamma_{\rm max}$ begins to~increase significantly from its~initial value after at~least $ct/L \gtrsim R_{\rm B\phi}/R_0$.
For~$f_{\rm mix} = 1$, it~typically shows a~single phase of~rapid growth to~the~energy limit $\gamma_{\rm lim}$ at~the~rates corresponding to~acceleration by~electric fields~$E_{\rm acc} \sim (0.05\,\text{---}\,0.08)B_0$.
In~cases where $\gamma_{\rm lim}$ is~exceeded, further increase of~$\gamma_{\rm max}$ slows down significantly
($E_{\rm acc} < 0.01B_0$ for~$\alpha_{B\phi} = -1.5,-1$ and~$\xi_n = 10$).
The~evolution of~$\gamma_{\rm max}$ is~more complex in~the~$f_{\rm mix} < 1$ cases; in~some of~them the~$\gamma_{\rm lim}$ limit is~reached in~two stages, with~$E_{\rm acc}$ only up~to~$\sim 0.02B_0$.
The~results for~high density contrast~$\xi_n = 100$ suggest that the~$\gamma_{\rm lim}$ is~also relevant to~those cases.
%
Among the~$\alpha_{\rm B\phi} > -1$ cases, the~rescaled $\gamma_{\rm max}/(R_{\rm B\phi}/R_0)$ reaches the $\gamma_{\rm lim}$ limit for $\alpha_{\rm B\phi} = -0.5$, and for $\alpha_{\rm B\phi} = 0$ and $\xi_n = 100$,
although we~cannot say whether it~would flatten subsequently if~these simulations could~be continued.
Particle acceleration appears to~be faster for~higher values of~$f_{\rm mix}$, this is consistent with the~correspondingly shorter instability growth time scales $\tau_{\rm min}$, as~shown in~Figure~\ref{fig_tau_min}.

\begin{figure*}
\includegraphics[width=0.5\textwidth]{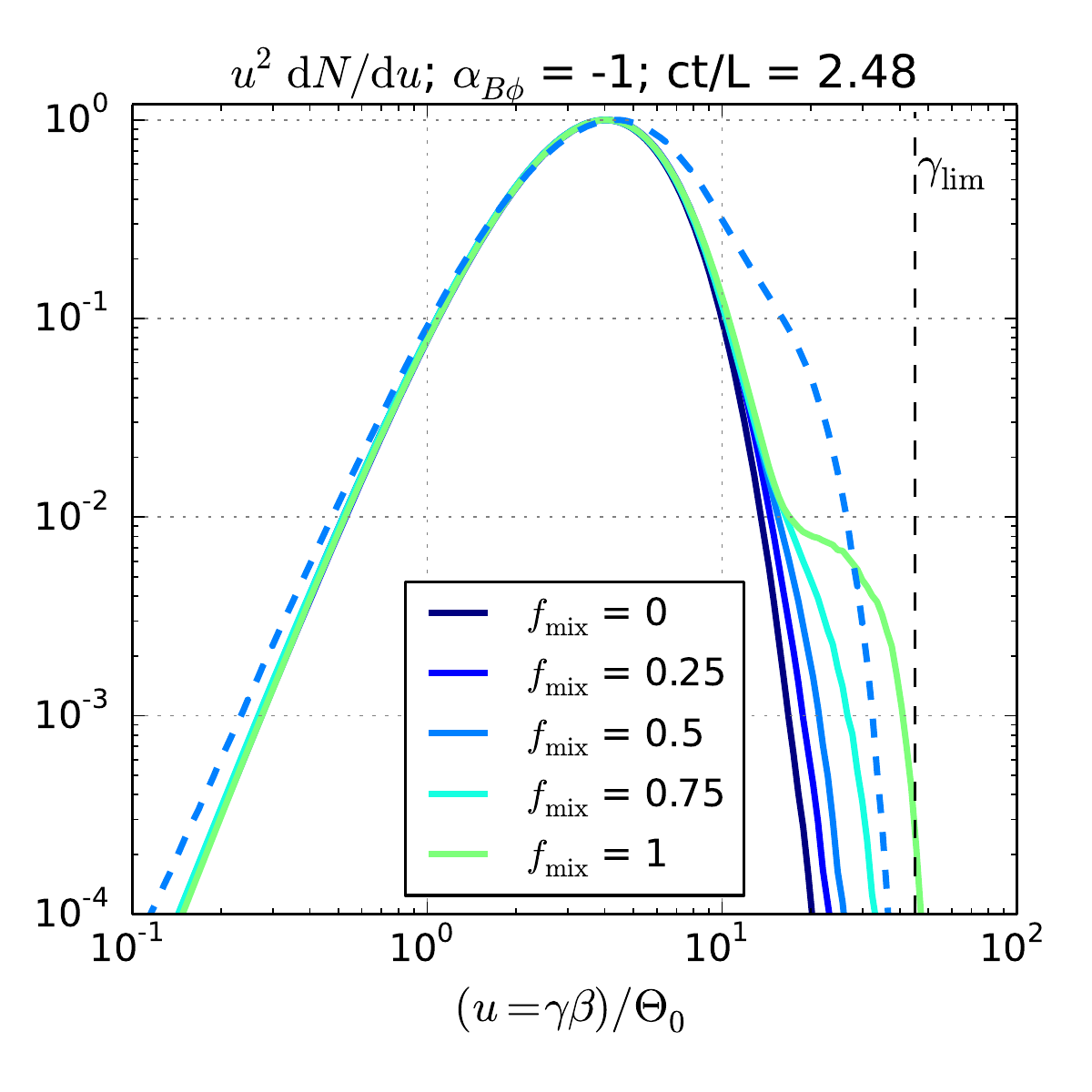}
\includegraphics[width=0.5\textwidth]{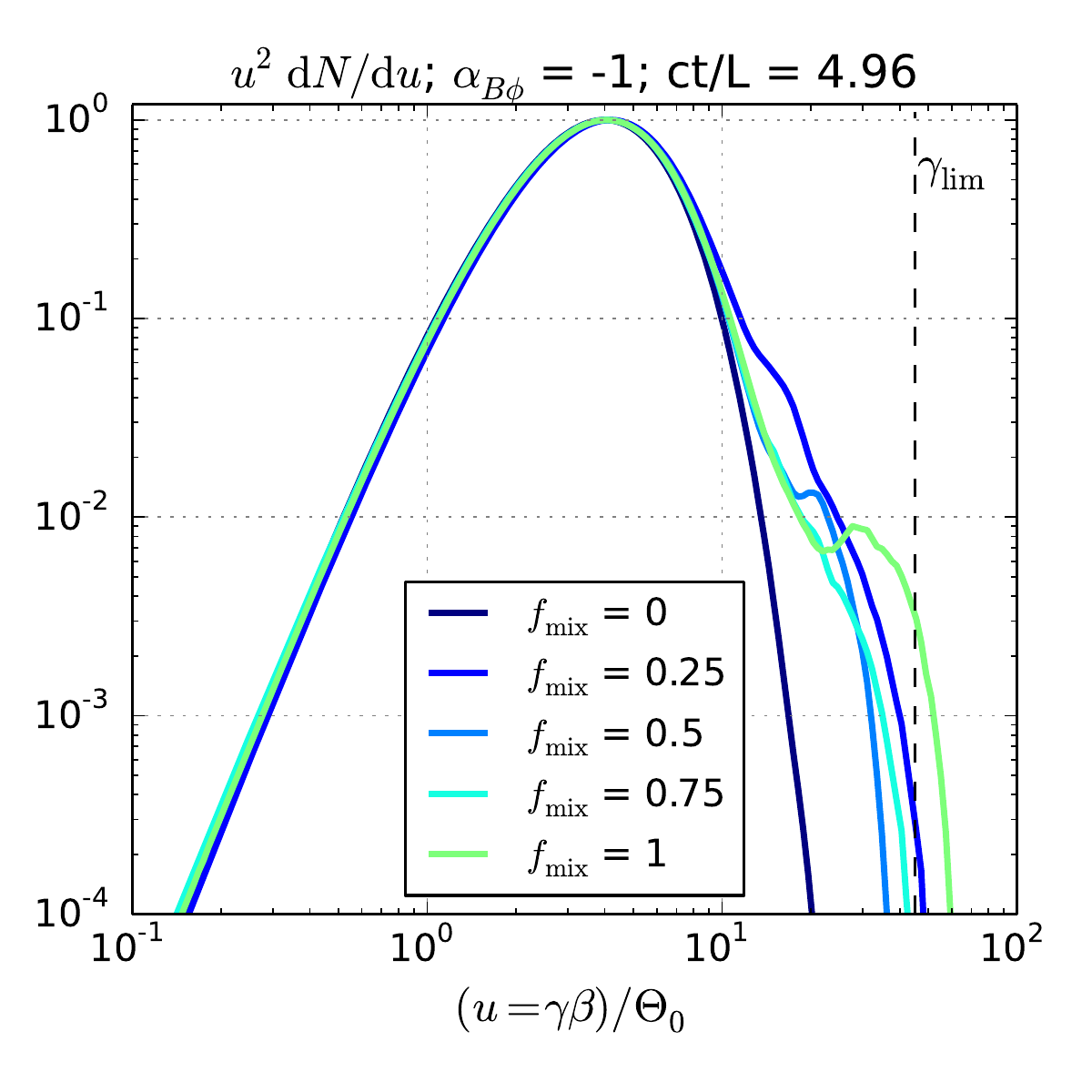}
\caption{Particle momentum distributions $u^2\,{\rm d}N/{\rm d}u$
(combined for both electrons and positrons and normalized to peak at unity)
compared at~two moments in~time (left/right panel) between simulations for $\alpha_{B\phi} = -1$, and different values of $f_{\rm mix}$.
The solid lines indicate the~cases of~$\xi_n = 10$, and the~dashed lines the~cases of~$\xi_n = 100$.
The vertical black dashed lines indicate the~\emph{confinement} energy limit $\gamma_{\rm lim} = 45\Theta_0$.}
\label{fig_spec_fmix}
\end{figure*}

Figure~\ref{fig_spec_fmix} compares the~particle momentum distributions between {\rm the}~simulations for~$\alpha_{B\phi} = -1$ (hence $R_{\rm B\phi} = R_0$), $\xi_n = 10$ (with one exception of $\xi_n = 100$), and different values of the pressure mixing parameter $f_{\rm mix}$.
This comparison is presented at two simulation times, because these simulations evolve at different rates.
At $ct/L \simeq 2.5$, when some $f_{\rm mix} = 1$ simulations are already in the slow acceleration phase while the $f_{\rm mix} = 0.25$ case is just at the onset of fast acceleration, the high-energy distribution tails appear fairly regular ($\gamma_{\rm max}$ increases with increasing $f_{\rm mix}$), even if very weak for $f_{\rm mix} \le 0.5$.
At~$ct/L \simeq 5$, when most simulations are in their final stages, the~high-energy tails are~clearly present for~$f_{\rm mix} \ge 0.25$ (in~the~case $f_{\rm mix} = 0$ particle acceleration begins only after $ct/L=5$), although they are~rather irregular with~bumps instead of~power-laws.
At~$ct/L \simeq 5$, the~fractions of~particles contained in~these high-energy tails are~roughly $\simeq 1\%$ and they carry $\simeq 2\%$ of the total particle energy.
 The~case of~high density contrast $\xi_n = 100$ for~$f_{\rm mix} = 0.5$ at~$ct/L \simeq 2.5$ shows significantly higher fractions: $\simeq 5\%$ of~particles carrying $\simeq 14\%$ of~particle energy.

\begin{figure*}
\includegraphics[width=\textwidth]{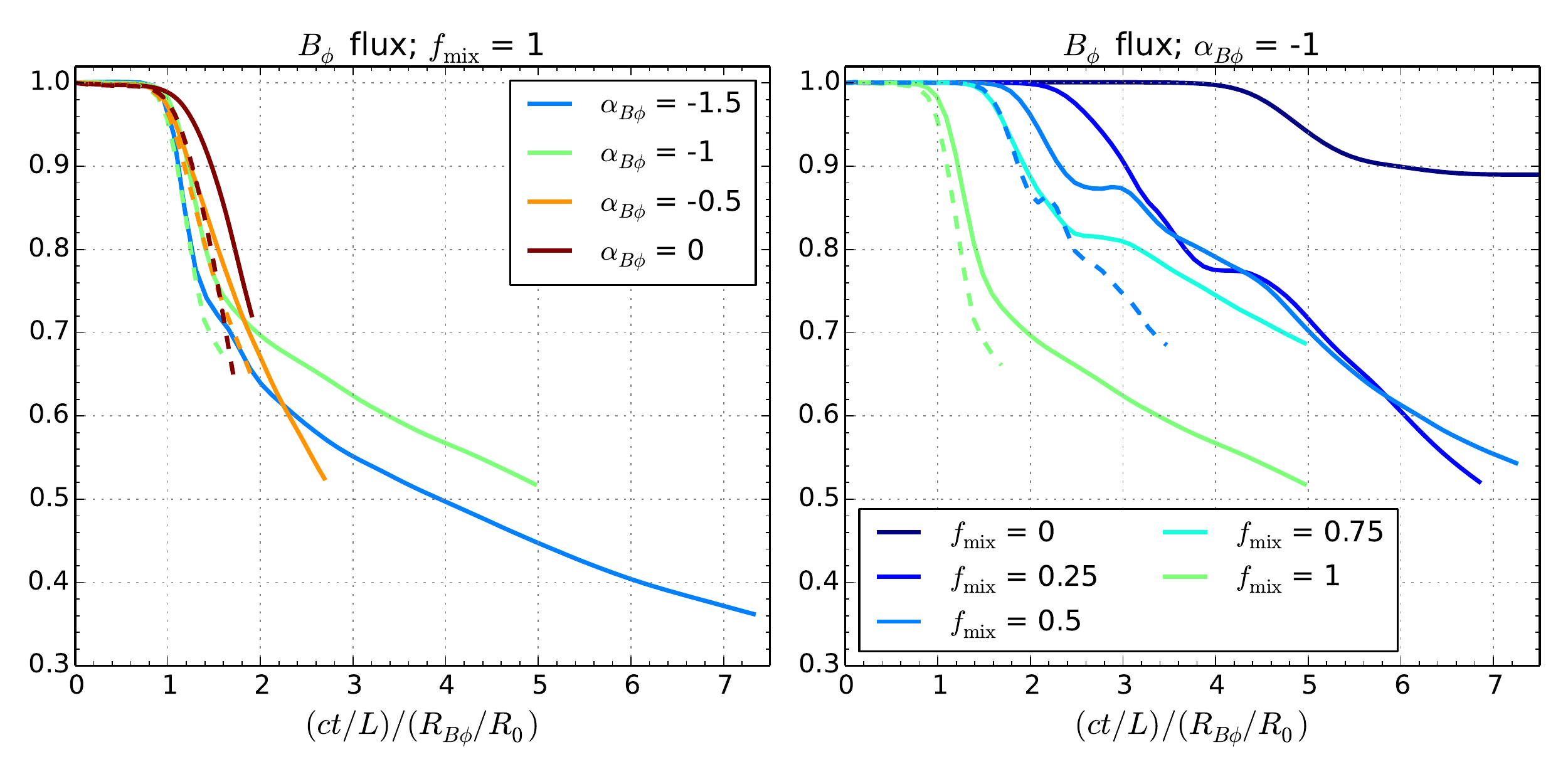}
\caption{Time evolutions of~the~toroidal magnetic field flux $\Psi_{\rm B\phi}$, normalized to~unity at $t=0$, compared for two series of~simulations,  using simulation time $t$ scaled by~the~characteristic radius $R_{\rm B\phi}$.
The line types are the~same as in~Figure~\ref{fig_gamma_max}.}
\label{fig_flux_Bphi}
\end{figure*}

Figure~\ref{fig_flux_Bphi} compares
the~time evolutions of~the~toroidal magnetic flux $\Psi_{\rm B\phi}$ (normalized to its initial value)
 for~all simulations.
In~all cases we~observe an~initial period of~almost constant $\Psi_{\rm B\phi}$ followed by~the~onset of~the~fast magnetic dissipation phase,
which in~some cases is~followed by~a~transition to~a~slow magnetic dissipation phase.
The~simulation time has been scaled by~$R_{\rm B\phi}$, so that for $f_{\rm mix} = 1$ the~fast magnetic dissipation phases begin roughly at~$t \simeq (L/R_0)R_{\rm B\phi}/c \simeq 20R_{\rm B\phi}/c$.
On the~other hand, for $\alpha_{\rm B\phi} = -1$,
these onsets are delayed for the~$B_z$-balanced cases of~$f_{\rm mix} < 1$.
The overall relative decrease of~$\Psi_{\rm B\phi}$ is in~the~range of~$\sim (30\,\text{---}\,60)\%$.
Transitions from  the~fast magnetic dissipation phase to~the~slow magnetic dissipation phase can be seen in~most cases of~$\alpha_{\rm B\phi} \le -1$ (although in~some $B_z$-balanced cases the~histories of~$\Psi_{\rm B\phi}(t)$ are more complicated).
Such a~transition is not seen in~the~cases of~$\alpha_{\rm B\phi} > -1$ (although there is a~hint of~that in~the~case f1\_$\alpha$-05\_$\xi$10).
Comparing Figure~\ref{fig_flux_Bphi} with Figure~\ref{fig_gamma_max}, a~connection between the~evolutions of~$\gamma_{\rm max}$ and $\Psi_{\rm B\phi}$ can be noticed.
Simulations in~which $\gamma_{\rm max} \gg \gamma_{\rm lim}$ are the~same in~which the~fast magnetic dissipation phase is not
 complete
before the~perturbations reach the~boundaries
 and the simulation is interrupted.
The episodes of~rapid increase of~$\gamma_{\rm max}$ are simultaneous with a~rapid decrease of~$\Psi_{\rm B\phi}$.

As an example, let us consider in~more detail the~case $f_{\rm mix} = 1$, $\alpha_{\rm B\phi} = -0.5$ and $\xi_n = 100$ ($R_{\rm B\phi} = 1.55R_0$; the dashed orange lines in~the~left panels of~Figures~\ref{fig_gamma_max} and~\ref{fig_flux_Bphi}).
In this case, the~fast magnetic dissipation phase begins at $t \simeq 1.4L/c$ and lasts till the~end of~the~simulation at $t \simeq 3.4L/c$.
During that time, $\gamma_{\rm max}$ increases linearly, reaching the~level of~$\gamma \simeq 90\Theta_0 = 2\gamma_{\rm lim}$.
The~right panels of~Figure~\ref{fig_rprof_Bphi_meanEz} show the~time evolutions of~the~radial profiles of~$\left<B_\phi\right>(r)$ and $\left<E_z\right>(r)$, extending until the~simulation ends.
The time separations between successive lines are $\Delta t \simeq 0.3L/c$, the~same as for the~reference simulation.
Compared with the~reference simulation, efficient dissipation of~toroidal magnetic field progresses towards larger radii, essentially until it reaches the~outer cut-off region.
This corresponds to~much more extended radial profiles of~the~net mean axial electric field $\left<E_z\right> > 0$.
The key difference from the~reference case is that the~axial electric field does not decay in~the~central core, settling at the~value of~$\left<E_z\right> \sim 0.04B_0$ for $r < R_0$, which is 4 times higher than in~the~reference case.
This is also reflected in~the~fact that dissipation of~toroidal magnetic field in~the~central core proceeds to~deeper levels.
The e-folding growth time scale of~$\left<E_z\right>(R_0)$ has been estimated as $\tau_{Ez} \simeq 0.11L/c \simeq 2.2R_0/c$ over the~period of~$1.1 < ct/L < 1.7$, about $2.4$ times longer than in~the~reference case.
This means that in~this simulation the~fast magnetic dissipation phase lasts for at least $\simeq 18\tau_{Ez}$, longer in~relation to~$\tau_{Ez}$ than in~the~reference case.

Having a~radially decreasing net axial electric field is key for efficient dissipation of~the~average toroidal magnetic field, which is governed by~the~Maxwell-Faraday equation:
\be
\frac{\partial\left<B_\phi\right>}{c\,\partial t} = \frac{\partial\left<E_z\right>}{\partial r} < 0\,.
\ee
For example, the~steepest radial gradients of~the~net axial electric field in~both presented cases are $\Delta\left<E_z\right> \simeq -0.04B_0$ over $\Delta r = R_0$, which corresponds to~the~peak magnetic dissipation rate of~$\Delta\left<B_\phi\right>/B_0 \simeq -0.8 c\Delta t/L$.
On the~other hand, in~the~final state of~the~simulation f1\_$\alpha$-05\_$\xi$100, we have $\Delta\left<E_z\right> \simeq -0.045B_0$ over $\Delta r = 9R_0$, which yields $\Delta\left<B_\phi\right>/B_0 \simeq -0.1 c\Delta t/L$.
These dissipation rates are consistent with the~results presented in~the~top panels of~Figure~\ref{fig_rprof_Bphi_meanEz}.

The~right panels of~Figure~\ref{fig_ion_histories} show the~acceleration history of~an energetic positron (denoted as `pos \#469') in~the~simulation f1\_$\alpha$-05\_$\xi$100.
This particle accelerates almost linearly from $\gamma \simeq 0.3\gamma_{\rm lim}$ at $t = 1.4L/c$ to~$\gamma \simeq 1.75\gamma_{\rm lim}$ at $t = 3.4L/c$, which means an energy gain of~$\Delta\gamma \simeq 1.45\gamma_{\rm lim}$ over the~period of~$\Delta t = 2L/c$.
This acceleration is dominated by~action of~the~axial electric field, which is sustained at the~level of~$E_z \simeq 0.04B_0$ throughout this time range.
For most of~the~acceleration period (until $t \simeq 2.4L/c$), the~particle is located within the~central core ($r < R_0$), and its confinement indicator is $\xi_{\rm conf} < 0.2$ even as it starts oscillating outside the~core at later times.
This particle is well confined by~the~toroidal magnetic fields, and yet it is able to~accelerate beyond~$\gamma_{\rm lim}$.
Other energetic particles in~this simulation reach $\xi_{\rm conf} \simeq 0.6$, and yet they do not escape their confinement and keep accelerating.

\subsection{Parallel vs. perpendicular acceleration}
\label{sec_res_accel_Epara}

We have addressed the~problem of~comparing the~relative importance of~parallel and perpendicular electric fields in~particle acceleration by~analyzing large samples of~individually tracked particles, denoted with index $i$, for which we recorded as functions of~simulation time their energy histories $\gamma_i(t)$, as well as the~local magnetic and electric field vectors $\bm{B}_i(t),\bm{E}_i(t)$.
Out of~these samples we have selected \emph{energetic particles} defined by~two criteria: $\max[\gamma_i(t)] \ge 15\Theta_0$ and $\max[\gamma_i(t)] - \min[\gamma_i(t)] \ge 10\Theta_0$.
Let ${\rm d}\gamma_i(t) = (q_i/mc)(\bm\beta_i\cdot\bm{E}_i)(t)\,{\rm d}t$ represent the~instantaneous energy change of~the~$i$-th particle between the~times $t$ and $t + {\rm d}t$,
where $q_i = \pm e$ is the~particle charge.
The corresponding contributions from perpendicular and parallel electric fields are ${\rm d}\gamma_{i,\perp}(t) = (q_i/mc)(\bm\beta_i\cdot\bm{E}_{i,\perp})(t)\,{\rm d}t$ and ${\rm d}\gamma_{i,\parallel}(t) = (q_i/mc)(\bm\beta\cdot\bm{E}_{i,\parallel})(t)\,{\rm d}t$, respectively.
The total energy gain of~the~$i$-th particle has been calculated as $\Delta\gamma_i = \int_{t=0}^{t_{{\rm peak},i}}{\rm d}\gamma_i$,
interrupting the~integration at the~moment $t_{{\rm peak},i}$ at which the~particle energy $\gamma_i(t)$ attains a~global peak.
The corresponding energy gains due to~the~perpendicular and parallel electric fields are $\Delta\gamma_{i,\perp} = \int_{t=0}^{t_{{\rm peak},i}}{\rm d}\gamma_{i,\perp}$ and $\Delta\gamma_{i,\parallel} = \int_{t=0}^{t_{{\rm peak},i}}{\rm d}\gamma_{i,\parallel}$, respectively.

\begin{figure*}
\includegraphics[width=\textwidth]{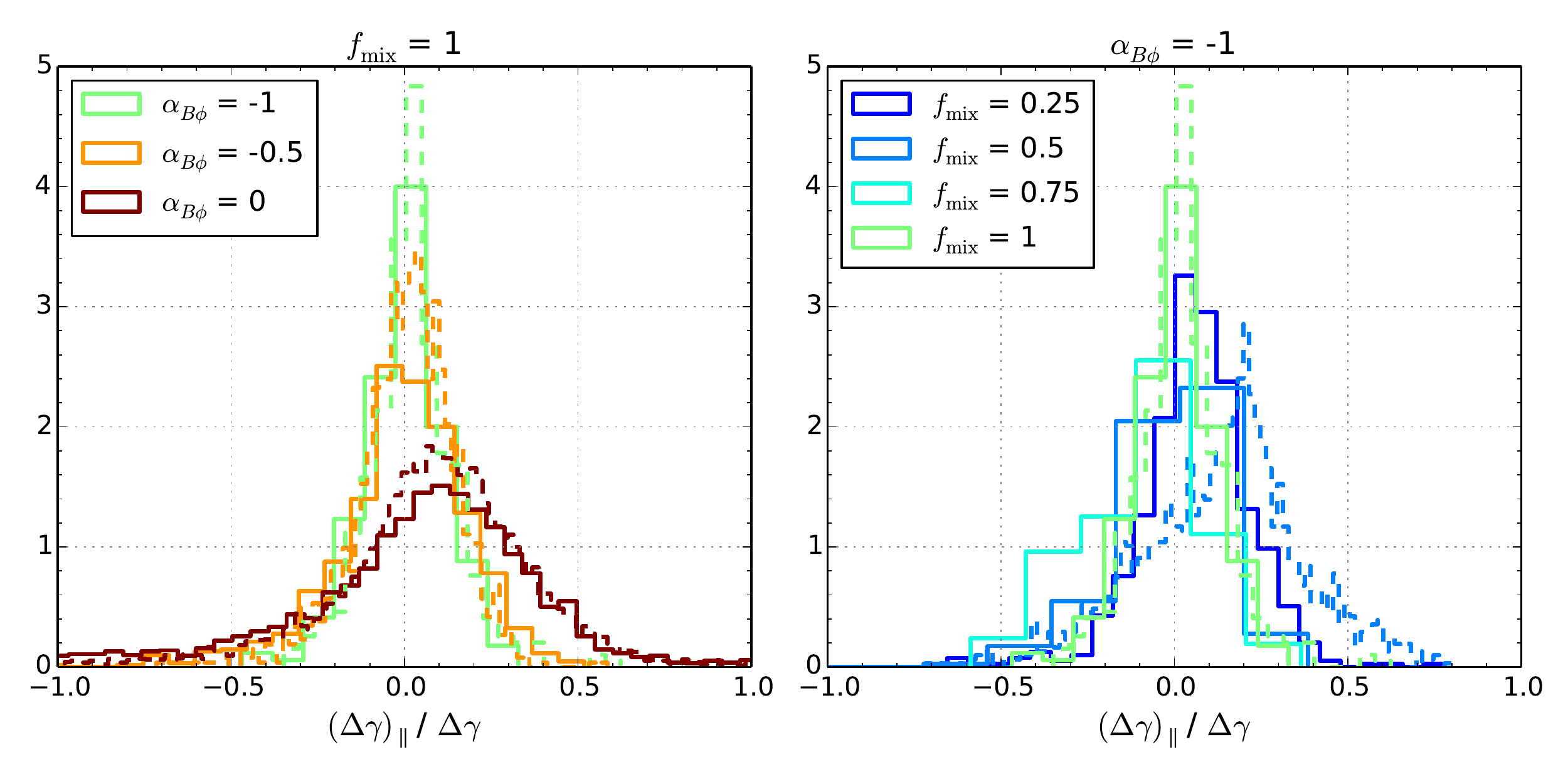}
\caption{Distributions of~the~relative contribution $(\Delta\gamma)_\parallel$ of~the~electric field component $\bm{E}_\parallel$ parallel to~the~local magnetic field to~the~total energy gain $\Delta\gamma$ for complete samples of~energetic particles.
The line types are the~same as in~Figure~\ref{fig_gamma_max}.}
\label{fig_orbit_Dgpara}
\end{figure*}

Figure~\ref{fig_orbit_Dgpara} compares the~distributions of~the~$\Delta\gamma_{i,\parallel} / \Delta\gamma_i$ ratio for multiple simulations.
In most cases, the~distributions peak at $\Delta\gamma_{i,\parallel} \simeq 0$, which means that particle acceleration is dominated by~perpendicular electric fields.
However,
 in~the~f05\_$\alpha$-1\_$\xi$100 case, the~distribution peaks at~$\Delta\gamma_{i,\parallel} \simeq 0.2\Delta\gamma_i$.
Additional analysis of~this case reveals various and complex histories of individual energetic particles that attain this level of parallel acceleration, an example of which will be presented further in this subsection.
In the cases of~$f_{\rm mix} = 1$ and~$\alpha_{B\phi} = 0$, the~distributions of~$\Delta\gamma_{i,\parallel}$ peak at~$\simeq 0.1\Delta\gamma_i$.
Here, the~reason is~the~relatively long duration of~the~initial simulation phase when random fluctuations of~electric field contribute roughly equally to~$\Delta\gamma_{i,\parallel}$ and~$\Delta\gamma_{i,\perp}$.

\begin{figure}
\includegraphics[width=\columnwidth]{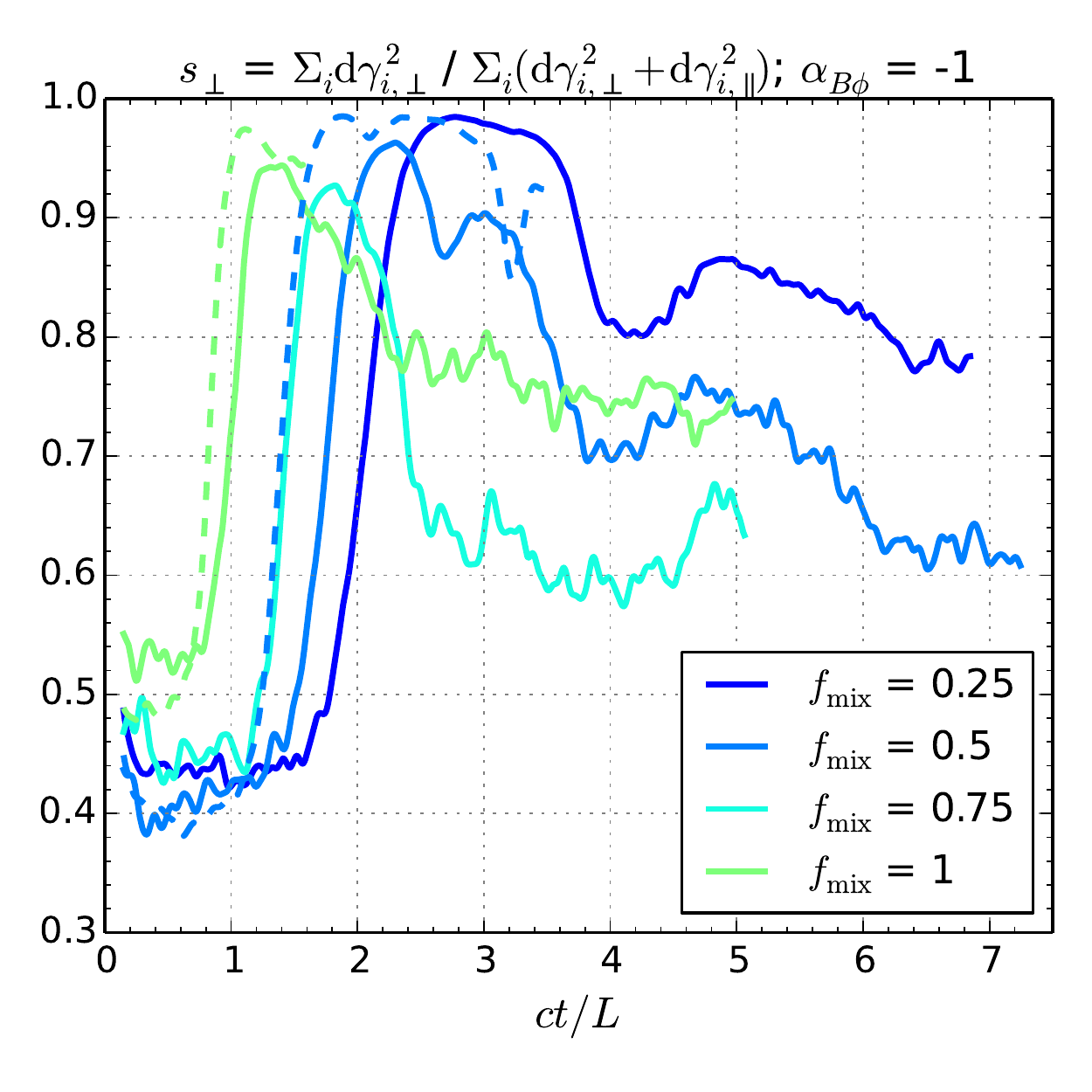}
\caption{Relative contribution of~perpendicular electric fields to~the~squared energy gains of~energetic particles (using the formula shown in~the~plot title) as~a~function of~simulation time for~simulations with~$\alpha_{\rm B\phi} = -1$  and~different values of~$f_{\rm mix}$ (indicated by~the~line color) and~$\xi_n$ (indicated by~the~line type, same as~in~Figure~\ref{fig_gamma_max}).}
\label{fig_orbit_Dgperp_t}
\end{figure}

Figure~\ref{fig_orbit_Dgperp_t} compares the~relative contributions
of perpendicular electric fields to~the~energization of~particles as a function of~simulation time for the~cases of~$\alpha_{\rm B\phi} = -1$ with different values of~$f_{\rm mix}$ and $\xi_n$.
These contributions have been summed over the~samples of~energetic particles as $S_\perp(t) = \sum_i{\rm d}\gamma_{i,\perp}^2(t)$ and $S_\parallel(t) = \sum_i{\rm d}\gamma_{i,\parallel}^2(t)$
 (summing over energy squares greatly reduces the~noise),
and~the~relative contribution has been calculated as $s_\perp(t) = S_\perp(t)/[S_\perp(t)+S_\parallel(t)]$.
Initially, for $t \lesssim L/c$, before the~fast magnetic dissipation phase, when particle energy changes are limited to~small random fluctuations, we find $s_\perp \sim 0.5$.
With the~onset of~the~fast magnetic dissipation phase, the~relative contribution of~perpendicular acceleration  increases to~$s_\perp \sim 0.95$,
followed by~a~slow irregular decrease to~the~level of~$s_\perp \sim 0.6\,\text{---}\,0.8$.
As~far as~we can~say, these results are~not sensitive to~the~density ratio $\xi_n$, but~they seem to~depend on~the~pressure mixing parameter $f_{\rm mix}$ in~the~non-linear phase.
For~$4.0 < ct/L < 4.5$, when the $s_\perp(t)$ functions for $f_{\rm mix} < 1$ achieve broad local minima, the~highest relative contribution of~perpendicular acceleration  is~$s_\perp \simeq 0.8$ for~$f_{\rm mix} = 0.25$, and~the~lowest is~$s_\perp \simeq 0.6$ for~$f_{\rm mix} = 0.75$~\footnote{ The~case f05\_$\alpha$-1\_$\xi$10 also achieves $s_\perp \simeq 0.6$, but~only for~$ct/L > 6$.}.
This dependence appears to~be~driven by~differences in~$S_\perp$ values ( during that time, it~is higher by~factor $\simeq 5$ 
 in~the~$f_{\rm mix} = 0.25$ case compared to~the~$f_{\rm mix} = 0.75$ case), rather than by~differences in~$S_\parallel$ values (higher by~only $\simeq 50\%$ for~the~same comparison).

\begin{figure}
\includegraphics[width=\columnwidth]{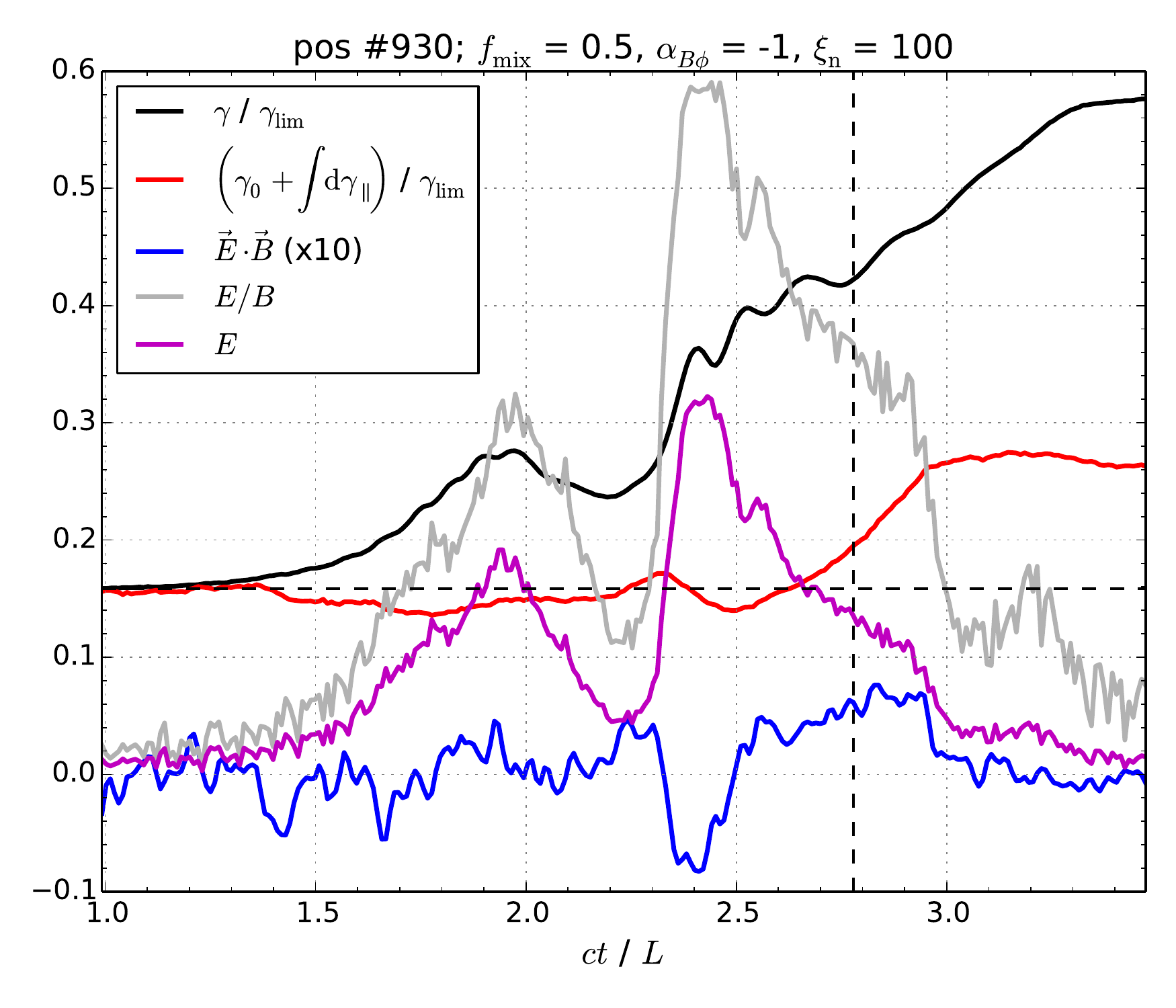}
\caption{Acceleration history of~the~energetic positron \#930 in~the~simulation f05\_$\alpha$-1\_$\xi$100 that shows a~significant contribution of~parallel electric fields to~its acceleration, with $\Delta\gamma_{i,\parallel} \simeq 0.25\Delta\gamma_i$.
The black solid line shows the~particle energy $\gamma_i(t)$ and the~red solid line shows the~integrated contribution of~parallel electric fields $\gamma_i(t=0) + \int{\rm d}\gamma_{i,\parallel}$, both normalized to~$\gamma_{\rm lim}$.
The magenta, gray and blue solid lines show the~local values of~$|\bm{E}|$, $|\bm{E}|/|\bm{B}|$ and $(\bm{E}\cdot\bm{B})\times 10$, respectively.
The vertical dashed line indicates the~moment presented in~Figure~\ref{fig_xzmaps_f05_aBp1_nr100}.}
\label{fig_history_ion930_f05_aBp1_nr100}
\end{figure}

We have searched the~sample of~energetic particles in~the~simulation f05\_$\alpha$-1\_$\xi$100 for 
an~illustrative example of~significant contribution of~parallel electric fields.
Figure~\ref{fig_history_ion930_f05_aBp1_nr100} shows the~acceleration history of~an energetic positron (denoted as `pos \#930' or with subscript `930'), which is characterized by the initial energy of $\gamma_{\rm 930,ini} \simeq 7.1\Theta_0$, and by $ct/L \simeq 3.5$ it reaches the peak energy of $\gamma_{\rm 930,peak} \simeq 25.9\Theta_0$, hence $\Delta\gamma_{\rm 930} \simeq 18.8\Theta_0 \simeq 0.42\gamma_{\rm lim} \simeq 2.6\gamma_{\rm 930,ini}$, of which $\simeq 25\%$ is the contribution from parallel electric fields.
Systematic acceleration by~parallel electric fields with $\bm{E}\cdot\bm{B} \sim 0.005 B_0^2$ is observed mainly in~the~period of~$2.5 < ct/L < 2.95$~\footnote{ This period of~time coincides with~the~second phase of~rapid increase of~the~maximum particle energy $\gamma_{\rm max}$ in~this simulation, taking it~beyond the~$\gamma_{\rm lim}$ limit (see the~dashed blue line in~the~right panel of~Figure~\ref{fig_gamma_max}).}.
During this time, the~total electric field strength at~the~position of~this particle is~$|\bm{E}| \sim (0.1\,\text{---}\,0.2)B_0 \sim (0.3\,\text{---}\,0.5)|\bm{B}|$.
As~it~happens, these are not the~strongest electric fields that this particle experiences.
At~$t \simeq 2.4L/c$, the~corresponding values are~$\bm{E}\cdot\bm{B} \simeq -0.008B_0^2$ and~$|\bm{E}| \simeq 0.3B_0 \simeq 0.58|\bm{B}|$.
At~that time our~particle experiences rapid acceleration, but the~contribution of~parallel electric fields up~to~that point is~not important.

\begin{figure*}
\includegraphics[width=\textwidth]{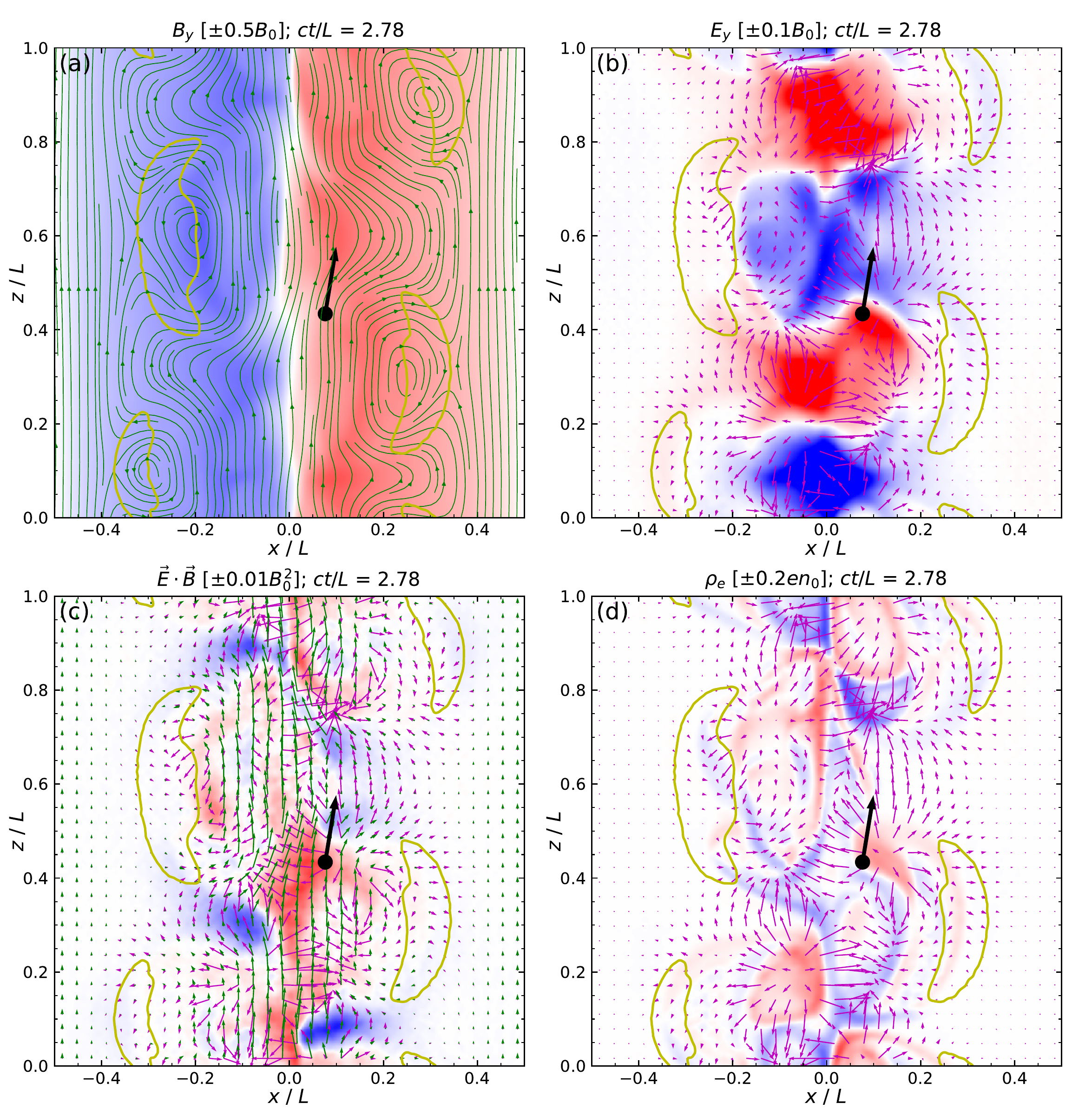}
\caption{A snapshot from the~simulation f05\_$\alpha$-1\_$\xi$100 at the~time $t = 2.78L/c$, when the~energetic positron \#930 introduced in~Figure~\ref{fig_history_ion930_f05_aBp1_nr100} (marked with the~black circles, with the~black arrows indicating the~in-plane velocity direction) experiences systematic acceleration by~electric field parallel to~the~local magnetic field.
Each panel shows an $(x,z)$-map  in~the~$y = 0$ plane, which  also contains the~energetic positron.
The color maps (red is positive, blue is negative) show:
(a) out-of-plane magnetic field component $B_y$,
(b) out-of-plane electric field component $E_y$,
(c) $\bm{E}\cdot\bm{B}$,
(d) charge density $\rho_e$.
The magenta vector fields show the~in-plane electric field $(E_x,E_z)$,
and the~green vector field (panel (c))  shows the~in-plane magnetic field $(B_x,B_z)$.
Panel (a) also includes a~streamplot of~$(B_x,B_z)$ with solid green lines.
The yellow contours indicate where $B_z = 0$.}
\label{fig_xzmaps_f05_aBp1_nr100}
\end{figure*}

Figure~\ref{fig_xzmaps_f05_aBp1_nr100} provides a~detailed context for the~parallel acceleration of~positron \#930.
This particle is located not far from the~central axis ($r \simeq 1.5R_0$), within a~large patch of~positive $\bm{E}\cdot\bm{B}$.
Additional analysis shows that this particle interacts with the~same patch for the~entire period of~parallel acceleration while propagating along a~helical trajectory (since the~dominant magnetic field components are still $B_z$ and $B_\phi$).
In~the~$(x,y)$ plane following the~particle along the~$z$ coordinate, the~patch is~seen to~rotate around the~central axis, and~the~particle motion appears to~be synchronized with~this rotation.
The~origin of~that $\bm{E}\cdot\bm{B} > 0$ patch is~ discussed in~Section~\ref{sec_disc_Epara}.

\subsection{Relation to current density and electric field structures}
\label{sec_res_accel_current}

We have considered the possible relation between acceleration of energetic particles and their location with respect to structures of current density and electric field.
In general, this is a very complex problem, because it requires characterizing a small sample of individual particles that succeed in achieving high energies by interacting with dynamical 3D electric fields and currents.
We are attempting to~illustrate this relation with~only a~few snapshots out~of~many that we have produced and examined.

\begin{figure}
\includegraphics[width=0.495\columnwidth]{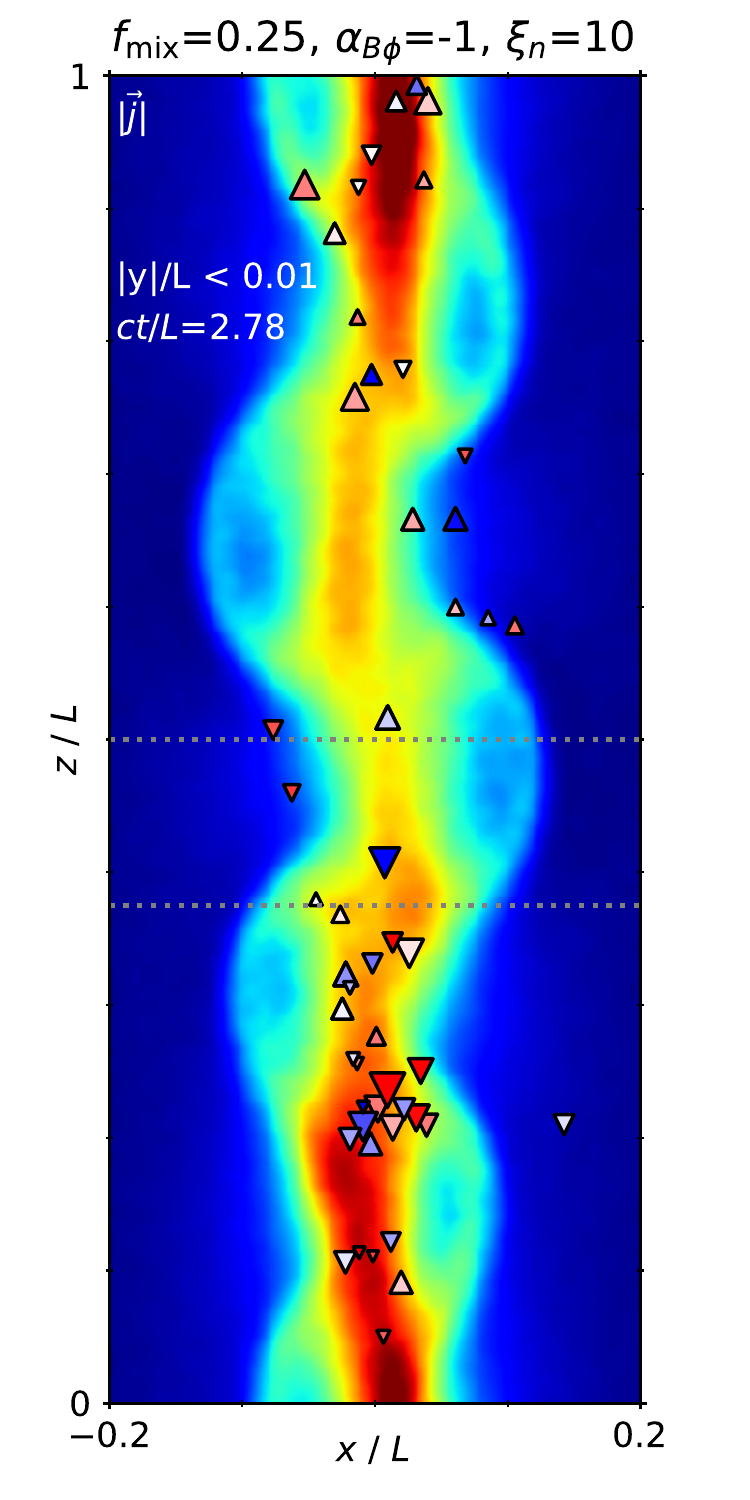}
\includegraphics[width=0.495\columnwidth]{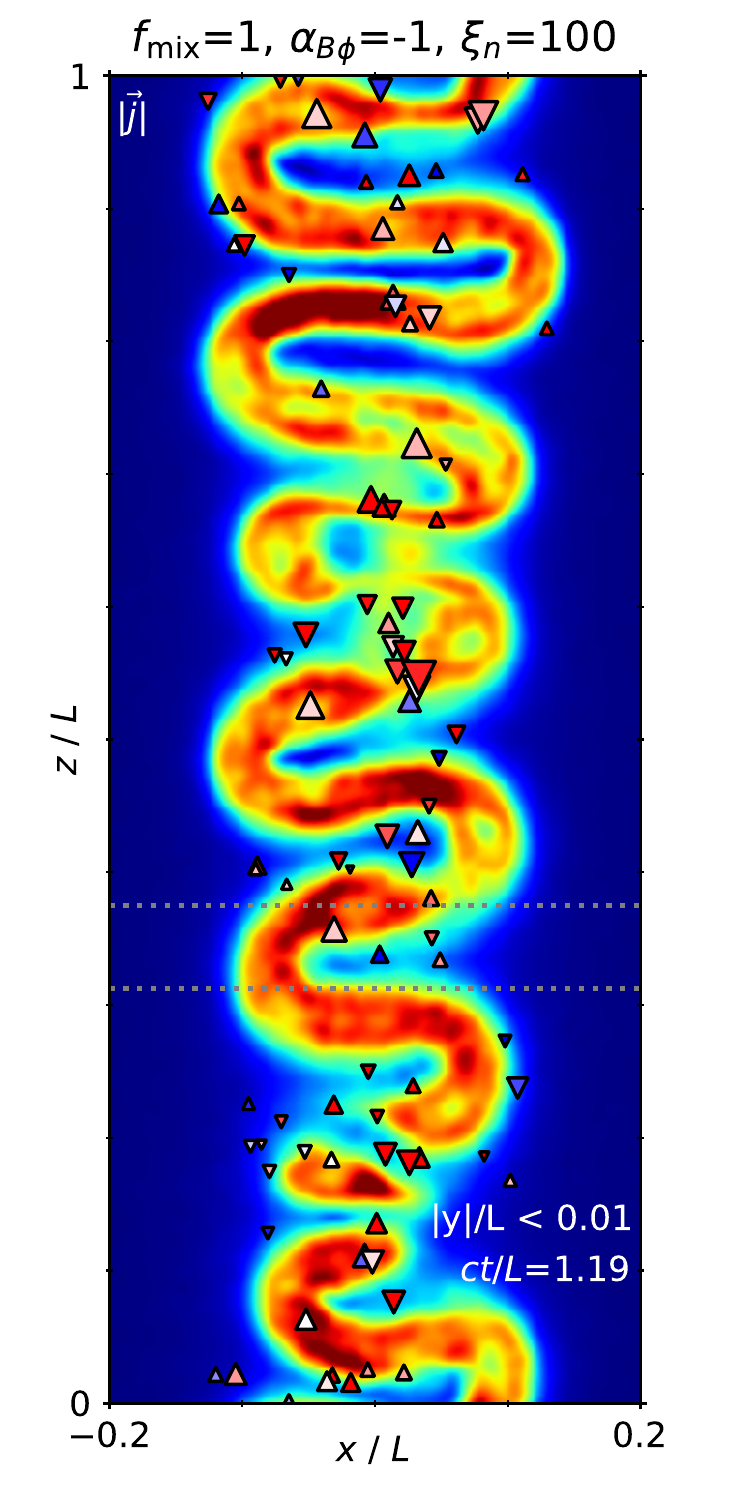}
\includegraphics[width=0.495\columnwidth]{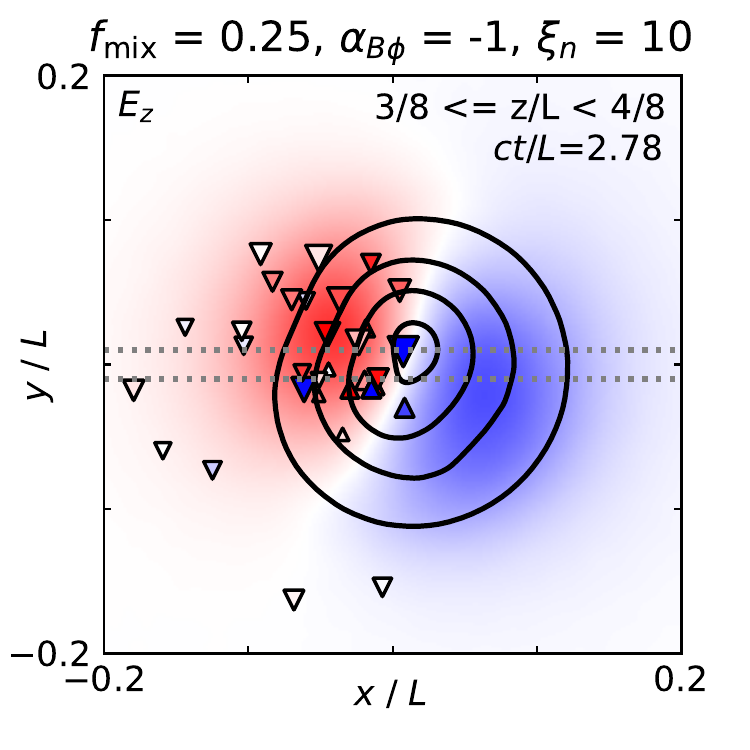}
\includegraphics[width=0.495\columnwidth]{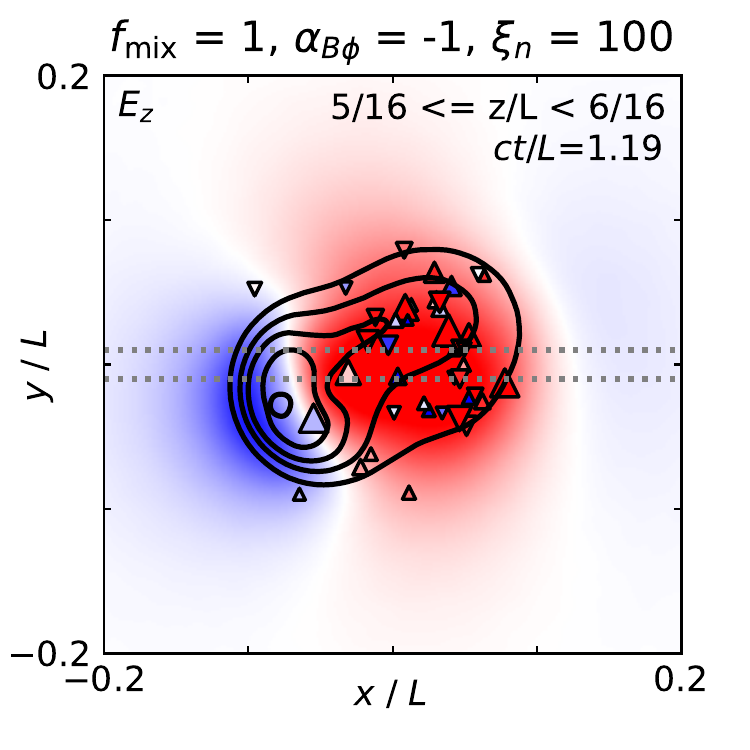}
\caption{The \emph{upper panels} show $(x,z)$ maps of total current density $|\bm{j}|$ (arbitrary units; the color scale is the same as in Figure \ref{fig_phimodes_r0-10}) in the $y/L \simeq 0$ plane for the simulations f025\_$\alpha$-1\_$\xi$10 (cf. the upper left panel of Figure \ref{fig_xzmaps_rmsEz_peak}) and f1\_$\alpha$-1\_$\xi$100.
The symbols indicate the positions of individual energetic particles: triangles pointing up for the positrons, triangles pointing down for the electrons, the sizes indicate the current particle energy, the colors indicate the current particle energization rate (red means energy gain, blue means energy loss).
The gray dotted lines indicate the range of $z$ used for integrating the $(x,y)$ maps shown below.
The \emph{lower panels} show $(x,y)$ maps of axial electric field $E_z$ (red means $E_z > 0$, blue means $E_z < 0$) averaged over the indicated range of $z/L$.
The gray dotted lines indicate the range of $y$ used for integrating the $(x,z)$ maps shown above.
}
\label{fig_xy_xz_j_Ez}
\end{figure}

Figure \ref{fig_xy_xz_j_Ez} presents single snapshots for~two simulations (f025\_$\alpha$-1\_$\xi$10 and f1\_$\alpha$-1\_$\xi$100, chosen because they probe different values of~$f_{\rm mix}$ and~produce relatively large samples of~energetic particles), each illustrated by~an~$(x,z)$ map (in~the~$y/L \simeq 0$ plane) of~total current density~$|\bm{j}|$,
and~by~an~$(x,y)$ map (averaged over a~range of~$z/L$) of~axial electric field~$E_z$ with overlaid contours of~$|\bm{j}|$.
In~all maps we~indicate the~locations, energies and~acceleration rates of~individual particles that are~present in~the~probed volume regions and~that are~going to~become energetic (according to~the~definition given in~Section~\ref{sec_res_accel_Epara}).

One question that we are attempting to address is~whether there are thin current layers that could~be sites of~magnetic reconnection.
We~find in~general that structures of~current density are~not sharp on~kinetic scales ($\sim \rho_0$).
A~particularly complex meandering $(x,z)$ structure of~$|\bm{j}|$ can~be seen in~the~f1\_$\alpha$-1\_$\xi$100 simulation, it~results from shearing of~an~initially cylindrical current core while retaining its~initial thickness scale ($\sim R_0$).

Another question is whether the~locations of~energetic particles are correlated with these current density structures.
We do not find any evidence for~that.
However, the~$(x,y)$ maps suggest that energetic particles are~preferentially located in~regions of~$E_z > 0$ while avoiding regions of~$E_z < 0$.
We~know already from Figure~\ref{fig_xzmaps_rmsEz_peak} that there is no~symmetry between regions of~positive and~negative $E_z$.
The~former ($E_z>0$) dominate and are~better connected, which allows local particles to~spend more time in~these acceleration zones.
We~thus find that the locations of~particles undergoing successful acceleration are~related more strongly to~the~structures of~electric field, rather than current density.

\section{Discussion}
\label{sec_disc}

The first motivation for this project has been to~bridge the~diverse magnetic pinch configurations investigated recently with 3D kinetic numerical simulations: the~Z-pinch case (with the~toroidal magnetic fields  balanced entirely by~the~gas pressure) studied by~\cite{2018PhRvL.121x5101A,2019PhPl...26g2105A}, and the~FF screw-pinch case (with the~toroidal magnetic fields balanced entirely by~the~axial magnetic field) studied by~\cite{2020ApJ...896L..31D}.
To this end, we introduced the~pressure mixing parameter $f_{\rm mix}$ such that $f_{\rm mix} = 0$ corresponds to~the~FF screw-pinch limit, and $f_{\rm mix} = 1$ corresponds to~the~Z-pinch limit.
This allowed us to~investigate the~effect of~$f_{\rm mix}$ for exactly the~same radial profiles of~toroidal magnetic field $B_\phi(r)$.
In the~case $f_{\rm mix} = 0$, we have found particle acceleration to~be inefficient, with~$\gamma_{\rm max}$ increasing by~only $\simeq 10\%$ (see Figure~\ref{fig_gamma_max}; note that this simulation ran until $t \simeq 15L/c$ without any further increase of~$\gamma_{\rm max}$).
A~key difference from the~setup of~\cite{2020ApJ...896L..31D} is~that they initialized the~plasma as~relativistically cold with $\Theta_0 = 10^{-2}$, and~our plasma is~initialized as~relativistically hot with~$\Theta_0 = 10^4$.
In~their simulations, particles reach Lorentz factors $\gamma \sim \sigma$ with magnetizations $\sigma \sim 10\,\text{---}\,40$.
In~our case $f_{\rm mix} = 0$ we~have peak magnetization of~only $\sigma_{\rm hot}(r=0) \simeq 2.8$,
much less than the~initial Maxwell-J\"{u}ttner value of~$\gamma_{\rm max}/\Theta_0 \simeq 20$;
we~think that this is~the~reason for~inefficient particle acceleration.
Nevertheless, we achieve higher initial magnetizations (cf. the~right panels of~Figure~\ref{fig_config_fmix1}),
and hence efficient particle acceleration, in~other cases.
Already for $f_{\rm mix} = 0.25$ (and higher), the most energetic particles achieve the Hillas-type energy limit $\gamma_{\rm lim}$ introduced by \cite{2018PhRvL.121x5101A}.

Our second motivation has been to~include a~flexible power-law section in~the~radial profile of~toroidal magnetic field $B_\phi(r)$.
Both \cite{2018PhRvL.121x5101A,2019PhPl...26g2105A} and \cite{2020ApJ...896L..31D} investigated steeply decaying $B_\phi(r)$ profiles beyond the~core radius $R_0$.
In the~case of~\cite{2018PhRvL.121x5101A,2019PhPl...26g2105A} it was an exponential tail $B_\phi(r \gg R_0) \propto \exp(-r/R_0)$,
and in~the~case of~\cite{2020ApJ...896L..31D} it was a~family of~profiles with approximately $\propto r^{-1}$ tails
(cf. the~left panel of~Figure~\ref{fig_config_fmix1}).
In the~Z-pinch limit ($f_{\rm mix} = 1$), analytical predictions of~\cite{1998ApJ...493..291B} and \cite{2019MNRAS.482.2107D} are that power-law profiles $B_\phi(r) \propto r^{\alpha_{\rm B\phi}}$ should be locally unstable for $\alpha_{\rm B\phi} > -1$.
Such shallow-decay (or even flat) profiles of~$B_\phi(r)$  have not been studied before by~means of~kinetic simulations.\footnote{A `sinusoidal' profile of~$B_\phi(r) \propto [1-\cos(2\pi r/R_{\rm out})]$ that was investigated with RMHD simulations by~\cite{2012MNRAS.422.1436O} is~included in~the~left panel of~Figure~\ref{fig_config_fmix1}. It~has a~similar symmetry to~our case $\alpha_{\rm B\phi} = 0$, but very different asymptotics at $r < R_0$.}

Despite the~modest numerical scale separation of~our simulations, we were able to~confirm most of~the~previous results, especially those of~\cite{2018PhRvL.121x5101A} in~the~Z-pinch limit --- the~structure of~electric fields in~the~linear instability phase and the~existence of~the~particle energy limit $\gamma_{\rm lim}$.
In the~case of~shallow $B_\phi(r)$ profiles
($\alpha_{\rm B\phi} > -1$),
this limit needs to~be redefined to~$(R_{\rm B\phi}/R_0)\gamma_{\rm lim}$, introducing a~new characteristic radius $R_{\rm B\phi}(\alpha_{\rm B\phi}) > R_0$.
In~Section~\ref{sec_config} we~suggested preliminary values of~this radius: $R_{\rm B\phi} \simeq 1.55R_0$ for $\alpha_{\rm B\phi} = -0.5$, and $R_{\rm B\phi} \simeq 2.5R_0$ for $\alpha_{\rm B\phi} = 0$.
However, our simulations for $\alpha_{\rm B\phi} > -1$ had to~be interrupted early, since the~perturbations reached the~domain boundaries
before the~fast magnetic dissipation phase and the~associated particle acceleration were complete.
Simulating a~complete fast magnetic dissipation phase would require, e.g., shifting the~outer cut-off to~an intermediate radius.

We argue that the~particle energy limit $\gamma_{\rm lim}$ should not be interpreted as resulting directly from particle confinement by~toroidal magnetic fields, because we have not found any example of~an energetic particle (out of~$6\times 10^5$ individually tracked particles per simulation), the~acceleration of~which would be interrupted by~its escape from the~inner radii.
Efficient particle acceleration coincides with the~fast magnetic dissipation phase, which in~the~cases of~$\alpha_{\rm B\phi} \le -1$ is of~well-defined duration, and~is also confined to~the~inner region of~$r \lesssim 2R_0$ (see Figures~\ref{fig_rprof_Bphi_meanEz} and~\ref{fig_phimodes_r0-10}).
 However, in~the~cases of~$\alpha_{\rm B\phi} > -1$, magnetic dissipation can propagate to~larger radii ($r > 2R_0$), and~induce widespread and~sustained electric fields.

In the~most extreme Z-pinch case f1\_$\alpha$0\_$\xi$100, which provides the~highest magnetization in~the~outer regions, we find a~weak signature of~a~local pinch mode at the~intermediate radii of~$(5\,\text{---}\,7)R_0  \simeq R_{\rm out} / 2$ (see Figure~\ref{fig_phimodes_r0-10}).
This is~the~first numerical confirmation that the~Z-pinch modes identified analytically by~\cite{1998ApJ...493..291B} can~be truly local in~the~sense that $k_z r_{\rm c} \gg 1$ and~$\sigma_r \ll r_{\rm c}$ (for a~mode centered at~$r=r_{\rm c}$ with~radial dispersion $\sigma_r$; this was predicted in~the~linear limit by~\citealt{2019MNRAS.482.2107D}).
 The fact that this mode has~been identified in~this particular case at~these particular radii is~consistent with the~solutions of~the~local dispersion relation of~\cite{1998ApJ...493..291B} presented in~Figure~\ref{fig_disp_begelman98} and~described in~Appendix~\ref{app_disp_begelman98}.
That weak local mode is eventually dominated by~a~stronger pinch mode propagating outward from the~central core region.

\begin{figure*}
\includegraphics[width=\textwidth]{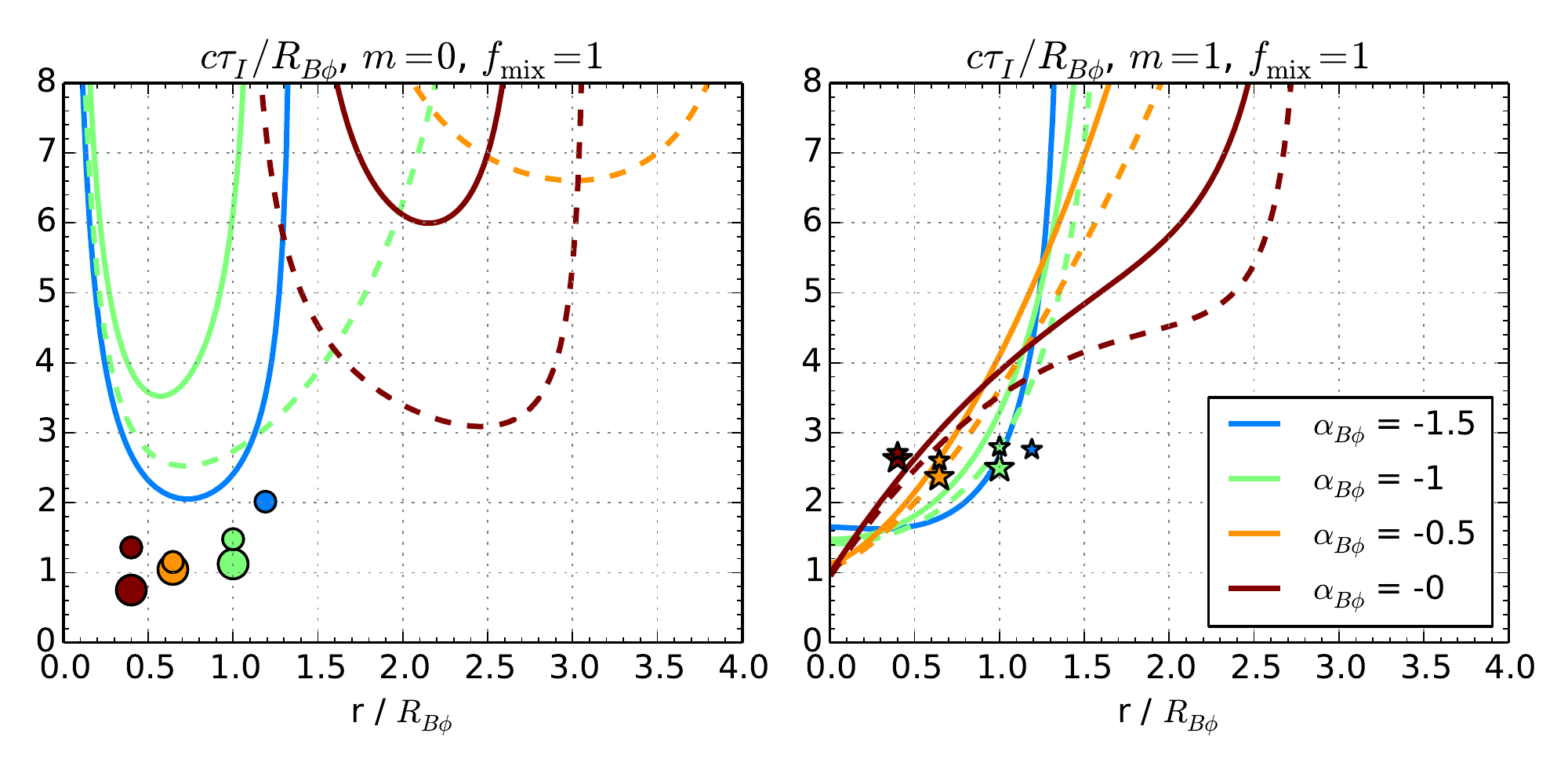}
\caption{Growth timescales $\tau_{\rm I}$ of~the~local instability modes for the~initial configurations of~the~$f_{\rm mix} = 1$ Z-pinch cases calculated according to~the~dispersion relation of~\cite{1998ApJ...493..291B}.
The left panel shows the~results for the~$m = 0$ pinch mode, and the~right panel shows the~results for the~$m = 1$ kink mode.
The line colors indicate the~value of~$\alpha_{\rm B\phi}$.
The solid lines correspond to~$\xi_n = 10$, and the~dashed lines correspond to~$\xi_n = 100$.
 Minimum growth time scales $c\tau_{\rm min}/R_{\rm B\phi}$ measured at the~$1 < r/R_0 < 2$ shell region of~our simulations are indicated at $r = R_0$ using the~same symbols as~in~Figure~\ref{fig_tau_min} (the~smaller symbols correspond to~$\xi_n = 10$ and the~larger ones to~$\xi_n = 100$).}
\label{fig_disp_begelman98}
\end{figure*}

\subsection{Parallel vs. perpendicular acceleration}
\label{sec_disc_Epara}

Particle acceleration by~parallel electric fields in~strongly magnetized jets with axial magnetic flux has been demonstrated by~\cite{2020ApJ...896L..31D}, who attribute these fields to~magnetic reconnection.
They show examples of~magnetic X-points in~the~$(x,z)$ plane along the~outer fronts of~perturbation, where the~axial magnetic field component is~reversed on~the~perturbation side ($B_z < 0$), interacting with~$B_z > 0$ in~the~unperturbed medium;
and~other magnetic irregularities in~the~$(x,y)$ plane.

Such X-points can~be seen clearly in~our simulation f05\_$\alpha$-1\_$\xi$100, where Figure~\ref{fig_xzmaps_f05_aBp1_nr100} panel~(a)  shows closed  yellow contours  in the $(x,z)$ plane, along which $B_z = 0$, meaning that $B_z > 0$ outside (like in~the~entire domain in~the~initial configuration) and $B_z < 0$ inside.
We also show using the~{\tt matplotlib.pyplot.streamplot} tool (which by~default does not illustrate the~field strength) that the~inner (with~respect to~the~central axis) sections of~those contours include a~magnetic O-point in~the~$(x,z)$ plane, and~the~outer sections include a~magnetic X-point, as~has~been shown by~\cite{2020ApJ...896L..31D}.
These magnetic X-points are potential sites of~magnetic reconnection.
Since the~$B_y \equiv B_\phi$ component is~smooth across the~X-points (a~finite guide field~$B_{\rm g}$),
 one would expect the~reconnection-induced non-ideal out-of-plane electric field to~have a~component parallel to~the~local magnetic field.
However, Figure~\ref{fig_xzmaps_f05_aBp1_nr100} shows that $E_y$ is~very weak along the~outer sections of~the~$B_z = 0$ contours, and~moreover, that $\bm{E}\cdot\bm{B}$ is~also insignificant there, as~compared with the~inner regions.
We~have also checked the~total non-ideal electric field $\bm{E}_{\rm nonid}$ and the~current density $\bm{j}$, finding that~they are~all very weak at~the~magnetic X-points.
We~therefore conclude that these magnetic X-points are not sites of~active magnetic reconnection.
Note that the~relative strength of~the~reconnecting field component is $|B_z| \sim 0.2|B_\phi|$ inside the~$B_z=0$ contours, which corresponds to~a~very strong guide field of~$B_{\rm g} \sim 5|B_z|$, sufficient to~suppress magnetic reconnection and particle acceleration \citep[e.g.,][]{2016PhPl...23l0704D,2017ApJ...843L..27W,2021JPlPh..87f9013W}.

Let us also discuss the~origin of~the~positive $\bm{E}\cdot\bm{B}$ region that accelerates positron \#930 in~the~non-linear stage of~simulation f05\_$\alpha$-1\_$\xi$100,  presented in~Figure~\ref{fig_xzmaps_f05_aBp1_nr100} and~described in~Section~\ref{sec_res_modes_fftz}.
Additional analysis reveals that non-zero $\bm{E}\cdot\bm{B}$ first appears during the~saturation of~the~linear instability stage around $t \simeq 1.8L/c$, with positive $\bm{E}\cdot\bm{B}$ aligned with the~deformed column of~strong electric current, and with negative $\bm{E}\cdot\bm{B}$ outside that column.
In this simulation we also observe a~second generation of~the~linear central instability, starting at $t \simeq 2.3L/c$, which is evidenced in~the~history of~the~energetic positron \#930 (Figure~\ref{fig_history_ion930_f05_aBp1_nr100}) as~a~major increase of~the~total electric field following that moment.
This~second instability also generates the~inner structures of~charge density (within $-0.15 < x/L < 0.15$) that can be seen in~the lower right panel of~Figure~\ref{fig_xzmaps_f05_aBp1_nr100}.
We~conjecture that the~centrally located patch of~positive $\bm{E}\cdot\bm{B}$ seen for $2.5 < ct/L < 2.95$ results from the~non-linear saturation of~the~second-generation instability.
The energetic positron \#930 experiences the~most efficient parallel acceleration because it happens to~interact with that patch for as long as possible.

\subsection{Instability growth timescales}
\label{sec_disc_tau}

In the~FF case $f_{\rm mix} = 0$, the~measured minimum instability time scales $\tau_{\rm min}$ reported in~the~right panel of~Figure~\ref{fig_tau_min} can be compared with the~analytical predictions of~\cite{2000A&A...355..818A}.
In the~case of~constant magnetic pitch $\mathcal{P}_0 = R_0$,
the shortest growth time scale is theoretically predicted for the~$|m| = 1$ kink mode: $c\tau/R_0 \simeq 7.52/\beta_{\rm A} \simeq 8.78$, where $\beta_{\rm A} \simeq 0.857$ is the~Alfv\'en velocity at $r = 0$ in~our $f_{\rm mix} = 0$ configuration.
This is very close to~our measurement of~$c\tau/R_0 \simeq 8.45$~\footnote{ The~corresponding analytical wavelength is~$\simeq 8.43 R_0$, very close to~our effective wavelength of~$\lambda_z \simeq 8.8R_0$ reported in~Section~\ref{sec_res_modes_fftz}.}.
 However, we also measure shorter time scales for the~$m = 0$ and $m = 2$ modes in~this case, which is inconsistent with the~analytical predictions, according to~which the~pinch mode should be stable, and the~$m = 2$ mode should have a~longer growth time scale (and a~wavelength that would hardly fit in~our domain).
This suggests that the~measured $m = 0$ and $m = 2$ modes are not linear, but instead they are secondary modes triggered non-linearly by~the~linear kink mode.
This interpretation is supported by~the~fact that the~onsets of~the~$m = 0$ and $m = 2$ modes are delayed with respect to~the~onset of~the~kink mode, and this is actually evident in~most other cases reported in~Figure~\ref{fig_phimodes_r1-2}.

In the~Z-pinch cases of~$f_{\rm mix} = 1$, the~measured values of~$\tau_{\rm min}$ can be compared with the~solutions of~the~local dispersion relation of~\cite{1998ApJ...493..291B}, reported in~Figure~\ref{fig_disp_begelman98} and described in~Appendix~\ref{app_disp_begelman98}.
Although the~analytical solutions can be very steep functions of~$r$, a~fairly close agreement is found for the~$m = 1$ kink mode between the~measured values extracted from the~$1 < r/R_0 < 2$ shell and the~theoretical solutions evaluated at $r = R_0$.
For the~$m = 0$ pinch mode, the~measured values are shorter than the~theoretical solutions even for $\alpha_{\rm B\phi} \le -1$.
This again suggests that only the~kink mode measured in~our simulations is linear.

\subsection{Astrophysical implications}
\label{sec_disc_astro}

The potential for~development of~instabilities due to~toroidal magnetic field has~been an~important question in~the~theoretical picture of~magnetized astrophysical jets.

The~toroidal component of~ordered magnetic fields is~an~essential ingredient of~relativistic jets.
As~the~jets are~rooted in~rotating structures (e.g., spinning black holes), toroidal fields are generated by azimuthal shearing of~poloidal fields, they provide the~magnetic pressure that accelerates the~jet, and~they carry the~outgoing Poynting flux \citep[e.g.,][]{2020ARA&A..58..407D}.
Toroidal fields provide a~tension force that in~principle would allow to~collimate (pinch) the~jet\footnote{Although, in~the~acceleration stage of~relativistic jets, collimation by~external pressure is~more important in~determining the~final jet opening angle \citep{2010NewA...15..749T,2010MNRAS.407...17K}.}, and~to~maintain a~structure of~radially decreasing (away from the~jet axis) total pressure and~energy density \citep[e.g.,][]{1984RvMP...56..255B}.
In~the~lateral jet expansion, toroidal fields decay more slowly than poloidal fields, hence, the~pinching effect of~toroidal fields can~be expected to~increase with~distance along the~jet \citep[e.g.,][]{1999MNRAS.305..211B}.

The~relative importance of~current-driven and~pressure-driven modes in~astrophysical jets depends crucially on~the~strength, lateral distribution, and~evolution along the~jet of~the~poloidal magnetic field.
\cite{2020ApJ...896L..31D} argued that FF configurations with significant poloidal fields are~naturally expected in the relativistic jets emerging from~the~bulk-acceleration and~collimation zone, protected from external modes by~the~lack of~causal contact across the~jet \citep{2016MNRAS.456.1739B,2016MNRAS.461L..46T}.
Causality would~be regained due to~recollimation once the~external pressure becomes important, and~that would make FF~jets unstable, at~first to~the~current-driven modes.

It~should~be noted, however, that whether causality is~lost in~a~relativistic jet depends on~the~scaling of~external pressure $P_{\rm ext}$ with distance $z$ along the~jet.
\cite{2015MNRAS.452.1089P} showed that causality is~lost only when $P_{\rm ext}(z)$ is~steeper than~$z^{-2}$, and even in~such case the~jet core can~be pinched by~toroidal magnetic field, triggering an~internal instability.


In order for the Z-pinches to~operate in~jets, sufficient gas pressure needs to~build~up, presumably due to~other heating mechanisms, e.g., internal shocks \citep[e.g.,][]{2001MNRAS.325.1559S,2018MNRAS.477.2376P}, recollimation shocks \citep[e.g.,][]{2009ApJ...699.1274B,2009MNRAS.392.1205N}, magnetic reconnection due to global field reversals \citep[e.g.,][]{1997ApJ...484..628L,2011MNRAS.413..333N,2019MNRAS.484.1378G}, or~non-linear saturation of~the~current-driven modes, in~effect boosting their dissipation efficiency.
 Gas pressure can also be~reduced by~radiative cooling, especially in~the~jets hosted by~powerful quasars and~GRBs, or~due to~adiabatic expansion.
This suggests that Z-pinches can~only operate in~the~vicinity of~gas pressure sources.

At any distance $z$ along the~jet, the strength of toroidal magnetic field must peak at~some radius $R_{B\phi}$, intermediate when compared with the~jet radius: $0 < R_{B\phi}(z) < R_{\rm j}(z)$.
Our study shows that the $R_{B\phi}(z)$ function is of~considerable interest, because it~largely determines the minimum growth time scale $\tau_{\rm min}$ of the instabilities (see Figure \ref{fig_tau_min}).
Let~us then consider qualitatively development of~an~initially FF jet --- as~it expands, internal pinching is~expected to~reduce the~$R_{B\phi}/R_{\rm j}$ ratio.
In~the~FF jet core, the~instability growth time scale is roughly $\tau(z) \sim 9R_{B\phi}(z)/c$, it~would evolve much slower than the~jet crossing time scale $R_{\rm j}(z)/c$.
It~is then quite likely that this first instability will be able to evolve non-linearly and to~saturate.
At~this point, a~fraction of~the~inner toroidal field will~be dissipated, tending to~increase~$R_{B\phi}$, most likely in~a~fashion similar to~the~results shown in~Figure~\ref{fig_rprof_Bphi_meanEz} for two Z-pinch cases.
Another effect of~the~first instability is~generation of~gas pressure, effectively increasing $f_{\rm mix}$, so~that the~jet core is no~longer FF.
If~the~gas pressure is~significant compared with the~axial magnetic pressure, it~would reduce the~instability growth time scale noticeably (Figure~\ref{fig_tau_min} shows that already for~$f_{\rm mix} = 0.25$ the~growth time scale of~the~kink mode is~reduced by~$\sim 40\%$ compared with~the~FF limit~$f_{\rm mix} = 0$).
The~overall outcome of~this scenario depends on~the~relative importance of: (1)~pinching of~the~jet core by~outer toroidal fields, (2)~dissipation of~the~inner toroidal fields, and~(3)~production of~gas pressure.
This problem should be addressed by~future numerical simulations of~initially FF jets with a significantly larger separation of scales $L/R_0$.




%

We~have also demonstrated numerically the~presence of~a~weak pinch mode localized at~intermediate radii in~the~case $f_{\rm mix} = 1$ and~$\alpha_{B\phi} = 0$, as~predicted by~\cite{1998ApJ...493..291B} and~\cite{2019MNRAS.482.2107D}.
 Such local modes can~operate in~the~outer jet regions even when their cores are~relatively stable.
 In~particular, flat $B_\phi(r)$ profiles, decreasing with~$r$ more slowly than~$r^{-1}$ ($\alpha_{B\phi} > -1$), would~be susceptible to~these modes.
Such flat $B_\phi(r)$ profiles may~develop in~relativistic jets at~large distances, as~was found in~global 3D~RMHD simulations by~\cite{2016MNRAS.456.1739B}\footnote{Although, it~appears that instability is~suppressed in~that case by~a~strong poloidal field \citep{2019MNRAS.482.2107D}.}.

\section{Conclusions}
\label{sec_conc}

We~have presented the~results of~3D kinetic numerical simulations of~cylindrical static jets with toroidal magnetic fields in~relativistic pair plasma.
Our simulations were initiated from configurations based on power-law profiles of $B_\phi(r) \propto r^{\alpha_{B\phi}}$, with~the~toroidal field index $-1.5 \le \alpha_{B\phi} \le 0$, modified by~inner and~outer cut-offs.
The~toroidal field was balanced by a~combination of~axial magnetic field $B_z(r)$ and~gas pressure $P(r)$,  whose relative importance was parametrized by~the~pressure mixing parameter $0 \le f_{\rm mix} \le 1$,
such that $f_{\rm mix} = 0$ corresponds to~the~force-free screw-pinch case with~uniform gas pressure,
and~$f_{\rm mix} = 1$ corresponds to~the~Z-pinch case with~$B_z(r) = 0$.
The~initial hot magnetizations were up~to~$\sigma_{\rm hot} \simeq 8$ locally.

We~found that all investigated cases were unstable,
with the~$m=1$ kink mode being either dominant (for~$f_{\rm mix} < 1$)
or~comparable to~the $m=0$ pinch mode (for~$f_{\rm mix} = 1$).
The~minimum linear growth time scale $\tau_{\rm min}$ for~the~kink mode in~$E_z$, as~well~as the~effective axial wavelength $\lambda_z$, were found to~decrease systematically with~increasing $f_{\rm mix}$.
In the~case $f_{\rm mix} = 1$ and~$\alpha_{\rm B\phi} = 0$, we~have found a~weak $m = 0$ pinch mode localized at~intermediate radial distances, consistent with the~local dispersion relation of~\cite{1998ApJ...493..291B}.

These instabilities are~associated with~dissipation of~toroidal magnetic flux $\Psi_{B\phi}$, which typically proceeds in~two~phases: a~fast magnetic dissipation phase followed by~a~slow one.
The~fast magnetic dissipation phase drives~efficient particle acceleration.
For~shallow toroidal field profiles ($\alpha_{B\phi} \ge -0.5$), magnetic dissipation proceeds more slowly, and~the~fast magnetic dissipation phase is~typically not~complete before the~perturbations reach the~outer boundaries.

Particle acceleration is dominated by~electric fields perpendicular to~the~local magnetic fields.
Acceleration by~parallel electric fields is possible in~the~central core region ($r < R_0$) in~the~presence of~axial magnetic fields ($f_{\rm mix} < 1$), however, strong guide fields suppress the~efficiency of~magnetic reconnection in~the~outer regions. 
While current density $|\bm{j}|$ forms complex volumetric structures, we have not identified kinetically sharp current layers.
 For~steep toroidal field profiles ($\alpha_{B\phi} \le -1$), the~most energetic particles reach the~confinement energy limit $\gamma_{\rm lim} = eB_0R_0/mc^2$.
On the other hand, for~shallow toroidal field profiles ($\alpha_{B\phi} \ge -0.5$), the~most energetic particles approach a~rescaled energy limit of~$(R_{B\phi}/R_0)\gamma_{\rm lim}$, where $R_{B\phi}$ is~related to~the~peak radius of~the~initial $B_\phi(r)$ function ($R_{B\phi} = R_0$ for~$\alpha_{B\phi} = -1$).

We have thus confirmed most of~the~results of~previous kinetic simulations in~the~Z-pinch limit \citep{2018PhRvL.121x5101A} and in~the~FF screw-pinch limit \citep{2020ApJ...896L..31D}.
 These previous studies represent two special cases among many possible internal jet configurations.
 In~contrast, our work is~more general, as~we demonstrated how these previous results can be~bridged by~investigating the~cases of~mixed pressure balance.
Thanks to~the~capabilities of~the~{\tt Zeltron} code \citep{Cer13},
this investigation can be extended to~include the~effect of~radiative cooling due to~synchrotron and inverse Compton processes on~particle acceleration \citep{2018JPlPh..84c7501N,2019MNRAS.482L..60W,2020MNRAS.493..603Z},
and to~calculate the~radiative output including multiwavelength light curves and linear polarization \citep{2016ApJ...828...92Y}.
This could potentially identify a~unique signature of~these instabilities in~the~vast observational data on~blazars \citep[e.g.,][]{2016ARA&A..54..725M}, a~subclass of~AGNs dominated by non-thermal emission from relativistic jets.

\acknowledgments
These results are based on numerical simulations performed at the~supercomputer {\tt Prometheus} located at the~Academic Computer Centre `Cyfronet' of~the~AGH University of~Science and Technology in~Krakow, Poland (PLGrid grants {\tt pic19,plgpic20,plgpic21,ehtsim});
and at the~computing cluster {\tt Chuck} located at the~Nicolaus Copernicus Astronomical Center of~the~Polish Academy of~Sciences in~Warsaw, Poland.
This work was supported by~the~Polish National Science Centre grants 2015/18/E/ST9/00580 and 2021/41/B/ST9/04306, and~by~the~U.S. National Science Foundation grants AST~1903335 and~AST~1806084.
B.M.~acknowledges support from~DOE through the~LDRD program at~LANL and~from~the~NASA Astrophysics Theory Program.


\appendix

\section{Particle confinement by~toroidal magnetic field}
\label{app_conf}

Consider a~radial profile of~toroidal magnetic field $B_\phi(r)$ in~the~form given by~Equation (\ref{eq_Bphi}), but without an outer cutoff.
In the~absence of~electric fields, a~relativistic particle of~mass $m$, charge $q = \pm e$, constant Lorentz factor $\gamma \gg 1$,
 dimensionless velocity $\bm\beta = \bm{v}/c = [\beta_r,\beta_\phi,\beta_z]$
(parametrized as $\beta_r \simeq \sin\theta$, $\beta_\phi = 0$, $\beta_z \simeq \cos\theta \equiv \mu$),
 and momentum $\bm{p} = \gamma\bm\beta mc$
would propagate
 under the Lorentz force ${\rm d}\bm{p}/{\rm d}t = q(\bm\beta\times\bm{B}) = qB_\phi(r)[-\cos\theta,0,\sin\theta]$.
Noting that ${\rm d}\bm{p}/{\rm d}t = \gamma mc({\rm d}\bm\beta/{\rm d}t) = \gamma mc^2\beta_r({\rm d}\bm\beta/{\rm d}r) = \gamma mc^2\beta_r[\cos\theta,0,-\sin\theta]({\rm d}\theta/{\rm d}r)$ and ${\rm d}\mu/{\rm d}r = -\sin\theta({\rm d}\theta/{\rm d}r)$,
the particle trajectory can be
described by~the~equation
\be
\frac{{\rm d}\mu}{{\rm d}(r/R_0)} = \pm\left(\frac{\gamma}{\gamma_{\rm lim}}\right)^{-1}\frac{B_\phi(r)}{B_0}\,,
\ee
where $\gamma_{\rm lim} = eB_0R_0/(mc^2)$.
For $\alpha_{\rm B\phi} = -1$, this equation can be solved analytically, yielding
\be
\mu(r) = \mu_0 \pm \frac{1}{2}\left(\frac{\gamma}{\gamma_{\rm lim}}\right)^{-1}\ln\left[1+\left(\frac{r}{R_0}\right)^2\right]\,,
\ee
where $\mu_0 = \mu(r=0)$.
In the~specific case $q = +e$ (a positron), the~maximum radius $r_{\rm max}$ is~given by~$\mu(r_{\rm max}) = 1$.
If~we~adopt a~confinement criterion $r_{\rm max} < r_{\rm conf}$, this~can be~expressed as
\be
\gamma - u_{z,0} < \frac{\gamma_{\rm lim}}{2}\ln\left[1+\left(\frac{r_{\rm conf}}{R_0}\right)^2\right]\,,
\ee
where $u_{z,0} \simeq \gamma\mu_0$ is the~axial 4-velocity component at $r=0$.
This criterion simplifies to~$\gamma - u_{z,0} < \gamma_{\rm lim}$ for $r_{\rm conf} \simeq 2.53R_0$, which is a~reasonable threshold.
Note that $\gamma_{\rm lim}$ is relevant as confinement energy limit only for particles with $u_{z,0} \simeq 0$, i.e., crossing the~symmetry axis at the~right angle.
Particles propagating along the~axis can reach energies beyond $\gamma_{\rm lim}$ without escaping.
For analysis of~acceleration histories of~individual particles, we introduce the~\emph{particle confinement indicator} defined as $\xi_{\rm conf} = (\gamma - |u_z|)/\gamma_{\rm lim}$, so that a~particle is considered confined if $\xi_{\rm conf} < 1$.

\section{Linear growth time scales in~the~Z-pinch cases}
\label{app_disp_begelman98}

Instability growth time scales in~the~$f_{\rm mix} = 1$ Z-pinch cases can be calculated analytically using the~local dispersion relation expressed by~the~Eq. (3.32) of~\cite{1998ApJ...493..291B} in~the~limit of~$B_z = 0$ and $l \ll k$ (negligible radial wavenumber).
The dispersion relation is solved for the~initial equilibria used in~our simulations as function of~radius $r$.
Figure~\ref{fig_disp_begelman98} presents the~resulting exponential growth time scales $\tau_{\rm I} = 1/\omega_{\rm I}$ for the~$m = 0$ pinch and $m = 1$ kink modes
scaled by~the~characteristic radius $R_{\rm B\phi}$.
Higher azimuthal modes are found to~be stable in~all considered cases.
All configurations are unstable to~the~kink mode with  the~shortest growth time scales ($c\tau_{\rm I}/R_{\rm B\phi} \sim 1.0\;\text{---}\;1.6$) found within $r < R_{\rm B\phi}$.
For $\alpha_{\rm B\phi} > -1$, the~unstable region extends towards intermediate radii,  up to~$r \simeq 2.5R_{\rm B\phi}$ for the~case f1\_$\alpha$0\_$\xi$100, but with significantly longer growth time scales.
The solutions are more diverse for the~pinch mode --- the~central core region is found to~be unstable only for $\alpha_{\rm B\phi} \le -1$, and for $\alpha_{\rm B\phi} > -1$ a~distinct instability region is located at intermediate radii.
In the~case f1\_$\alpha$0\_$\xi$100, the~pinch mode shows a~growth time scale of~$c\tau_{\rm I}/R_{\rm B\phi} \simeq 3$ at $r = 2.5R_{\rm B\phi}$, which is shorter than the~local growth time scale for the~kink mode, but longer than that for the~kink mode at the~central core.

\end{document}